\documentclass[11pt,letterpaper,
onecolumn,noarxiv,accepted=2025-08-22]{quantumarticle}

\usepackage{subcaption,color,fancyvrb}
\usepackage{algorithm,algpseudocode}
\usepackage{amsmath,amsthm,amsfonts}
\usepackage[colorlinks,bookmarks=true, citecolor=blue!90!black, urlcolor=blue!90!black,linkcolor=blue!90!black, pdfencoding=auto, psdextra]{hyperref}	
\usepackage{bbm}
\usepackage{graphicx}
\usepackage{nicematrix}
\usepackage{array,braket,times,amssymb,amsmath,amsfonts,amsthm,bbm,
mathrsfs,color,enumitem,booktabs,cite,bm,tikz,eczoo}
\usetikzlibrary{quantikz2}
\newtheorem{theorem}{Theorem}[section]
\newtheorem{lemma}[theorem]{Lemma}

\newtheorem{proposition}[theorem]{Proposition}

\newtheorem{corollary}[theorem]{Corollary}
\theoremstyle{remark}
\newtheorem{definition}{Definition}[section]
\newtheorem{remark}{Remark}[section]
\newtheorem{example}[theorem]{Example}

\allowdisplaybreaks
\newcommand\nc\newcommand
\nc\oA{{\overline{A}}}
\nc\oB{{\overline{B}}}
\nc\oC{{\overline{C}}}
\nc\oD{{\overline{D}}}
\nc\oE{{\overline{E}}}
\nc\oF{{\overline{F}}}
\nc\oG{{\overline{G}}}
\nc\oH{{\overline{H}}}
\nc\oI{{\overline{I}}}
\nc\oJ{{\overline{J}}}
\nc\oK{{\overline{K}}}
\nc\oL{{\overline{L}}}
\nc\oM{{\overline{M}}}
\nc\oN{{\overline{N}}}
\nc\oO{{\overline{O}}}
\nc\oP{{\overline{P}}}
\nc\oQ{{\overline{Q}}}
\nc\oR{{\overline{R}}}
\nc\oS{{\overline{S}}}
\nc\oT{{\overline{T}}}
\nc\oU{{\overline{U}}}
\nc\oV{{\overline{V}}}
\nc\oW{{\overline{W}}}
\nc\oX{{\overline{X}}}
\nc\oY{{\overline{Y}}}
\nc\oZ{{\overline{Z}}}
\nc\oCZ{{\overline{CZ}}}
\nc\oCNOT{{\overline{CNOT}}}

\nc\ffa{{\mathbb a}}\nc\ffA{{\mathbb A}}\nc\cA{{\mathcal A}}\nc\sfA{{\mathsf{A}}}
\nc\ffb{{\mathbb b}}\nc\ffB{{\mathbb B}}\nc\cB{{\mathcal B}}\nc\sfB{{\mathsf{B}}}
\nc\ffc{{\mathbb c}}\nc\ffC{{\mathbb C}}\nc\cC{{\mathscr C}}\nc\sfC{{\mathsf{C}}}
\nc\ffd{{\mathbb d}}\nc\ffD{{\mathbb D}}\nc\cD{{\mathcal D}}\nc\sfD{{\mathsf{D}}}
\nc\ffe{{\mathbb e}}\nc\ffE{{\mathbb E}}\nc\cE{{\mathcal E}}\nc\sfE{{\mathsf{E}}}
\nc\fff{{\mathbb f}}\nc\ffF{{\mathbb F}}\nc\cF{{\mathscr F}}\nc\sfF{{\mathsf{F}}}
\nc\ffg{{\mathbb g}}\nc\ffG{{\mathbb G}}\nc\cG{{\mathscr G}}\nc\sfG{{\mathsf{G}}}
\nc\ffh{{\mathbb h}}\nc\ffH{{\mathbb H}}\nc\cH{{\mathcal H}}\nc\sfH{{\mathsf{H}}}
\nc\ffi{{\mathbb i}}\nc\ffI{{\mathbb I}}\nc\cI{{\mathcal I}}\nc\sfI{{\mathsf{I}}}
\nc\ffj{{\mathbb j}}\nc\ffJ{{\mathbb J}}\nc\cJ{{\mathcal J}}\nc\sfJ{{\mathsf{J}}}
\nc\ffk{{\mathbb k}}\nc\ffK{{\mathbb K}}\nc\cK{{\mathcal K}}\nc\sfK{{\mathsf{K}}}
\nc\ffl{{\mathbb l}}\nc\ffL{{\mathbb L}}\nc\cL{{\mathcal L}}\nc\sfL{{\mathsf{L}}}
\nc\ffm{{\mathbb m}}\nc\ffM{{\mathbb M}}\nc\cM{{\mathcal M}}\nc\sfM{{\mathsf{M}}}
\nc\ffn{{\mathbb n}}\nc\ffN{{\mathbb N}}\nc\cN{{\mathcal N}}\nc\sfN{{\mathsf{N}}}
\nc\ffo{{\mathbb o}}\nc\ffO{{\mathbb O}}\nc\cO{{\mathcal O}}\nc\sfO{{\mathsf{O}}}
\nc\ffp{{\mathbb p}}\nc\ffP{{\mathbb P}}\nc\cP{{\mathcal P}}\nc\sfP{{\mathsf{P}}}
\nc\ffq{{\mathbb q}}\nc\ffQ{{\mathbb Q}}\nc\cQ{{\mathcal Q}}\nc\sfQ{{\mathsf{Q}}}
\nc\ffr{{\mathbb r}}\nc\ffR{{\mathbb R}}\nc\cR{{\mathcal R}}\nc\sfR{{\mathsf{R}}}
\nc\ffs{{\mathbb s}}\nc\ffS{{\mathbb S}}\nc\cS{{\mathcal S}}\nc\sfS{{\mathsf{S}}}
\nc\fft{{\mathbb t}}\nc\ffT{{\mathbb T}}\nc\cT{{\mathcal T}}\nc\sfT{{\mathsf{T}}}
\nc\ffu{{\mathbb u}}\nc\ffU{{\mathbb U}}\nc\cU{{\mathcal U}}\nc\sfU{{\mathsf{U}}}
\nc\ffv{{\mathbb v}}\nc\ffV{{\mathbb V}}\nc\cV{{\mathscr V}}\nc\sfV{{\mathsf{V}}}
\nc\ffw{{\mathbb w}}\nc\ffW{{\mathbb W}}\nc\cW{{\mathscr W}}\nc\sfW{{\mathsf{W}}}
\nc\ffx{{\mathbb x}}\nc\ffX{{\mathbb X}}\nc\cX{{\mathcal X}}\nc\sfX{{\mathsf{X}}}
\nc\ffy{{\mathbb y}}\nc\ffY{{\mathbb Y}}\nc\cY{{\mathscr Y}}\nc\sfY{{\mathsf{Y}}}
\nc\ffz{{\mathbb z}}\nc\ffZ{{\mathbb Z}}\nc\cZ{{\mathcal Z}}\nc\sfZ{{\mathsf{Z}}}
\nc{\bb}{{\mathbbm{1}}}
\nc\reals{{\mathbb R}}
\nc{\ff}{{\mathbb F}}
\nc{\PP}{{\mathbb P}}
\nc{\red}[1]{{\color{red}#1}}

\DeclareMathOperator{\Img}{Im}
\DeclareMathOperator\Cliff{Cliff}
\DeclareMathOperator{\Span}{span}
\DeclareMathOperator{\Supp}{supp}

\usepackage[normalem]{ulem}
\newcommand\redout{\bgroup\markoverwith{\textcolor{red}{\rule[0.5ex]{2pt}{0.8pt}}}\ULon}

\usepackage{xargs}
\usepackage[colorinlistoftodos,prependcaption]{todonotes}
\usepackage{todonotes}
\newcommandx{\rednote}[2][1=]{\todo[linecolor=red,backgroundcolor=red!25,bordercolor=red,#1]{#2}}
\newcommandx{\bluenote}[2][1=]{\todo[linecolor=blue,backgroundcolor=blue!25,bordercolor=blue,#1]{#2}}
\newcommandx{\yellownote}[2][1=]{\todo[linecolor=yellow,backgroundcolor=yellow!25,bordercolor=yellow,#1]{#2}}
\newcommandx{\greennote}[2][1=]{\todo[inline,linecolor=olive,backgroundcolor=green!25,bordercolor=olive,#1]{#2}}

\begin{document}
	
	\title[Clifford gates for HGP codes]{Targeted Clifford Logical Gates for Hypergraph Product Codes}

\author{Adway Patra}
\affiliation{Department of ECE and Institute for Systems Research, University of Maryland, College Park, MD 20742}
\email{apatra@umd.edu}
\author{Alexander Barg}
\affiliation{Department of ECE and Institute for Systems Research, University of Maryland, College Park, MD 20742}
\email{abarg@umd.edu}
\orcid{0000-0002-8972-4413}
	
	\begin{abstract}
Starting with an explicit framework for designing logical Clifford circuits for CSS codes, 
we construct logical gates for Hypergraph Product Codes. 
We first derive symplectic matrices for CNOT, CZ, Phase, and Hadamard operators, which together generate the Clifford group. This enables us to design explicit transformations that result in targeted logical gates for arbitrary codes in this family. As a concrete example, we 
give logical circuits for the $[[18,2,3]]$ toric code.
	\end{abstract}
	\maketitle
	
	\section{Introduction}
In the class of stabilizer qubit codes, a large body of work has been devoted to the construction
and analysis of various families of quantum codes that have the additional property of low
  stabilizer weight, known as quantum low-density parity-check, or qLDPC 
codes \eczoo[]{errorcorrectionzoo.org/c/qldpc}. These codes have been drawing much attention in quantum
coding research because of theoretical results that assert the possibility of fault-tolerant computing with constant overhead \cite{gottesman2014fault,kovalev2013fault,fawzi2020constant,breuckmann2021quantum}.
This restriction implies that the code stabilizers act nontrivially only on a constant number of physical qubits, which enables low-complexity stabilizer measurement circuits. Among the early constructions in this class are the toric codes \eczoo[]{toric} \cite{kitaev1997quantum} with stabilizers defined by the vertices or faces of a square grid with periodic boundaries, their close relatives known as surface codes \cite{bravyi1998quantum}, and other topological codes \cite{Bombin2007,raussendorf2007fault}. These families rely on geometrically simple constructions and have low-weight stabilizers, but they only encode a small number of logical qubits (two for the toric code), which results in a low information rate relative to the number of physical qubits. More general families of qLDPC codes afford higher rates while still relying on stabilizers of bounded support.
Recent research on qLDPC codes has yielded spectacular advances, including code families of asymptotically nonvanishing rate and a nonvanishing proportion of correctable errors \cite{PK21b,LZ2022,DHLV2022}. 

Fault-tolerant quantum computations rely on error-correction procedures together with circuit
design that protects the computational procedure from noise propagation through faulty gates and
measurements. Logical operations performed on the encoded data should be compatible with
the overall noise protection of the computation. This gives rise to two related, equally important research problems, designing elementary logical operations in the form of explicit circuits, and constructing fault-tolerant implementations that can be realized in hardware. 
The problem of designing logical operators for quantum 
error-correcting codes has been at the forefront of quantum algorithms research since
the inception of quantum computing \cite{gottesman1998theory,steane1999efficient}.

The task of supporting universal quantum computation involves implementing a class of operators in the logical Clifford group, which together with a single non-Clifford gate suffice to provide universality.
Since this group is generated by a small number of simple gates, researchers have viewed it as a viable path toward designing practical quantum computations. 
In this paper we focus on the design of Clifford logical gates for a class of quantum stabilizer codes known as homological product codes \eczoo[]{homological_product} \cite{bravyi2014homological}, and more specifically, the hypergraph product (HGP) codes 
 \eczoo[]{hypergraph_product} of \cite{Tillich2014}. 
 These codes belong to the qLDPC family, combining simple description with good error-correcting properties. They encode a number of logical qubits that scales as the constant fraction of the code length $n$ and correct the order of $\sqrt n$ errors in the encoding. HGP codes also have a number of favorable properties such as a
 simple algebraic description, low stabilizer weight, and planar hardware implementation, which 
 places them among the leading candidates for implementing fault-tolerant computations in practice \cite{tremblay2022constantoverhead,xu2024fast}.
	
Notable prior works on the constructions of logical gates for HGP codes
include the works \cite{Krishna2021,Quintavalle2023,Breuckmann2022,xu2024fast}.
The first of these, \cite{Krishna2021}, provided simple descriptions for logical Pauli operators for HGP codes,  and \cite{Quintavalle2023} found an explicit basis for these operators.  Using it, the authors of \cite{Quintavalle2023} designed transversal global Clifford logicals\footnote{A {\em global} operator
acts nontrivially on all the logical qubits. As opposed to this, logicals with action limited to a subset of qubits are called {\em targeted}. Their structure is of interest because it provides
added flexibility in algorithm design.} for a subclass of HGP codes called symmetric codes. Their results for such codes also included a circuit implementation for certain targeted CZ and CNOT gates as well as single-qubit Hadamard and Phase gates using state injection.
The authors of \cite{Breuckmann2022} designed transversal gates for global Clifford operators for a 
certain class of CSS codes which includes symmetric HGP codes. Finally, \cite{xu2024fast} constructed targeted Clifford logicals by combining the global gates of \cite{Quintavalle2023} with two separate ancilla codes for teleporting some of the logical qubits out of the target code.  
This design approach leads to increased space complexity because it requires the order of $2n$ ancilla qubits for performing the logic. A very recent work \cite{lin2024transversalnonclifford} aims at designing non-Clifford targeted  logical operators such as CCZ for a general code family of sheaf codes, which contains many recent product constructions of qLDPC codes. Other recent papers \cite{zhu2024,breuckmann2024cupsgatesicohomology} use cohomology invariants and cup products to construct constant-depth circuits for a class of Clifford and non-Clifford logical operations in several copies of homological qLDPC codes.

An earlier paper \cite{Rengaswamy2020} on CSS code logic gave an algorithm for finding all symplectic matrices, and hence circuits, for any logical Clifford operator for general CSS codes. Their approach is built upon the well-known representation of logical Clifford operators by binary symplectic matrices \cite{calderbank1997quantum,gottesman2010introduction}. 
The algorithm of \cite{Rengaswamy2020} finds a set of linear equations based on the constraints imposed by the target logical operator and then finds {\em all} symplectic matrices that perform the action of this operator. Thereafter, one can convert any of these symplectic matrices into circuits using elementary Clifford gates \cite{Dehaene2003,Maslov2018}. However, the number of solutions to the linear system grows exponentially with the number of stabilizer generators, making it difficult to optimize the circuit parameters even for moderate-size codes.
	
In this work we take an alternative approach to finding these symplectic matrices that does not require searching through multiple possibilities, resulting in a general framework for finding 
explicit circuits for any logical Clifford operations. It is well known that the stabilizer generators of an $[[n,k]]$ CSS code form a part of a symplectic basis of $\ffF_2^{2n}$. The logical Clifford operators
perform a linear transformation between symplectic bases. By specifying this mapping of basis 
vectors we derive an explicit form of the matrix of this transformation, and these matrices
precisely implement the Clifford logical operators for the code (up to global phases).
	
Generally, this technique applies to any CSS code (or, in fact, any stabilizer code with an explicit choice of logical Pauli operators), but we rely on the structure of HGP codes to
construct circuits for single-qubit/two-qubit logical operators that span the logical Clifford group. 
This structure translates into a special form of the symplectic matrices for the logical Cliffords, allowing us to simplify the circuit building algorithm, resulting in simpler explicit procedures
for the targeted logicals compared to the state of the art. Our design works for any HGP code
and for all logical qubits of the code, lifting the limitations inherent to the procedures of 
\cite{Quintavalle2023,Breuckmann2022,xu2024fast}. We illustrate the construction by explicitly designing logical Clifford gates for a prominent example of the HGP code family, namely the toric code. The circuits that we construct do not require any additional ancilla qubits or protocols such as state injection and are built purely out of basic single and two-qubit gates. At the same time, they are not fault-tolerant by design, requiring additional error-correction ancillas to support reliable computation. We leave this challenge for future work.

The structure of the paper is as follows. After covering notation and preliminaries in the next section, we derive 
symplectic matrices for the logical operators in Sec.~\ref{sec:symplectic_framework}. Our main results, presented in Sec.~\ref{sec:FTlogical}, are concerned with an explicit set of transformations from the symplectic matrices
to the logical circuits for each of the CNOT, CZ, Phase, and Hadamard targeted (single- or two-qubit) logical operators. In Sec~\ref{sec:toric_logical} we present explicit Clifford circuits for the [[18,2,3]] toric code. Some proofs and further examples are collected in Section~\ref{sec:appnx}.
 	
\section{Preliminaries}\label{sec:prelims}
We begin with the basics on quantum stabilizer and CSS codes, their logical operators, and a description of the HGP code family in the form used in the paper.  In the paper for any natural number $n$, $[n]$ will denote the set 
$\{1,2,\dots,n\}$ while for two natural numbers $a, b$, $a \le b$, $[a:b] = \{a,a+1,\dots,b-1,b\}$. Given an $r \times n$ matrix $M$, we denote by $M(i,:)$ and $M(:,j)$ its $i$-th row and $j$-th column, respectively, and write $M(i_1:i_2,j_1:j_2)$ to refer to the submatrix of $M$ indexed by the rows in 
$[i_1:i_2]$ and columns in $[j_1:j_2]$. Finally, we denote by $|x|$ the Hamming weight of the binary vector $x$.

	\subsection{Stabilizer and CSS Codes} Let $\cS$ be an Abelian subgroup of the Pauli group $\cP_n$ on $n$ qubits.
A stabilizer code $\cC(\cS)$ is a subspace of the Hilbert space $(\ffC^2)^{\otimes n}$ that forms the common eigenspace of $\cS$ with eigenvalue $+1$ \cite{gottesman1997stabilizer}.  Below we use the notation $\cS$ to denote the stabilizer group or a set of its generators interchangeably. Let $\sfN(\cS)=\{M\in \cP_n| SM=MS\text{ for all }S\in \cS\}$ be the normalizer group of $\cS$ in the Pauli group. Of course, $\cS\subset \sfN(\cS)$, so we can consider the quotient group $\cL=\sfN(S)/\cS$. We denote by $L \in \cL$ a representative operator of its coset, which can be chosen arbitrarily. Any element $L\in\cL, L\ne I$ has the property that, for any $\ket{\psi} \in \cC(\cS)$:
     $$
     SL \ket{\psi} = LS\ket{\psi} = L\ket{\psi}, \hspace*{.2in}\;\text{for all } S \in \cS
     $$
and so $L\ket{\psi}$ is also a codeword of $\cC(\cS)$. We observe that $L$ maps codewords to codewords, i.e., it preserves
the codespace. Its elements
are Pauli operators, and they perform operations on the encoded (logical) qubits, while not taking the result
outside of the codespace. For this reason, they are called {\em logical Pauli operators}. Characterizing 
logical Paulis and similar logical operators outside the Pauli group is of primary importance to quantum 
computing as it allows operations on the encoded data that are protected from noise by the properties of the code.

CSS codes \eczoo[]{qubit_css} \cite{CS1996,S1996} form a subclass of stabilizer codes \eczoo[]{qubit_stabilizer} whose stabilizer set is 
constructed using a pair of classical binary linear codes $(C_X,C_Z)$ such that $C_X^{\perp} \subseteq C_Z$. Letting $H_X$ be an $m_X\times n$ parity-check matrix of the classical code $C_X$, we form the set of
$X$ stabilizers of the quantum code as $\otimes_{j \in [n]}X^{(H_X)_{ij}}$ for each row $H_X(i,:), i=1,\dots,m_X$ of the code $C_X$, and we do the same for the $Z$ stabilizers and the $m_Z\times n$ matrix $H_Z$.
The $X$ and $Z$ stabilizers commute by design because of the subcode assumption, which is equivalent to $H_XH_Z^\intercal  = 0$.

Equivalently, a CSS code $\cQ$  can be described in homological terms. Consider the following 3-term chain complex
	\begin{equation}\label{eq:3term}
 \ffF_2^{m_{\!{_{\tiny Z}}}} \stackrel{H_Z^\intercal }{\longrightarrow}\ffF_2^n \stackrel{H_X}{\longrightarrow} \ffF_2^{m_{\!{_{\tiny X}}}}
    \end{equation}
with boundary operators $\partial_2$ and $\partial_1$ given by the matrices $H_Z^\intercal$ and $H_X$, respectively.
By the assumption, $C_Z^{\perp} \subseteq C_X$, which translates to $H_XH_Z^\intercal  =0$, assuring
that the complex is well defined. The elements of $\ker(H_X)$ correspond to $Z$-type Paulis that commute with all $X$ stabilizers (cycles), while the elements of $\Img(H_Z^\intercal )$ correspond to $Z$-type Paulis given by
products of the $Z$-type stabilizer generators (trivial boundaries). Hence, the elements of the (first) homology group 
$H_1(\cQ,\ff_2)\cong\ker(H_X)/\Img(H_Z^\intercal )$ form nontrivial cycles (cycles without boundary), giving rise to $Z$-type logical Pauli operators on $\cQ.$ In other words, these are the $Z$-type Paulis that commute with all $X$-type stabilizers but are not themselves $Z$ stabilizers.

Flipping the arrows, we also consider the co-complex
   $$
   \ff_2^{m_{\!{_{\tiny Z}}}}\stackrel{H_Z}\longleftarrow \ff_2^n\stackrel{H_X^\intercal}\longleftarrow\ff_2^{m_{\!{_{\tiny X}}}}.
   $$
where the matrices $H_X^\intercal$ and $H_Z$ define the coboundary operators $\delta_0$ and $\delta_1$. The elements of $\ker(H_Z)$, $\Img(H_X^{\intercal})$ and the first cohomology group $H^1(\cQ,\ff_2)\cong \ker(H_Z)/\Img(H_X^\intercal )$, respectively, denote cocycles, coboundaries and nontrivial or essential cocycles, which give rise to $X$-type Pauli logical operators. This rephrasing offers a convenient way to find the logical Pauli operators of a CSS code, as fixing a basis of $H_1$ and $H^1$ fixes a basis for the logical Pauli space.

	\subsection{Clifford Groups}
	Let $\ffU_n$ denote the group of unitary matrices acting on $(\ffC^2)^{\otimes n}$. The Clifford group $\Cliff_n \subset \ffU_n$ (more precisely, the {\em physical Clifford group}) is the normalizer of the $n$-qubit Pauli group $\cP_n$, 
	$$
 \Cliff_n = \{U \in \ffU_n: UPU^{\dagger} \in \cP_n, \text{ for all } \; P \in \cP_n\}.
 $$
Recall that the Clifford group  $\Cliff_n$ can be generated by a small set of
operators, for instance, by $\{H,S,CNOT\}$ (the Hadamard, Phase, and $CNOT$ (or $CZ$) gates), which 
form a set of elementary gates that can serve as building blocks for Clifford circuits. Below we denote logical operators by adding a bar above the operator symbol, for instance, $\overline P$ refers to a
logical Pauli operator. The logical  Clifford group of an $[[n,k]]$ quantum code $\cQ$ is defined as the set of those $n$-qubit unitary operators that normalize the stabilizer group, $\cS$, and normalize the subgroup generated by the logical Paulis, $\overline{\cP}_k$, of $\cQ$,
 $$
 \overline{\Cliff}_k := \{U \in \cU^n: USU^{\dagger} \in \cS, U\overline{P}U^{\dagger} \in \overline{\cP}_k, \text{ for all } S\in \cS, \overline{P}\in 
 \overline{\cP}_k\}.
 $$
For every $U \in \overline{\Cliff}_k$ and every code state $\ket\psi$, we have $U\ket{\psi}\in\cQ$ 
and $U\overline{P}\ket{\psi} = (U\overline{P}U^{\dagger})U\ket{\psi}$ for every $\overline{P} \in \overline{\cP}_k$.

As an outcome of this discussion, consider the operators $U\in\overline{\Cliff}_k \cap \Cliff_n$. 
It is this operator set that we study in this work because  any such $U$  performs a logical operation on 
the code state (manipulates the encoded qubits) while at the same time affording a quantum circuit that can be formed
only of Clifford gates.

 \subsection{Hypergraph Product Codes}\label{subsec:HP}

A hypergraph product code of \cite{Tillich2014}, $\cQ(C_a,C_b)$, is a CSS code constructed from two classical codes $C_a[n_a,k_a,d_a]$ and $C_b[n_b,k_b,d_b]$ with parity-check matrices $H_a \in \ff_2^{m_a \times n_a}$ and $H_b \in \ff_2^{m_b\times n_b}$, where $m_a=n_a-k_a, m_b=n_b-k_b$. 
The $X$ and $Z$ stabilizers of an HGP code are given by the rows of the matrices $H_X$ and $H_Z$ as shown below:
	\begin{equation}\label{eq:hp_stab}
		H_X = \left[H_a \otimes I_{n_b} | I_{m_a} \otimes H_b^\intercal\right],\hspace{0.2in}
		H_Z = \left[I_{n_a} \otimes H_b | H_a^\intercal \otimes I_{m_b}\right].
	\end{equation}
It is easy to check that $H_X\cdot H_Z^\intercal = 2(H_a \otimes H_b^\intercal) = 0 \;(\bmod \;2)$, showing that 
$\cQ$ is a well-defined CSS code. The number of physical qubits (the length) of the code $\cQ$
equals $n=n_an_b +m_am_b$. To define the dimension and the distance, recall that a
transpose code $A^\intercal$ of a classical code $A$  is the space orthogonal over $\ff_2$ to the column space of its parity-check matrix $H$; in other words, the columns of $H$ serve as the parity relations
for $A^\intercal$. Then the dimension and distance of the code $\cQ$ equal $k=k_ak_b+k_a^\intercal k_b^\intercal$ and $d=\min\{d_a,d_b,d_a^\intercal,d_b^\intercal\}.$
As shown in \cite{Tillich2014}, with a proper choice of the constituent codes, the 
number of logical qubits $k=\Theta(n)$ and the distance scales as $\Theta(\sqrt{n})$. If the matrices $H_a$ and $H_b$ have small rows weight (give rise to classical LDPC codes), then the code $\cQ$ is itself a qLDPC code.
	
It will be convenient to view the physical qubits of the HGP code arranged in two grids (sectors) of size $n_a \times n_b$ and $m_a \times m_b$ respectively,
calling them the left and the right sector. In this convention, we can identify individual qubits by indices $(i,h,L), 1 \le i \le n_a, 1 \le h \le n_b$ and $(j,l,R), 1 \le j \le m_a, 1\le l \le m_b$, 
which refer to the row and column indices and to one of the two sectors as appropriate.  
From the definition of the stabilizers in Eq. \eqref{eq:hp_stab}, we see that in this view, each stabilizer, whether $X$ or $Z$, is supported on a single column or row of left sector and a single column or row of right sector, and within them it coincides with the support of a row or a column of $H_a$ or $H_b$.  Specifically, the factors of $X$ stabilizer generators can be indexed by a pair $(h,j)$ such that the generator $S_X(h,j)$ 
acts on a subset of qubits in column $h$ of the left sector and row $j$ of the right sector. The $Z$ stabilizer generators can similarly be indexed by pairs $(i,l)$ such that $S_Z(i,l)$ has support on row $i$ of the left sector and column $l$ of 
the right sector.
	\begin{center}
		\begin{figure}[ht]
			\includegraphics[scale=0.75]{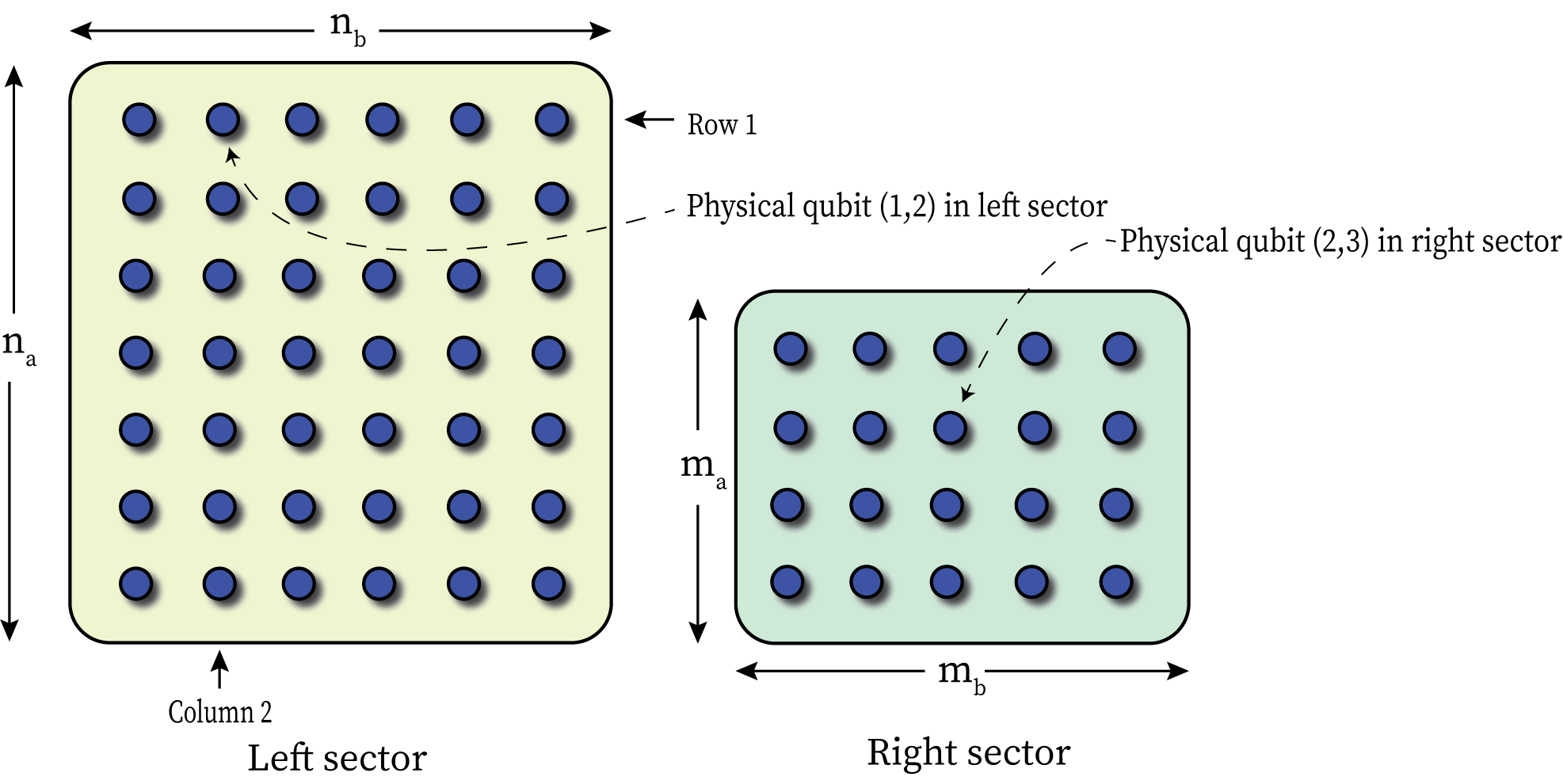}
			\centering
			\caption{Two-grid view of physical qubits in HGP codes}
			\label{fig:hp_grid_view}
		\end{figure}
	\end{center}

A convenient way of thinking of HGP codes in linear-algebraic terms arises from considering the tensor product of 1-complexes. This construction, called {\em homological product} \cite{bravyi2014homological}, generalizes the above combinatorial approach as follows. We view the classical codes that form the base of the construction as 1-complexes $\ff_2^{n_i}\stackrel{H_i}\longrightarrow \ff_2^{m_i}, i\in\{a,b\}$ and consider the tensor product spaces
   $$
   \cC_{2}=\ff_2^{n_a}\otimes\ff_2^{m_a}, \cC_1=\ff_2^{n_a}\otimes\ff_2^{n_b}\oplus \ff_2^{m_a}\otimes\ff_2^{m_b}, \cC_0=\ff_2^{m_a}\otimes\ff_2^{n_b}.
  $$
The homological (hypergraph) product code $\cQ$ defined by the matrices $H_X,H_Z$ \eqref{eq:hp_stab} can be described by the following diagram:	
\begin{equation} \label{eq:total}
{\begin{tikzcd}[column sep=.2in, row sep=.3in]
\setwiretype{n}&\ff_2^{n_a}\otimes\ff_2^{n_b} \arrow{dr}{\hspace*{0in} H_a\otimes I_{n_{_b}}}\\[.1in]
\setwiretype{n}\ff_2^{n_a}\otimes \ff_2^{m_b} \arrow{dr}[below]{\hspace*{-.4in} H_a\otimes I_{_{m_b}}}\arrow{ur}{\hspace*{-.3in} I_{n_{_a}}\otimes H_b^\intercal}&& \ff_2^{m_a}\otimes\ff_2^{n_b}\\[.1in]
\setwiretype{n}&\ff_2^{m_a}\otimes\ff_2^{m_b} \arrow{ur}[below]{\hspace*{.4in}I_{m_{_a}}\otimes H_b^\intercal}\\[.05in]
\cC_{2} \arrow{r}{\partial_{2}}&\cC_1 \arrow{r}{\partial_1} &\cC_0
\end{tikzcd}}
\end{equation}
where $\partial_2=H_Z^\intercal$ and $\partial_1=H_X$ in accordance with Eq. \eqref{eq:3term}.
The bottom part of this diagram describes the {\em total complex} of the 1-complexes $\cA=\ff_2^{n_a}\stackrel{H_a}\longrightarrow\ff_2^{m_a}$
and $\cB=\ff_2^{m_b}\stackrel{H_b^\intercal}\longrightarrow\ff_2^{n_b}$, written explicitly as
 \begin{equation}\label{eq:th}
    \ff_2^{n_a}\otimes \ff_2^{m_b} \stackrel{H_Z^\intercal}\longrightarrow
    (\ff_2^{n_a}\otimes\ff_2^{n_b})\oplus(\ff_2^{m_a}\otimes\ff_2^{m_b})
    \stackrel{H_X}\longrightarrow \ff_2^{m_a}\otimes\ff_2^{n_b},
 \end{equation}
and it is this complex that gives rise to a hypergraph product code. 
Physical qubits of the code $\cQ$ correspond to the basis of the space $\cC_1,$ and thus the length of the code $\cQ$ 
again is given by $n=n_an_b+m_am_b.$ 

 \section{Constructing Clifford circuits for CSS codes}\label{sec:symplectic_framework}
Suppose that $\cQ$ is a CSS code with $k$ logical qubits. The {\em logical Pauli operators} map code states to code states non-identically, while preserving the entire codespace. For the code $\cQ$ there exist $k$ logical $X$ operators 
$\overline X_i, i=1,\dots,k$ and $k$ logical $Z$ operators $\overline Z_i,i=1,\dots,k$, all of which commute with the stabilizers. 
Furthermore, the logical $\overline X_i$ operator anticommutes with $\overline{Z}_i$ and commutes with all the other logical Paulis. As discussed above, a basis of logical $Z$ operators $\{\overline{Z} _i\}_{i=1}^k$ can be found by fixing a basis of the homology space $H_1$ \cite{gottesman1997stabilizer, wilde2009}. By construction, 
this choice is equivalent to choosing a representative for each coset in $\ker(H_X)/\Img(H_Z^\intercal)$. Similarly, a choice of the basis for the cohomology space $H^1$ gives a set of logical $X$ operators $\{\overline{X}_i\}_{i=1}^k$. Note that all the logical Paulis together generate a non-Abelian subgroup  
$\overline{\cP}_k := \langle \{\overline{X}_i,\overline{Z}_i: i=1,\dots,k\}\rangle$ of $\cP^n$ of order $2k$, called the {\em logical Pauli group}.
	
It is clear that since these logical operators are products of physical $X$ or $Z$ operators, they can be implemented in parallel in a circuit. However, to support universal quantum computations
we also need to be able to implement any logical (non-Pauli) Clifford operation on any subset of logical qubits (even this is not enough for universal computations, which additionally require a non-Clifford operator, but this is outside the scope of this paper). Below we present a procedure to construct circuits for logical Clifford operators for a general CSS code. 

	\subsection{Symplectic representation and circuit design algorithm}
Designing logical operators for the Clifford group can be phrased relying on concepts from symplectic geometry over $\ff_2$. 
Given binary vectors $a,b \in \ffF_2^n$, let 
   $$
   D(a,b): = X^{a_1}Z^{b_1} \otimes \cdots \otimes X^{a_n}Z^{b_n}.
   $$ 
 The elements of the $n$-qubit Pauli group $\cP_n$ are then given by the matrices $\pm D(a,b)$ and $\pm iD(a,b)$ for all $a,b \in \ffF_2^n$. 
 The standard symplectic form is defined as
	$$([a,b],[a',b'])_s = a'b^\intercal+b'a^\intercal = [a,b]\Omega[a',b']^\intercal$$ 
where $\Omega = \begin{bmatrix}
		0 & I_n \\
		I_n & 0
	\end{bmatrix}.$ 
Note that two $n$-qubit Paulis $D(a,b)$ and $D(a',b')$ commute if and only if their binary representations are symplectically orthogonal $([a,b],[a',b'])_s=0$. 
Hence, the mapping $\phi : \cP_n \rightarrow \ffF_2^{2n}$, defined 
by $\phi(i^{\kappa}D(a,b)) = [a,b]$, for $\kappa \in \{0,1,2,3\}$, allows an efficient representation of any $n$ qubit Paulis up to the global phase. Let $\tilde{\cP}_n$ refer to the group $\cP_n/\langle iI_n\rangle$, i.e, the Pauli group with the center quotiented out. With a minor abuse of notation, $\phi : \tilde{\cP}_n \to \ffF_2^{2n}$ is an isomorphism. The usefulness of this representation for the construction of quantum codes was realized in \cite{calderbank1997quantum}; see also \cite{kitaev2002classical,gottesman2010introduction} for an overview.

Now recall that elements of $\Cliff_n$ are those unitaries that normalize the Pauli matrices and, hence, $\Cliff_n\cong\text{Aut}(\cP_n)$. Furthermore, the action of any element $g \in \Cliff_n$ is determined by the action of $g$ on the set $\{D(e_i,0):i \in [n]\}\cup  \{D(0,e_i):i\in [n]\}$, 
where $e_i$ is the standard basis vector, simply because the collection of these sets forms a basis of $\tilde{\cP}_n$. This establishes a direct relationship between Clifford operators and binary symplectic matrices.
	
	\begin{definition}
A symplectic matrix is a $2n\times 2n$ binary matrix that preserves the symplectic form. Symplectic matrices form the binary symplectic group 
$Sp(2n,\ffF_2)$. Equivalently, $F = \begin{bmatrix}
			A & B \\
			C & D
		\end{bmatrix} \in Sp(2n,\ffF_2)$ if and only if
		$F\Omega F^\intercal = \Omega$, or in other words
		$$
		AB^\intercal = BA^\intercal ,\hspace*{0.2in} CD^\intercal = DC^\intercal,\hspace*{0.2in}AD^\intercal+BC^\intercal = I_n.
		$$
	\end{definition}
	
The relationship between symplectic matrices and Clifford operators is formalized in the following theorem.
	\begin{theorem}[\cite{Rengaswamy2020}]\label{thm:symplectic_rep}
		The automorphism of $\cP_n$ induced by $g \in \Cliff_n$ satisfies
		$$gE(a,b)g^{\dagger} = \pm E([a,b]F_g)$$
		where $E(a,b) = i^{ab^{\intercal}}D(a,b)$ and $F_g = \begin{bmatrix}
			A_g & B_g\\
			C_g & D_g
		\end{bmatrix} \in Sp(2n,\ffF_2).$
	\end{theorem}
	\par\noindent\begin{remark} Note that Pauli operators, which are themselves in $\Cliff_n$, in this formulation are represented by the identity matrix.\end{remark} 
\subsection{From symplectic matrices to circuit design}\label{sec: F I F}
Once the symplectic representation of a Clifford operator is found, we can convert it to a Clifford circuit via decomposing the matrix into elementary row operations. It was shown in \cite{Dehaene2003,aaronson2004improved} that any symplectic matrix can be decomposed into elementary matrices corresponding to single- and two-qubit operations in the set $\{H,S,CNOT,CZ,SWAP\}$. We list these operations in Table \ref{table:circuit_algo} along with their effect on an arbitrary symplectic matrix by left and right multiplication. For $g \in \Cliff_n$ with $F_g$ being its corresponding symplectic matrix, let $\{V_{k,L}:k=1,\dots,g_L\}$ 
and $\{V_{k,R}:k=1,\dots,g_R\}$ be two sequences of symplectic matrices from this elementary set such that
	$$
	\left(\prod_{k=1}^{g_L}V_{k,L}\right) F_g \left(\prod_{k=1}^{g_R}V_{k,R}\right) = I.
	$$ 
Then, utilizing the fact that these elementary symplectic matrices are invertible, we get the following decomposition
	\begin{equation}\label{eq:symp-decomposition}
	F_g = \left(\prod_{k=g_L}^{1}V_{k,L}^{-1}\right) \left(\prod_{k=g_R}^{1}V_{k,R}^{-1}\right).
	\end{equation}
 Translating each of these elementary matrices to the corresponding elementary Clifford gate gives a circuit that performs the desired effect of the unitary $U$ (up to possible Pauli corrections, refer to Sec.~\ref{subsec:Pauli_correc} for details). Alternatively, we could also find a sequence of operations that, when performed on the $I_{2n \times 2n}$ matrix, give us the symplectic matrix $F$, and then translate this sequence to Clifford gates in the reverse order. We illustrate this 
 process in Example~\ref{exmpl:symplectic_circuit} elsewhere in the paper.
 
{\small		\begin{table}
			\begin{center}
			\begin{tabular}{|m{4em}|m{6cm}|m{6cm}|}
			\hline
			Clifford Element & Left Multiplication & Right Multiplication\\
			\hline
			$H_i$ & Switch the $i$th row of $(A|B)$ and $(C|D)$ & Switch the $i$th column of $(\frac{A}{C})$ and $(\frac{B}{D})$\\[0.2in]
			
			$S_i$ & Add the $i$th row of $(C|D)$ to the $i$th row $(A|B)$ & Add the $i$th column of $(\frac{A}{C})$to the $i$th column of $(\frac{B}{D})$\\[0.2in]
			
			$CNOT_{i,j}$ & Add the $j$th row of $(A|B)$ to the $i$th row of $(A|B)$ and adds $i$th row $(C|D)$ to the $j$th row of $(C|D)$ & Add the $i$th column of $(\frac{A}{C})$ to the $j$th column of $(\frac{A}{C})$ and adds $j$th column of $(\frac{B}{D})$ to the $i$th column of $(\frac{B}{D})$\\[0.3in]
			
			$CZ_{i,j}$ & Add the $j$th row of $(C|D)$ to the $i$th row of $(A|B)$ and adds $i$th row $(C|D)$ to the $j$th row of $(A|B)$ & Add the $i$th column of $(\frac{A}{C})$ to the $j$th column of $(\frac{B}{D})$ and adds $j$th column of $(\frac{A}{C})$ to the $i$th column of $(\frac{B}{D})$\\[0.3in]
			
			$SWAP_{ij}$ & Swaps rows $i$ and $j$ of $(A|B)$ and rows $i$ and $j$ of $(C|D)$  & Swaps columns $i$ and $j$ of $(\frac{A}{C})$ and columns $i$ and $j$ of $(\frac{B}{D})$\\
			\hline
		\end{tabular}
		\caption{Elementary circuit operations generating the Clifford group and their action on symplectic matrices}
		\label{table:circuit_algo}
	\end{center}
	\end{table}
}
	
	\subsection{Building an arbitrary logical Clifford circuit}\label{subsec:building_cliffords}
	We now focus on the case when the unitary $g \in \Cliff_n$ is also a logical Clifford operator of an $[[n,k]]$ CSS code. By definition, such a $g$ has to separately normalize the stabilizer space and the logical Pauli space, i.e., 
	\begin{gather*}
	\langle\{gD(H_X(i,:),0)g^{\dagger}\}_{i=1}^{m_X}\cup \{gD(0,H_Z(i,:))g^{\dagger}\}_{i=1}^{m_Z}\rangle = \cS,\\
	\langle \{g\overline{X}_ig^{\dagger}\}_{i=1}^k\cup\{g\overline{Z}_ig^{\dagger}\}_{i=1}^k\rangle = \overline{\cP}_k.
	\end{gather*}
Translating these requirements to the symplectic domain using the map $\phi$ and Theorem \ref{thm:symplectic_rep}, we obtain
constraints on the corresponding symplectic matrix $F_g$. Solving the resulting system of linear equations leads to finding the matrices representing logical operators, which can also be converted to circuits \cite{Rengaswamy2020}. However, 
as we show below, these matrices can also be found in a more explicit and efficient manner.  
	
For a logical operator $g \in \overline{\Cliff}_k\cap \Cliff_n$, let $\Gamma_g: \overline{\cP}_k \rightarrow \overline{\cP}_k$ be 
the corresponding automorphism of the logical Pauli group. Denote the operation enacted by $g$ on the stabilizer group
by $\Upsilon_g$,  $\Upsilon_g: \cS \rightarrow \cS$. As mentioned before, the $Z$-type Pauli logical operators can be chosen by fixing a basis of the homology space $H_1 = \ker(H_X)/\Img(H_Z^\intercal)$; let us denote them by $\{l_{i,Z}\}_{i=1}^k$. Similarly, a basis for the cohomology space $H^1=\ker(H_Z)/\Img(H_X^\intercal )$ gives the $X$-type Pauli logical operators;
denote them 
$\{l_{i,X}\}_{i=1}^k$. More precisely:
	$$
    \phi(\overline{X}_i) = [l_{i,X},0], \hspace*{.2in} \phi(\overline{Z}_i) = [0,l_{i,Z}]\hspace*{.2in}\text{ for all }\;i=1,\dots,k.
    $$
	Let $L$ be the $2k\times 2n$ block-diagonal matrix with rows $1$ to $k$ given by $\{\phi(\overline{X}_i)\}$ and 
rows $k+1$ to $2k$ given by $\{\phi(\overline{Z}_i)\}$. 
 Since  $\tilde{\cP}_n \stackrel{\phi}\cong \ffF_2^{2n}$ is an isomorphism, $\phi(\overline{\cP}_k) = \Span(L)$, and the automorphism $\Gamma_g$ can be described by a $2k \times 2k$ invertible binary matrix $M_g^{(L)}$ acting on $L$ from the left. Similarly, if we define $H = \begin{bmatrix}
		H_X & 0\\
		0 & H_Z
	\end{bmatrix}$, then the action of $g$ on the stabilizer space, $\Upsilon_g$, can be described by a $(n-k) \times (n-k)$ invertible binary matrix $M_g^{(S)}$, acting on $H$ from the left.
	
	\begin{example}
		For a set $I \subseteq [k]$, the logical Clifford operation $g :=\otimes_{i \in I}\overline{H}_i$ that fixes the stabilizers pointwise is described by the matrices 
        $$
        M_g^{(L)} = \sum_{i \in I}\left(e_i^\intercal e_{i+k}+e_{i+k}^\intercal e_i\right)
        +\sum_{i \in [k]\setminus I}\left(e_i^\intercal e_i +e_{i+k}^\intercal e_{i+k}\right),\hspace*{.2in}M_g^{(S)} = I_{n-k}.
        \qed$$
	
\end{example}
While $M_g^{(S)}$ can be any matrix from $\text{GL}(n-k,\ffF_2)$, the matrix $M_g^{(L)}$ in addition 
to being invertible, also needs to respect the fact that, for all $i$, $\overline X_i$ and $\overline Z_i$ anticommute. We describe these conditions below. Define
	\begin{align*}
		I_{i,X} &= \Supp(M_g^{(L)}(i,1:k)),\hspace*{0.45in}I_{i,Z} = \Supp(M_g^{(L)}(k+i,1:k))\\
		J_{i,X} &= \Supp(M_g^{(L)}(i,k+1:2k)),\hspace*{0.1in}J_{i,Z} = \Supp(M_g^{(L)}(k+i,k+1:2k)).
	\end{align*}
Expressing this in a different but equivalent form, 
	\begin{align*}
		\phi(\Gamma_g(\overline{X}_i)) &= \Big[\sum_{m \in I_{i,X}}l_{m,X},\sum_{n \in J_{i,X}}l_{n,Z}\Big]\\
		\phi(\Gamma_g(\overline{Z}_i)) &= \Big[\sum_{m \in I_{i,Z}}l_{m,X}, \sum_{n \in J_{i,Z}}l_{n,Z}\Big].
	\end{align*} 
Below for two sets $A,B$ we write $|A|_2=|B|_2$ as a short notation for $|A|=|B|\pmod{2}$. Further for $P \in \tilde{\cP}_n$, define $\phi_X(P) := \phi(P)(1:k), \phi_Z(P) := \phi(P)(k+1:2k)$.
\begin{proposition}\label{prop:sym1}
		(1) For $i,j \in [k]$ and $P \in \{X,Z\}$,
		\begin{align*}
			|I_{i,P} \cap J_{j,P}|_2= |I_{j,P} \cap J_{i,P}|_2.
		\end{align*}
        (2) For $i,j \in [k]$,
		\begin{align*}
			|I_{i,X} \cap J_{j,Z}|_2 = \delta_{ij}+|I_{j,Z} \cap J_{i,X}|_2.
		\end{align*}
	\end{proposition}
	\begin{proof}
(1) For $i=j$ the claim is trivially satisfied. Assume that $i\ne j$ and let $P=X$. 
For any two logical $X$ operators $\overline{X}_i, \overline{X}_j$,
$(\phi(\overline{X}_i),\phi(\overline{X}_j))_s = 0$. This implies that
  $$
  (\phi(\Gamma_g(\overline{X}_i)),\phi(\Gamma_g(\overline{X}_j)))_s =0,
  $$ 
  which, by definition of the symplectic inner product, implies
  $$
    (\phi_X(\Gamma_g(\overline{X}_i)),\phi_Z(\Gamma_g(\overline{X}_j))) = (\phi_Z(\Gamma_g(\overline{X}_i)),\phi_X(\Gamma_g(\overline{X}_j))),
  $$
where $(\cdot, \cdot)$ is the standard inner product. Writing this relation explicitly,
  $$
	\Big(\sum_{m \in I_{i,X}}l_{m,X},\sum_{n \in J_{j,X}}l_{n,Z}\Big) = \Big(\sum_{n \in J_{i,X}}l_{n,Z},\sum_{m \in I_{j,X}}l_{m,X}\Big),
  $$
which implies that $|I_{i,X} \cap J_{j,X}|_2 = |I_{j,X} \cap J_{i,X}|_2$. The case of $P=Z$ is fully analogous.	\\
(2) The proof is similar to the previous one. Let $i,j \in [k]$ and let $\overline X_i,\overline Z_j$ be a pair of logical operators.
Them $(\phi(\overline{X}_i),\phi(\overline{Z}_j))_s = \delta_{ij}$, implying that
    $
(\phi(\Gamma_g(\overline{X}_i)),\phi(\Gamma_g(\overline{Z}_j)))_s =\delta_{ij},
   $
   or
     $$
		(\phi_X(\Gamma_g(\overline{X}_i)),\phi_Z(\Gamma_g(\overline{Z}_j))) = \delta_{ij}+(\phi_Z(\Gamma_g(\overline{X}_i)),\phi_X(\Gamma_g(\overline{Z}_j))).
    $$
Writing this explicitly,	
   $$ 
   \Big(\sum_{m \in I_{i,X}}l_{m,X},\sum_{n \in J_{j,Z}}l_{n,Z}\Big) = \delta_{ij}+\Big(\sum_{n \in J_{i,X}}l_{n,Z},\sum_{m \in I_{j,X}}l_{m,X}\Big),
   $$
which implies that	$|I_{i,X} \cap J_{j,Z}|_2 = \delta_{ij}+|I_{j,Z} \cap J_{i,X}|_2$.
	\end{proof}
We see that the symplectic representation of Paulis gives us a convenient way to analyze the actions of logical Clifford operators on $\overline{\cP}_k$ and $\cS$;
more precisely, commutation relations of the Pauli operators translate into certain restrictions on the matrices $M_g^{(S)}$ and $M_g^{(L)}$. We further note that the generators of $\cS$ and $\overline{\cP}_k$ in their binary symplectic form, i.e., the rows of $H$ and $L$, constitute a part of a symplectic basis of $\ffF_2^{2n}$. 
	\begin{definition}
		A symplectic basis for $\ffF_2^{2n}$ is a set of pairs $\{(u_i,v_i)\}_{i=1}^n$ of length-$2n$ binary vectors such that $(u_i,v_j)_s = \delta_{ij}, (u_i,u_j)_s = (v_i,v_j)_s =0$ for all $i,j \in [n]$. 
	\end{definition}
	Note that the rows of any matrix in $Sp(2n,\ffF_2)$ form a symplectic basis for $\ffF_2^{2n}$. There exists a symplectic Gram-Schimdt orthogonalization procedure that converts the standard basis of $\ffF_2^{2n}$ to a symplectic basis.
	
	Now let 
	\begin{equation}\label{eqtn:basis_assign}
\begin{aligned}
		&u_i= \phi(\overline{X}_i) = L(i,:), i \in [k],& &v_i = \phi(\overline{Z}_i) = L(k+i,:), i \in [k]\\
		&u_{i+k} = H(i,:), i \in [m_X], &&u_{i+k+m_X} = H(i+m_X,:), i\in [m_Z],
	\end{aligned} 
	\end{equation}
 where $m_X$ and $m_Z$ are the numbers of rows of $H_X$ and $H_Z$, respectively.
	Clearly, $(u_i,v_j)_s = \delta_{ij}$ for $i \in [n],j \in [k]$ and $(u_i,u_j)_s =0, i,j \in [n]$, $(v_i,v_j)_s = 0, i,j\in [k]$. Hence, fixing a CSS code and fixing a basis of its logical Pauli space gives us a total of $n+k$ vectors that span a subspace of $\ffF_2^{2n}$. A symplectic matrix $F_g$ corresponding to a specified Clifford logical operator, maps this set of vectors to another set while preserving the symplectic inner products. The new set of vectors also forms a part of a basis (another basis) of $\ffF_2^{2n}$. Using this, we can explicitly construct $F_g$ in terms of these two sets of vectors. To build an intuition of this process, let us recall the definition of symplectic transvections.
	\begin{definition}
		Given a vector $x \in \ffF_2^{2n}$, a symplectic transvection is a map $T_x: \ffF_2^{2n} \rightarrow \ffF_2^{2n}$ defined by $T_x(y) = y+ (y,x)_sx $. Equivalently, $T_x$ is given by the symplectic matrix $F_x = I+ \Omega x^\intercal x$.
	\end{definition}
	It is well known that symplectic transvections generate $Sp(2n,\ffF_2)$ \cite[Thm.~3.4]{grove2002classical}. Furthermore, it is known that physical Clifford gates have very simple transvection decompositions.
 Namely, for $n$ physical qubits, the Hadamard, Phase, CNOT and CZ gates have the following form:
	\begin{equation}
    \begin{aligned}\label{eqtn: eq2}
			H_i &\equiv T_{[e_i, e_i]},\\
			S_i &\equiv T_{[0,e_i]},\\
			CNOT_{i\rightarrow j} &\equiv T_{[e_j,0]}T_{[e_j,e_i]}T_{[0,e_i]},\\
            CZ_{i,j} &\equiv T_{[0, e_i]}T_{[0,e_i+e_j]}T_{[0,e_j]},
		\end{aligned}
	\end{equation}    
where $e_i$ is a vector of the standard basis of $\ff_2^n$. Since $H,S,$ and $CNOT$ generate the physical Clifford group on
$n$ qubits (up to Paulis), the above (products of) transvections generate the group of symplectic matrices corresponding to the physical Clifford operators on $n$ qubits. 
We shall later see that the logical Clifford operator space is also generated by logical operators whose symplectic matrix representations take the form similar to Eq. (\ref{eqtn: eq2}).
	
While a symplectic transvection $T_x$ transforms any vector $y$ not orthogonal to $x$, i.e., $(y,x)_s =1$, to $y+x$, we shall require a slightly different variation of this fact.
Let $x,w \in \ffF_2^{2n}$ be such that $(w,x)_s =0$ and define 
    $$
   T_{w,x}(y) = y+ (y,x)_sw.
    $$
The operator $T_{w,x}(\cdot)$ maps any vector $y$ non-orthogonal to $x$ to a vector $y+w$, and since
$w\bot_s x$, it preserves the symplectic form. In other words, the matrix $F_{w,x} = I+\Omega x^\intercal w$ is symplectic.
This implies that, for a symplectic basis $\{(u_i,v_i)\}_{i=1}^n$, the operator $T_{u_i+u_i',v_i}$ maps $u_i$ to $u_i'$ while leaving all other basis vectors unchanged. 
We obtain the following proposition.
\begin{proposition}\label{prop:symp_const}
Let $\{(u_i,v_i)\}_{i=1}^n$ and $\{(u_i',v_i')\}_{i=1}^n$ be two sets of vectors each constituting a symplectic basis for $\ffF_2^{2n}$.  Then the following matrix
	$$F =I + \sum_{i=1}^n\Omega\left( v_i^\intercal (u_i + u_i') + u_i^\intercal (v_i + v_i')\right).$$
  satisfies $u_iF = u_i', v_iF = v_i', i =1,\dots,n$.
\end{proposition}
	\begin{proof}
		For $r =1,\dots,n$, 
		$$
  u_rF  = u_r + \sum_{i=1}^n\left(u_r\Omega v_i^\intercal (u_i + u_i') + u_r\Omega u_i^\intercal (v_i + v_i')\right) 
  = u_r + (u_r+u_r') = u_r'.
  $$ 
Similarly,  $v_rF = v_r'$ for all $r=1,\dots,n$.
	\end{proof}

We are now ready to state our first theorem.

\begin{theorem}\label{thm:symplectic}
Let $g$ be a Clifford logical operator for an $[[n,k]]$ CSS code that acts on $\overline{\cP}_k$ by the matrix $M_g^{(L)}$. 
Let $L' = (M_g^{(L)}+I_{2k})L$. Assume that $g$ fixes the stabilizers pointwise, i.e., $M_g^{(S)} = I_{n-k}$, and define
the matrices
\begin{align*}
		A_g &= I + \sum_{i=1}^kl_{i,Z}^\intercal L'(i,1:n), \quad
		B_g = \sum_{i=1}^kl_{i,Z}^\intercal L'(i,n+1:2n)\\[-.05in]
		C_g &= \sum_{i=1}^kl_{i,X}^\intercal L'(k+i,1:n),\quad
		D_g = I+ \sum_{i=1}^kl_{i,X}^\intercal L'(k+i,n+1:2n).
	\end{align*}
Then the matrix
    $$
    F_g = \begin{bmatrix}
		A_g & B_g \\
		C_g & D_g
	\end{bmatrix},
    $$ 
 is a valid symplectic matrix for the logical operation $g$.	
\end{theorem}
\begin{proof}
Let $\{(u_i,v_i)\}_{i=1}^k \cup \{u_i\}_{i=k+1}^n$ be the vectors defined in Eq. (\ref{eqtn:basis_assign}). To complete the basis, choose $v_i \in \ffF_2^{2n}, i\in [k+1:n]$ such that 
    $$
    (u_i,v_j)_s = \delta_{ij}, \hspace*{.2in}i\in [n], j \in [k+1:n],\quad  (v_i,v_j)_s =0,\hspace*{.2in} i\in[n],j \in [k+1:n].
    $$ 
    Then $\{(u_i,v_i)\}_{i=1}^n$ form a symplectic basis for $\ffF_2^{2n}$. Now let $\hat{L} = M_g^{(L)}L$ and let 
        \begin{gather*}
        u_i' = \hat{L}(i,:),\quad v_i' = \hat{L}(k+i,:),\quad i=1,\dots,k,\\
        u_i' =u_i,\quad v_i' = v_i, \quad i = k+1,\dots,n.
        \end{gather*}
  By Proposition \ref{prop:symp_const}, the matrix 
	\begin{equation}\begin{split}
		F_g &= I + \sum_{i \in [k]}\Omega\left( v_i^\intercal [u_i + u_i'] + u_i^\intercal [v_i + v_i']\right)\\
		 &= I + \sum_{i \in [k]}\left(\begin{bmatrix}
		 	l_{i,Z}^\intercal \\
		 	0
		 \end{bmatrix}L'(i,:) + \begin{bmatrix}
		 0\\
		 l_{i,X}^\intercal 
	 \end{bmatrix}L'(k+i,:)\right)\\
 & = I + \sum_{i \in [k]}\begin{bmatrix}
 	l_{i,Z}^\intercal L'(i,1:n) & l_{i,Z}^\intercal L'(i,n+1:2n)\\
 	l_{i,X}^\intercal L'(k+i,1:n) & l_{i,X}^\intercal L'(k+i,n+1:2n)
 \end{bmatrix}.
		\end{split}
	\end{equation}
satisfies $u_iF_g = u_i', v_iF_g = v_i', i \in [k]$ and fixes the other vectors in the basis.	
\end{proof}

Specializing Theorem \ref{thm:symplectic} to the logical $\overline{H}_i,\overline{P}_i,\overline{CNOT}_{i\rightarrow j}$ and $\overline{CZ}_{i,j}$, we obtain the following representations
(note the similarities with Eq. \eqref{eqtn: eq2}).
	\begin{corollary}\label{cor:tranvections}
    \phantom{lemma}\\[-.15in]\begin{enumerate}
		\item[{\rm(1)}] A single-qubit logical Phase gate acting on the $i$-th logical qubit is given by the following symplectic transvection: 
		$$\overline{S}_i \equiv T_{[0,l_{i,Z}]}.$$
        \item[{\rm(2)}] A single-qubit logical Hadamard gate acting on the $i$-th logical qubit is given by the following symplectic transvection:
		$$\overline{H}_i \equiv T_{[l_{i,X},l_{i,Z}]}.$$
		\item[{\rm(3)}] A two-qubit logical CNOT gate with the $i$-th and $j$-th logical qubits as control and target, respectively, is given by the following composition of symplectic transvections:
		$$\overline{CNOT}_{i \rightarrow j} \equiv T_{[l_{j,X},0]}T_{[l_{j,X},l_{i,Z}]}T_{[0 ,l_{i,Z}]}.$$
		\item[{\rm(4)}] A two-qubit logical CZ gate between the $i$-th and $j$-th logical qubits is given by the following composition of symplectic transvections:
		$$\overline{CZ}_{i, j} \equiv T_{[0,l_{i,Z}]}T_{[0,l_{i,Z}+l_{j,Z}]}T_{[0,l_{j,Z}]}.$$
        \end{enumerate}
	\end{corollary}
	\begin{proof}
		(1) From Theorem \ref{thm:symplectic},  the symplectic matrix for $\overline{S}_i$ is given by
		$$F_g = I +\begin{bmatrix}
			0 & l_{i,Z}^\intercal l_{i,Z}\\
			0 & 0
		\end{bmatrix} = I + \Omega [0,l_{i,Z}]^\intercal  [0,l_{i,Z}] = T_{[0,l_{i,Z}]}.$$\\
		(2) Similarly,  the symplectic matrix for $\overline{H}_i$ is given by
		$$F_g = I +\begin{bmatrix}
			l_{i,Z}^\intercal l_{i,X} & l_{i,Z}^\intercal l_{i,Z}\\
			l_{i,X}^\intercal l_{i,X} & l_{i,X}^\intercal l_{i,Z}
		\end{bmatrix} = I + \Omega [l_{i,X},l_{i,Z}]^\intercal  [l_{i,X},l_{i,Z}] =  T_{[l_{i,X},l_{i,Z}]}.$$\\
		(3) The symplectic matrix for $\overline{CNOT}_{i \rightarrow j}$  is given by
		\begin{align*}
			F_g = I +\begin{bmatrix}
				l_{i,Z}^\intercal l_{j,X} & 0\\
				0 & l_{j,X}^\intercal l_{i,Z}
			\end{bmatrix} &= I + \Omega [l_{j,X},0]^\intercal  [l_{j,X},0] + \Omega [l_{j,X},l_{i,Z}]^\intercal  [l_{j,X},l_{i,Z}]+ \Omega [0,l_{i,Z}]^\intercal  [0,l_{i,Z}]\\ &=T_{[l_{j,X},0]}T_{[l_{j,X},l_{i,Z}]}T_{[0 ,l_{i,Z}]}.
		\end{align*}\\
	(4) The symplectic matrix for $\overline{CZ}_{i ,j}$ is given by
	\begin{align*}
		F_g = I +\begin{bmatrix}
			0 & l_{i,Z}^\intercal l_{j,Z}+l_{j,Z}^\intercal l_{i,Z}\\
			0 & 0
		\end{bmatrix} &= I + \Omega [0,l_{i,Z}]^\intercal  [0,l_{i,Z}] + 
        \Omega [0,l_{j,Z}]^\intercal  [0,l_{j,Z}]\\
        &\hspace*{1in}+ \Omega [0,l_{i,Z}+l_{j,Z}]^\intercal  [0,l_{i,Z}+l_{j,Z}]\\ &=T_{[0,l_{i,Z}]}T_{[0,l_{i,Z}+l_{j,Z}]}T_{[0,l_{j,Z}]}.\qedhere
	\end{align*}
	\end{proof}
	The above corollary, combined with the fact that any Clifford logical operator can be decomposed into elementary logical Hadamard, Phase and CNOT (or CZ) operators, directly implies the following theorem.
	\begin{theorem}\label{thm:transvection_decomp}
For an $[[n,k]]$ CSS code, any logical Clifford operator can be represented as
a product of transvection operators from the following set: $$\{T_{[l_{i,X},l_{i,Z}]}\}_{i=1}^k\cup \{T_{[0,l_{i,Z}]}\}_{i=1}^k\cup \{T_{[l_{j,X},0]}T_{[l_{j,X},l_{i,Z}]}T_{[0 ,l_{i,Z}]}\}_{i,j \in [k],i \ne j}.$$
	\end{theorem}
This theorem implies that any Clifford logical operator can be implemented by first decomposing the $k$-qubit logical operator into elementary operators ($H,S,CNOT/CZ$) using the Bruhat or transvection \cite{Dehaene2003,pllaha2020weyl,Pllaha2021} decomposition and then performing the corresponding $n$-qubit transvections on the physical qubits. However, note that the decomposition might not be unique, and the process of circuit building using the elementary building blocks given in Corollary \ref{cor:tranvections} may be less efficient in terms of the circuit parameters compared to a direct implementation using Theorem \ref{thm:symplectic}. 
 
 The matrix $F_g$ as defined by Theorem \ref{thm:symplectic}, acts non-trivially only on the subspace of $\ffF_2^{2n}$ spanned by the $2k$ vectors $\{u_i,v_i\}_{i=1}^k$. 
The exact nature of this action is determined by the Clifford logical operation $g$, i.e., the operational description of the Clifford operator 
defines a nontrivial mapping between vectors $u_i \rightarrow u_i'$ and $v_i \rightarrow v_i', i \in[k]$. As for the remaining part of the basis $\{u_i,v_i\}_{i=k+1}^n$,
the action of $F_g$ fixes them pointwise.
 
Recall that generally, a logical Clifford operator normalizes the stabilizer group and does not necessarily fix the stabilizers pointwise. 
Moreover, so far we have only fixed $n+k$ basis vectors using the stabilizer generators and Pauli logical generators of the CSS code, while neither specifying the other $n-k$ basis vectors nor the action of the Clifford logical operator on them. 
 We shall do that now, which will allow us to construct all the possible symplectic matrices corresponding to a particular logical action on the logical Pauli space.
\begin{definition}[\cite{aaronson2004improved}]
Given a set of independent generators of the stabilizer group $\cS$, a \emph{destabilzer} for a generator $s \in \cS$ is a Pauli operator in $\cP_n$ that anticommutes with $s$ and commutes with every other generator and logical Pauli operator. 
For a CSS code with $n-k$ stabilizer generators defined by $H_X$ and $H_Z$, there exists a choice of $n-k$ destabilizers. The destabilizers jointly form the destabilizer group $\cD$.   
\end{definition} 

\begin{proposition}\label{prop:destabilizer}
	A choice of $n-k$ vectors $\{v_i\}_{i={k+1}}^n$ to complete the symplectic basis corresponds to a choice of destabilizers.
\end{proposition}
\begin{proof}
	Recall that by Eq. (\ref{eqtn:basis_assign}), $u_{k+1},\dots,u_n$ represent the stabilizers. For any $v_i, i=k+1,\dots,n$, we have $(u_i,v_i)_s=1$ and $(u_j,v_i)_s=0, j\ne i$, i.e., $D(v_i(1:n),v_i(n+1:2n))$ anticommutes with exactly one stabilizer generator $D(u_i(1:n),u_i(n+1:2n))$ and commutes with the others. Furthermore, $(u_j,v_i)_s =0, (v_j,v_i)_s=0, j=1,\dots,k$ implies that $D(v_i(1:n),v_i(n+1:2n))$ commutes with the logical Paulis. Hence $D(v_i(1:n),v_i(n+1:2n))$ is a valid destabilizer. 
\end{proof}

In the next lemma we find the number of choices for the set of destabilizers, each of which gives a different way to complete the basis described before. The proof follows the arguments in Theorem 7 of \cite{Rengaswamy2020}. 
\begin{lemma}\label{lemma:basis_count}
The set of vectors $\{v_i\}_{i=k+1}^n$ can be chosen in $2^{(n-k)(n-k+1)/2}$ different ways.
\end{lemma}
In addition to defining an automorphism of the stabilizer group $\cS$, a Clifford logical operator induces an automorphism $\Sigma_g: \cD \rightarrow \cD$ of the destabilizers.
Let $R$ be a $(n-k) \times 2n$ matrix with the $i$-th row equal to $v_{k+i}$ and let $M_g^{(R)}$ be the invertible binary matrix corresponding to the automorphism 
$\Sigma_g$. The following proposition shows that $\Sigma_g$ is completely described by the action, $\Upsilon_g$, of $g$ on $\cS$.
\begin{proposition}\label{prop:stab_destab}
    $
  (M_g^{(R)})^\intercal  = (M_g^{(S)})^{-1}.
    $
\end{proposition}  
\begin{proof}
We have that $H\Omega R^\intercal  = I_{n-k}$, and since the actions of $M_g^{(S)}$ and $M_g^{(R)}$ on $H$ and $R$ respectively preserve the symplectic inner product, we have 
    $$
    M_g^{(S)}H \Omega (M_g^{(R)}R)^\intercal  = I_{n-k},
    $$ 
 which implies $M_g^{(S)}(M_g^{(R)})^\intercal  = I_{n-k}$.
\end{proof}
Hence, it suffices to define the action of $g$ on the stabilizer space and the logical Pauli space. This gives the total number of symplectic matrices for a given $g$, as argued in the next
lemma.
\begin{lemma}\label{lemma:total_count}
The total number of symplectic matrices for a given logical operator specified by $\Gamma_g$ is at most $|GL(n-k,\ffF_2)|2^{(n-k)(n-k+1)/2}$. 
\end{lemma}
\begin{proof}
By Lemma \ref{lemma:basis_count}, there are $2^{(n-k)(n-k+1)/2}$ of ways to complete the symplectic basis. For each such choice, the number of ways of choosing $\Upsilon_g$ (and hence $\Sigma_g$) clearly equals $|GL(n-k,\ffF_2)|$. However, some of the resulting matrices may coincide, so the expression is an upper bound on the number of distinct matrices.
\end{proof}
We are now ready to state a generalized version of Theorem \ref{thm:symplectic}. While Theorem \ref{thm:symplectic} gives one very simple matrix representation for the operator $g$ (the one that acts trivially on all the stabilizers), the next theorem yields all the possible matrices. 
\begin{theorem}\label{thm:symplectic_general}
	Let $g$ be a Clifford logical operator for an $[[n,k]]$ CSS code with its respective actions on $\overline{\cP}_k$ and $\cS$ defined by the matrices $M_g^{(L)}$ and $M_g^{(S)}$ respectively. Let $R$ be a valid choice for the destabilizer matrix: $H\Omega R^\intercal  = I_{n-k}, L\Omega R^\intercal  =0$. Let 
	\begin{gather*}
	L' = (M_g^{(L)}+I_{2k})L, \quad H' = (M_g^{(S)}+I_{n-k})H\\
	 R' = (((M_g^{(S)})^{-1})^{T}+I_{n-k})R.
	 \end{gather*} 
Then the following matrix
	\begin{equation*}\begin{split}
			F_g  = &I + \sum_{i \in [k]}\begin{bmatrix}
				l_{i,Z}^\intercal L'(i,1:n) & l_{i,Z}^\intercal L'(i,n+1:2n)\\
				l_{i,X}^\intercal L'(k+i,1:n) & l_{i,X}^\intercal L'(k+i,n+1:2n).
			\end{bmatrix}\\
			& \hspace*{1in}+ \sum_{i=1}^{n-k}\Omega\left[R(i,:)^\intercal H'(i,:) + H(i,:)^\intercal R'(i,:)\right]
		\end{split}
	\end{equation*}
 is a valid symplectic matrix for the logical operation $g$.	
\end{theorem}
\begin{proof}
The proof is similar to Theorem \ref{thm:symplectic}. 
\end{proof}
Although this theorem may yield more efficient circuits compared to Theorem \ref{thm:symplectic}, explicit results appear difficult 
to come by because of the large number of symplectic matrices that implement the logical operator $g$.
	
\subsection{Applying Pauli Correction}\label{subsec:Pauli_correc}
Although circuits constructed using the above framework are guaranteed to perform the desired Pauli transformations with respect to their support, they can potentially introduce unwanted phase changes to some Paulis. This may happen because of two reasons, first
because symplectic representation ignores global phases and also because the elementary operations of Table \ref{table:circuit_algo} as listed there disregard potential Pauli corrections. Note that any global phases introduced in a logical Pauli operator by the circuit can be safely ignored because that would simply change the global phase of a codeword. However, phase changes in a stabilizer require special attention. In that case, the circuit does not perform a logical operation because it does not preserve the code space: namely, some 
	stabilizers $S$ are mapped to $-S$. Note that this is the only possible phase change because a Clifford automorphism cannot change the order of a Pauli in the group $\cP_n$, i.e., mapping $S$ to $iS$ is not allowed by definition. A small adjustment
	to the circuits ensures that the resulting operation performs the same logic and preserves all the stabilizers. As we show next, this is always possible.
	
	\begin{proposition}\label{prop:stabilizer_phase}
		Let $g$ be a Clifford logical operator of an $[[n,k]]$ CSS code, let $\sfC$ be a Clifford circuit on $n$ qubits performing the desired operation on the logical Pauli space, and let $U_{\sfC}$ be the corresponding unitary matrix. For a set of stabilizer generators $\cS$, let $\cS_F \subseteq \cS$ be the subset such that
		\begin{align*}U_{\sfC}SU_{\sfC}^{\dagger}& =(-1)^{{\mathbbm 1}_{\{S\in \cS_F\}}} S\hspace*{.2in}\text{for all } S \in \cS.
		\end{align*}
		Then there exists $P \in \cP_n$, such that $PU_{\sfC}$ performs the same logical operation $g$ while preserving all the stabilizer generators of $\cS$.
	\end{proposition}
	\begin{proof}
		For every $S\in \cS_F$, let $d_S$ be a destabilizer for $S$ such that the set $\{d_S: S \in \cS_F\}$ is formed of commuting operators. Let $\tilde{P} = \prod_{S \in \cS}d_S$, then 
		\begin{gather*}\tilde{P}U_{\sfC}SU_{\sfC}^{\dagger}\tilde{P}^{\dagger} = -\tilde{P}S\tilde{P}^{\dagger} = -d_SSd_S^{\dagger} = S,\hspace*{.2in}\text{for all } S \in \cS_F,\\
			\tilde{P}U_{\sfC}SU_{\sfC}^{\dagger}\tilde{P}^{\dagger} = \tilde{P}S\tilde{P}^{\dagger}= S,\hspace*{.2in}\text{for all } S \in \cS\setminus\cS_F.
		\end{gather*}
		Furthermore, for any logical Pauli $\overline{P}\in \overline{P}_k$, $\tilde{P}\overline{P}\tilde{P}^{\dagger} = \overline{P}$ by definition of destabilizers.
	\end{proof}
	Hence after constructing the circuits using the symplectic matrices, one may need to perform an additional layer of Pauli operators to make sure the stabilizer phases do not change. This added layer of operations increases the circuit depth by at most 1 while not affecting the other parameters.
	\section{Logical Gates for HGP Codes}\label{sec:FTlogical}
In this section, we use the above framework together with the matrix representations of Clifford logical operators
derived above to construct circuit realizations of these operators for HGP codes. Among the approaches
to implementing the Clifford group operators for CSS codes appearing in the literature are: state injection \cite{bravyi2005universal}, code deformation 
\cite{Krishna2021,dua2024clifford}, and others.  Our approach is more circuit-focused in the sense that we attempt to construct logical circuits using a small number of quantum gates on the physical qubits. Namely, our approach relies on an explicit form of the bases of the symplectic vector spaces spanned by the stabilizer generators and logical Pauli operators. This approach ties well with HGP codes for which these bases are recorded below.

We shall be interested in the following parameters for such physical Clifford circuits.
	\begin{definition}
For a Clifford circuit $\sfC$ acting on a code state of an $[[n,k]]$ CSS code, the support of $\sfC$, denoted $\chi(\sfC)$, is the  subset of physical qubits on which the circuit acts nontrivially. In other words, a circuit $\sfC$ enacting a unitary $U \in \cU^n$, has support $\chi(\sfC)$ if and only if $U$ acts as identity on every qubit in $[n]\setminus\chi(\sfC)$ and $\chi(\sfC)$ is the smallest such set.
	\end{definition}
	A Clifford circuit can be implemented by a sequential application of 
 elementary Clifford gates to a quantum state. A {\em layer} in a Clifford circuit is a set of gates that can be applied simultaneously, or equivalently, gates whose supports do not overlap. This leads us to our second circuit parameter of interest.
	\begin{definition}
		For a Clifford circuit $\sfC$ operating on an $n$-qubit state, the depth of $\sfC$, denoted $\delta(\sfC)$, is the minimum number of layers such that gates in the same layer do not overlap on qubits. 
	\end{definition}

The general goal of circuit design is to minimize both of these parameters, leading to faster implementation with lower complexity and reduced chance of error propagation while performing the desired logical operation on the codespace. We construct explicit circuits that perform single-qubit logical Hadamard, Phase, and two-qubit CNOT, CZ
operators for any family of HGP codes. If the length and dimension of the component codes of a HGP code
$\cQ(C_1,C_2)$ are taken to be about $\sqrt{n}$ (see also Sec.\ref{subsec:Clifford_HP} below), then the support and depth of the constructed
circuits scale as $\Theta(\sqrt n)$.
Combined with Theorem \ref{thm:transvection_decomp}, this implies that any Clifford logical operation on HGP codes that acts on a constant number of logical qubits, will have a Clifford circuit with $\chi$ and $\delta$ of order $\Theta(\sqrt{n})$.
	
	\subsection{Logical Paulis}\label{subsec:logical_paulis}
Understanding the circuitry of logical operators on a quantum code begins with implementing the 
logical Paulis on the code space, more specifically, being able to perform single qubit logical $X$ and $Z$ Paulis. The logical $X$ and $Z$ operators of an HGP code $\cQ(C_a,C_b)$ can be chosen to have support on a line of qubits in either of the two grids (see Fig.~\ref{fig:hp_grid_view}). The following lemma describes this special and very useful structure.
	
	\begin{lemma}[\cite{Krishna2021}, Lemma~1]\label{lemma:logicalP_HP}
Let $\cQ(C_a,C_b)$ be an HGP code with parity-check matrices $H_a$ and $H_b$ \eqref{eq:hp_stab}. The following is true for the logical Pauli operators of $\cQ$:
		\begin{enumerate}
			\item The logical $X$ operators are spanned by
			$$((\ffF_2^{n_a}/\Img(H_a^\intercal )) \otimes \ker(H_b)|0^{m_am_b})\cup (0^{n_an_b}|\ker(H_a^\intercal ) \otimes (\ffF_2^{m_b}/\Img(H_b)))$$
			\item The logical $Z$ operators are spanned by
			$$(\ker(H_a) \otimes (\ffF_2^{n_b}/\Img(H_b^\intercal ))|0^{m_am_b})\cup (0^{n_an_b}|(\ffF_2^{m_a}/\Img(H_a))\otimes \ker(H_b^\intercal ))$$
		\end{enumerate}
	\end{lemma}
To interpret this lemma, we refer to the description of HGP codes in Sec.~\ref{subsec:HP}. To connect the claims of the lemma to Fig.~\ref{fig:hp_grid_view}, choose a standard basis in the quotient space $\ff_2^{n_u}/H$, where $H$ is one of 
$H_a,H_b,H_a^\intercal,H_b^\intercal$, and $n_u$ matches this choice (i.e., is one of $n_a,n_b,m_a,m_b$). Note further that 
$\Img(H_a^\intercal)\cong C_a^\bot$ and $\ker(H_a^\intercal)\cong \ffF_2^n/C_a^\bot\cong C_a^\ast$, and similar claims hold for $H_b$. Then this lemma implies that any element of a basis of logical $X$'s 
can be supported on a single row of the left sector or a single column of the right sector. 
There are exactly $k_a$ logical $X$'s supported on a row (any row) of the left sector, and the 
corresponding binary vectors form a basis of $C_a$.  The same claims are true with respect to a basis of logical $Z$'s, with the roles of rows and columns interchanged, and $C_b$ swapped in for the code $C_a$. 

A further specialization of Lemma~\ref{lemma:logicalP_HP} was given in \cite{Quintavalle2023}. We begin with citing their definition.
	\begin{definition}\label{def: SLT}
		An $m \times n$ binary matrix $A$ is said to be \emph{strongly lower triangular} (SLT) if
		\begin{enumerate}
			\item Any column $j$ has an entry (a pivot) $p_j$ such that $A_{ij}=0$ for all $i>p_j$; 
			\item $p_{j_1}\ne p_{j_2}$ for $1\le j_1<j_2\le n$;
			\item Reordering the columns if necessary, if $A_{p_j,j} =1$, then $A_{p_j,t} = 0$ for $t > j$.  
		\end{enumerate}
	\end{definition}
We call a basis SLT if, once written in matrix form, this matrix can be brought to the SLT form upon performing row and column permutations. 

	\begin{theorem}[\cite{Quintavalle2023}]\label{thm:logical_paulis_hp}
Let $\cQ(C_a,C_b)$ be an HGP code with constituent parity-check matrices $H_a$ and $H_b$.
Suppose that $\{a_i\}, \{\alpha_j\}, \{b_h\}, \{\beta_l\}$ are SLT bases for $\ker(H_a), \ker(H_a^\intercal ),$
		$\ker(H_b), \ker(H_b^\intercal )$ respectively, where the pivot of $a_i$ is $i$ and similar for the other bases. Then
		\begin{enumerate}
			\item The logical $X$ operators are spanned by
			$$
            \{e_i \otimes b_h| 0^{m_am_b}\} \cup \{0^{n_an_b}|\alpha_j\otimes e_l\}
            $$ 
for all combinations of $e_i \in (\ffF_2^{n_a}/\Img(H_a^\intercal )), e_l \in (\ffF_2^{m_b}/\Img(H_b))$;
			\item  The logical $Z$ operators are spanned by
			$$\{a_i \otimes e_h| 0^{m_am_b}\} \cup \{0^{n_an_b}|e_j\otimes \beta_l\}$$ for all combinations of $e_h \in (\ffF_2^{n_b}/\Img(H_b^\intercal )), e_j \in (\ffF_2^{m_a}/\Img(H_a))$.
		\end{enumerate}
	\end{theorem}
With this choice of the representatives of the bases, $|\overline X_i\cap\overline Z_j|=\delta_{ij},$ i.e., every $\overline X_i$ overlaps with $\overline Z_i$ on a single qubit and is disjoint from all the over $\overline Z_j$'s.

\subsection{Logical Clifford operations}\label{subsec:Clifford_HP}
	In this section we give explicit symplectic representations and circuit-building
 procedures for logical single-qubit and two-qubit Clifford operators for general HGP codes. The first of these steps
 is performed  by specializing Theorem $\ref{thm:symplectic}$ and its corollary, Cor.~\ref{cor:tranvections}, to HGP 
 codes. 
 After that, we construct circuit representations of these matrices following the procedure described after
 Theorem~\ref{thm:symplectic_rep} using the operations from Table~\ref{table:circuit_algo}.

The circuits we construct perform the following logical tasks: a single-qubit logical 
Phase, Hadamard, two-qubit logical CNOT, and CZ. Without loss of generality, we will assume 
that the targeted logical operation is performed in the left sector 
for the Hadamard and Phase logical gates, noting that gates for the right sector can be derived similarly. We also assume that the targeted operator fixes the stabilizers pointwise, i.e, $\Upsilon_g$ is trivial. While this assumption, inherited from Theorem~\ref{thm:symplectic}, yields a concise descriptions of the symplectic matrices leading to explicit
physical circuits, we would like to reiterate that more efficient circuits might be possible by using Theorem \ref{thm:symplectic_general} (this was mentioned earlier at the end of Section \ref{subsec:building_cliffords}).

\vspace*{.1in}
{\em Notational conventions:} The following notation will be used throughout this section.
An HGP code constructed from classical codes defined by the matrices $H_a \in \ffF_2^{m_a\times n_a}, H_b \in 
\ffF_2^{m_b \times n_b}$ will be denoted by $\cQ(C_a,C_b)$. 
We denote by $\{a_i\}, \{\alpha_j\}, \{b_h\}, \{\beta_l\}$ SLT bases for $\ker(H_a), \ker(H_a^\intercal )$, $\ker(H_b), \ker(H_b^\intercal )$, respectively, and  denote by $\sfP(\cdot)$ the set of pivots for the basis argument (see Def. \ref{def: SLT}). Throughout, $n=n_an_n+m_am_b$ is the number of physical qubits (the length of the HGP code, cf.~\eqref{eq:hp_stab}), and in our statements about asymptotic scaling of the circuit parameters, we implicitly refer to a sequence of HGP codes with increasing $n$. We also assume throughout that the constituent codes $C_a,C_b$ in the HGP construction are chosen such that their dimension and distance scale as a constant proportion of their length, e.g.,
$k_a=\Theta(n_a), d_a=\Theta(n_a)$, and $n_a,n_b$ scale in
proportion to $\sqrt n$. As a result, the weights of the vectors in each of the 4 bases in Theorem \ref{thm:logical_paulis_hp} scale as $\Theta(\sqrt n)$. These 
assumptions will be factored into the statements in this section without further mention. 
  
\vspace*{.1in} The logical qubits in the two sectors can be naturally labeled as follows:
	$\{\overline{q}_{i,h,L}: i \in \sfP(\{a_i\}), h \in \sfP(\{b_h\})\}$ in the left sector and $\{\overline{q}_{j,l,R}: j \in \sfP(\{\alpha_j\}), l \in \sfP(\{\beta_l\})\}$ in the right sector. There are a total of $k=k_ak_b+k_a^\intercal k_b^\intercal $ logical qubits. We denote by $\overline{\Cliff}_k$ the logical Clifford group for this code and write the logical Paulis as
	\begin{gather*}
		\{\overline{X}_{i,h,L}:i \in \sfP(\{a_i\}), h \in \sfP(\{b_h\})\}, \hspace{0.2in}\{\overline{X}_{j,l,R}:j \in \sfP(\{\alpha_j\}), l \in \sfP(\{\beta_l\})\}\\
		\{\overline{Z}_{i,h,L}:i \in \sfP(\{a_i\}), h \in \sfP(\{b_h\})\}, \hspace{0.2in}\{\overline{Z}_{j,l,R}:j \in \sfP(\{\alpha_j\}), l \in \sfP(\{\beta_l\})\}.
	\end{gather*}
	The sets of physical qubits supporting these operators are given in Theorem \ref{thm:logical_paulis_hp}.
	
	\subsubsection{Targeted Logical Phase Gate}\label{subsub:phase_HP}
We limit ourselves to the logicals acting on left-sector qubits since the construction for the right ones is fully analogous.
Let $g=\bar{S}_{i,h,L}$ be the single-qubit logical Phase gate on the logical qubit $\overline{q}_{i,h,L}$. The automorphism $\Gamma_g$ 
corresponding to the operational description of $\bar{S}_{i,h,L}$ is described as follows:
	\begin{equation}
		\left\{\begin{aligned}
			&\Gamma_g(\overline{X}_{i,h,L}) = \overline{X}_{i,h,L}\overline{Z}_{i,h,L}\\
			&\Gamma_g(\overline{Z}_{i,h,L}) = \overline{Z}_{i,h,L}\\
			&\Gamma_g(\overline{P}_{i_1,h_1,A}) = \overline{P}_{i_1,h_1,A}\hspace*{.2in} \text{ for all }\hspace*{.2in} (i_1,h_1) \ne (i,h), \bar{P} \in \{\bar X,\bar Z\}, A \in \{L,R\}
		\end{aligned}\right.
	\end{equation}
Assume further that $g$ fixes the stabilizers pointwise. In the following proposition, we construct a symplectic matrix representation for $g$.
To simplify the notation, let $\tilde{n} = n_an_b, \tilde{m} = m_am_b$.
	\begin{proposition}\label{prop:logical_P}
		For $\cQ(C_a,C_b)$, a single-qubit logical targeted Phase gate that acts on the $(i,h,L)$-th logical qubit has the following symplectic representation:
		$$F^S_{i,h,L} = \begin{bmatrix}
			I_{\tilde{n}\times \tilde{n}} & 0_{\tilde{n} \times \tilde{m}} & \tilde{B} & 0_{\tilde{n} \times \tilde{m}}\\
			0_{\tilde{m} \times \tilde{n}} & I_{\tilde{m} \times \tilde{m}} & 0_{\tilde{m} \times \tilde{n}} & 0_{\tilde{m} \times \tilde{m}}\\
			0_{\tilde{n}\times \tilde{n}} & 0_{\tilde{n} \times \tilde{m}} &I_{\tilde{n}\times \tilde{n}} & 0_{\tilde{n} \times \tilde{m}}\\
			0_{\tilde{m} \times \tilde{n}} & 0_{\tilde{m} \times \tilde{m}} & 0_{\tilde{m}\times\tilde{n}} & I_{\tilde{m}\times\tilde{m}}
		\end{bmatrix}$$
		where \begin{align*}
			\tilde{B} &= (a_i \otimes e_h)^\intercal  (a_i \otimes e_h).
		\end{align*}
	\end{proposition}
	\begin{proof}
		By Corollary \ref{cor:tranvections}, the logical phase gate on the $(i,h,L)$-th logical qubit is given by $T_{[0, v_{i,h,L}]}$, where $v_{i,h,L} = [a_i \otimes e_h , 0]$. Hence, $F^S_{i,h,L}  = I + \begin{bmatrix}
			0 & v_{i,h,L}^\intercal v_{i,h,L}\\
			0 & 0
		\end{bmatrix}.$
	\end{proof}
	
	\begin{theorem}\label{thm:circuit_P}
Let $\cQ(C_a,C_b)$ be an HGP code. There exists a Clifford circuit $\sfC$, with $\chi(\sfC) = \Theta(\sqrt{n})$ and $\delta(\sfC) = \Theta(\sqrt{n})$  that implements $\bar{S}_{i,h,L}$. 
	\end{theorem}
	\begin{proof}

Let $v = (a_i \otimes e_h)$. We shall construct the matrix $F_{i,h,L}^S$
by performing the following row operations on the matrix $I_{2n\times 2n}$:
\begin{enumerate}
	\item Choose $x \in \Supp(v)$. For every $y \in \Supp(v)\setminus\{x\}$, add row $x$ to row $y$ and add row $(y+n)$  to row $(x+n)$.
	\item Add row $(x+n)$ to row $x$.
	\item For every $y \in \Supp(v)\setminus\{x\}$, add row $x$ to row $y$ and add row $(y+n)$  to row $(x+n)$.
\end{enumerate}

It is easy to see that the row operations produce the matrix $F$ from $I_{2n \times 2n}$; we include a detailed proof in Appendix \ref{appndx:phase_rowop}. The next step is to translate these row operations to physical gates using the operations in Table \ref{table:circuit_algo}. Let $I$ denote the 
physical qubit indices corresponding to the support of the $\overline Z_{i,h,L}$ logical Pauli. Steps (1) and (3) of row operations translate to 
applying CNOT gates and step (2) translates to applying a Phase gate. Since the circuit realizes a Clifford unitary having $F$ as its symplectic 
representation, the circuit is guaranteed to fix the support of the stabilizer generators and perform the desired transformations on the 
logical Pauli space. However, the circuit can still introduce undesired phase changes to some stabilizers, which may change the codespace. It turns out that adding a single $X$ gate to qubit $x \in I$ suffices to prevent this and preserve the stabilizer space with respective phases: we include a complete proof in Appendix \ref{appndx:phase_stab}. 

For reference, we give the complete circuit building procedure in Algorithm \ref{alg:algo_logical_phase_new}. 

\begin{algorithm}
	\caption{Logical phase gate circuit design algorithm}
	\label{alg:algo_logical_phase_new}
	\hspace*{-5.5in}\textbf{Input:} $I$
	\begin{algorithmic}[1]
		\State{Select any qubit $x \in I.$}
		\State \textbf{CNOT layer}
		\Statex \hspace{\algorithmicindent}  For {$y \in  I\setminus\{x\}$}, apply {CNOT from $y$ to $x$.}
		\State Apply Phase on qubit $x$.
		\State \textbf{CNOT layer}
		\Statex \hspace{\algorithmicindent}  For {$y \in  I\setminus\{x\}$}, apply {CNOT from $y$ to $x$.}
		\State Apply {$X$ on qubit $x$}
	\end{algorithmic}
\end{algorithm}

Finally we check the parameters of the circuit $\sfC$. It only acts on the set of qubits indexed by $I$, whose size equals the Hamming weight of $|a_i|$, hence $\chi = |a_i|$.  
Since there are two layers of $|a_i|-1$ CNOT gates each having the same qubit as their target, and the Phase gates and $X$ gate also operate on this qubit, the depth $\delta = 2|a_i|$. Since $\{a_i\}$ is an SLT basis of $\ker(H_a)$ then by Theorem \ref{thm:logical_paulis_hp}, $|a_i| = \Theta(\sqrt{n})$. Hence we conclude that $\chi = \Theta(\sqrt{n}), \delta = \Theta(\sqrt{n})$.
	\end{proof}
Note that the sequence of row operations to construct the desired symplectic matrix from the identity matrix is not unique. One can potentially find other sequences of row operations which, when translated to physical gates, could result in circuits with different parameters. As an example, we include an alternative proof of Theorem \ref{thm:circuit_P} in Appendix \ref{appndx:alt_proof}. While the circuit constructed there has a larger gate count, it has a simpler design since that all qubits in the support are acted upon in a symmetric way.

	\subsubsection{Targeted Logical Hadamard Gate}\label{subsub:hadamard_HP}
	Now let $g$ be the single-qubit logical Hadamard $\overline{H}_{i,h,L}$ acting on logical qubit $\overline{q}_{i,h,L}$, and let $F^H_{i,h,L}$ be its symplectic matrix representation as given in  Theorem \ref{thm:symplectic_rep}. The mapping $\Gamma_g$ 
exchanges $\overline X$ and $\overline Z$, i.e., $\Gamma_g(\overline{X}_{i,h,L})  = \overline{Z}_{i,h,L},$ 
$\Gamma_g(\overline{Z}_{i,h,L})= \overline{X}_{i,h,L}$ and acts identically on the logical Paulis outside the $(i,h)$ location.
 The matrix form of this gate is given in the next proposition.
\begin{proposition}\label{prop:logical_H}
For $\cQ(C_a,C_b)$, a single-qubit logical targeted Hadamard gate that acts on the $(i,h,L)$-th logical qubit has the following symplectic representation:
		$$F^H_{i,h,L} = \begin{bmatrix}
			\tilde{A} & 0_{\tilde{n} \times \tilde{m}} & \tilde{B} & 0_{\tilde{n} \times \tilde{m}}\\
			0_{\tilde{m} \times \tilde{n}} & I_{\tilde{m} \times \tilde{m}} & 0_{\tilde{m} \times \tilde{n}} & 0_{\tilde{m} \times \tilde{m}}\\
			\tilde{C} & 0_{\tilde{n} \times \tilde{m}} &\tilde{D} & 0_{\tilde{n} \times \tilde{m}}\\
			0_{\tilde{m} \times \tilde{n}} & 0_{\tilde{m} \times \tilde{m}} & 0_{\tilde{m}\times\tilde{n}} & I_{\tilde{m}\times\tilde{m}}
		\end{bmatrix}$$
		where \begin{align*}
			\tilde{A} &=I_{\tilde{n}\times\tilde{n}}+ (a_i \otimes e_h)^\intercal  (e_i \otimes b_h), \\\tilde{B} &= (a_i \otimes e_h)^\intercal  (a_i \otimes e_h),\\ \tilde{C} &= (e_i \otimes b_h)^\intercal   (e_i \otimes b_h), \\\tilde{D} &= I_{\tilde{n}\times\tilde{n}} + (e_i \otimes b_h)^\intercal   (a_i \otimes e_h).
		\end{align*}
\end{proposition}
	\begin{proof}
		By Corollary \ref{cor:tranvections}, the logical Hadamard gate on the $(i,h,L)$-th logical qubit is given by $T_{[u_{i,h,L}, v_{i,h,L}]}$ where $u_{i,h,L} = [e_i \otimes b_h, 0], v_{i,h,L} = [a_i \otimes e_h , 0]$. Hence, $$F^H_{i,h,L}  = I + \begin{bmatrix}
			v_{i,h,L}^\intercal u_{i,h,L} & v_{i,h,L}^\intercal v_{i,h,L}\\
			u_{i,h,L}^\intercal u_{i,h,L} & u_{i,h,L}^\intercal v_{i,h,L}
		\end{bmatrix}.$$
	\end{proof}
	\begin{theorem}\label{thm:circuit_H}
		Let $\cQ(C_a,C_b)$ be an HGP code. There exists a Clifford circuit $\sfC$, with $\chi(\sfC) = \Theta(\sqrt{n})$ and $\delta(\sfC) = \Theta(\sqrt{n})$ that implements $\overline{H}_{i,h,L}$. 
	\end{theorem}
	\begin{proof}
    Let us rewrite the matrix $F^H_{i,h,L}$ in a more convenient form. Denote  $u= (e_i \otimes b_h) , v= (a_i \otimes e_h)$, then
         $$
         F^H_{i,h,L}=\begin{bmatrix}
			I_{\tilde{n}\times \tilde{n}}+v^{\intercal}u & 0_{\tilde{n} \times \tilde{m}} & v^{\intercal}v & 0_{\tilde{n} \times \tilde{m}}\\
			0_{\tilde{m} \times \tilde{n}} & I_{\tilde{m} \times \tilde{m}} & 0_{\tilde{m} \times \tilde{n}} & 0_{\tilde{m} \times \tilde{m}}\\
			u^{\intercal}u & 0_{\tilde{n} \times \tilde{m}} &I_{\tilde{n}\times \tilde{n}}+u^{\intercal}v & 0_{\tilde{n} \times \tilde{m}}\\
			0_{\tilde{m} \times \tilde{n}} & 0_{\tilde{m} \times \tilde{m}} & 0_{\tilde{m}\times\tilde{n}} & I_{\tilde{m}\times\tilde{m}}
		\end{bmatrix}.
        $$ 
 Let $I=\Supp(u)$ and $J=\Supp(v)$ be the supports of the logical Paulis $\bar X_{i,h,L}$ and $\bar Z_{i,h,L}$ and note that, due to the remark after Theorem \ref{thm:logical_paulis_hp}, $I\cap J=(i,h,L)$. Let $\rho = n_a(i-1)+h$. 
        
Next, we give a sequence of row operations that transforms the $2n \times 2n$ identity matrix to 
$F^H_{i,h,L}$.
		\begin{enumerate}
			\item For each $x \in \Supp(v)\setminus \{\rho\}$, add row $(x+n)$ to row $\rho$ and add row $(\rho+n)$ to row $x$.
			\item For each $y \in \Supp(u)$, exchange rows $y$ and $(y+n)$.
			\item For each $y \in \Supp(u)\setminus \{\rho\}$, add row $(y+n)$ to row $\rho$ and add row $\rho+n$ to row $y$. 
			\item For each $y \in \Supp(u)$, exchange rows $y$ and $(y+n)$. 
			\item For each $y \in \Supp(u)\setminus \{\rho\}$, add row $y$ to row $\rho$ and add row $(\rho+n)$ to row $(y+n)$.
			\item For each $x \in \Supp(v)\setminus \{\rho\}$, add row $\rho$ to row $x$ and add row $(x+n)$ to row $(\rho+n)$.
			\item Exchange rows $\rho$ and $(\rho+n)$.
			\end{enumerate}
		It is not difficult to verify that these operations result in the correct symplectic matrix $F$. We give a detailed proof in Appendix \ref{appndx:hadamard_rowop}. 
        
Next, we construct the circuit $\sf C$ by translating each of the steps above to corresponding physical gates, reversing the order of the gates compared to the above steps  
in accordance with \eqref{eq:symp-decomposition}. Steps (1) and (3) correspond to CZ gates, steps (2), (4) and (7) correspond to Hadamard gates and steps (5), (6) correspond to CNOT gates. Although this circuit preserves the stabilizers up to their support, it does not preserve the phases of all stabilizers. However, we show in Appendix \ref{appndx:hadamard_stab} that adding a single $Y$ gate negates any undesired phase change.  The complete circuit design algorithm is described in Algorithm \ref{alg:algo_logical_hadamard} and a conceptual visualization diagram is included in Figure \ref{fig:hadamard_circuit_visual}.

        \begin{algorithm}
			\caption{Logical Hadamard gate circuit design algorithm}
			\label{alg:algo_logical_hadamard}
			\hspace*{-5.0in} \textbf{Input:} $I, J, \rho$
			\begin{algorithmic}[1]
            \State Hadamard on qubit $\rho$
            \State \textbf{CNOT layer}    
            \Statex \hspace{\algorithmicindent} For {$x \in J\setminus\{\rho\}$, apply $CNOT$ from $x$ to $\rho$}
            \State \textbf{CNOT layer}    
            \Statex \hspace{\algorithmicindent} For {$y \in I\setminus\{\rho\}$, apply $CNOT$ from $\rho$ to $y$}
            \State \textbf{Hadamard layer}    
            \Statex \hspace{\algorithmicindent} For {$y \in I$}, apply Hadamard to qubit $y$
            \State \textbf{CZ layer}    
            \Statex \hspace{\algorithmicindent} For {$y \in I\setminus\{\rho\}$, apply $CZ$ between $y$ and $\rho$}
            \State \textbf{Hadamard layer}    
            \Statex \hspace{\algorithmicindent} For {$y \in I$}, apply Hadamard to qubit $y$
            \State \textbf{CZ layer}    
            \Statex \hspace{\algorithmicindent} For {$x \in J\setminus\{\rho\}$, apply $CZ$ between $x$ and $\rho$}
            \State{Y gate on qubit $\rho$}
			\end{algorithmic}
		\end{algorithm}
\begin{center}
	\begin{figure}[h]
		\includegraphics[scale=0.5]{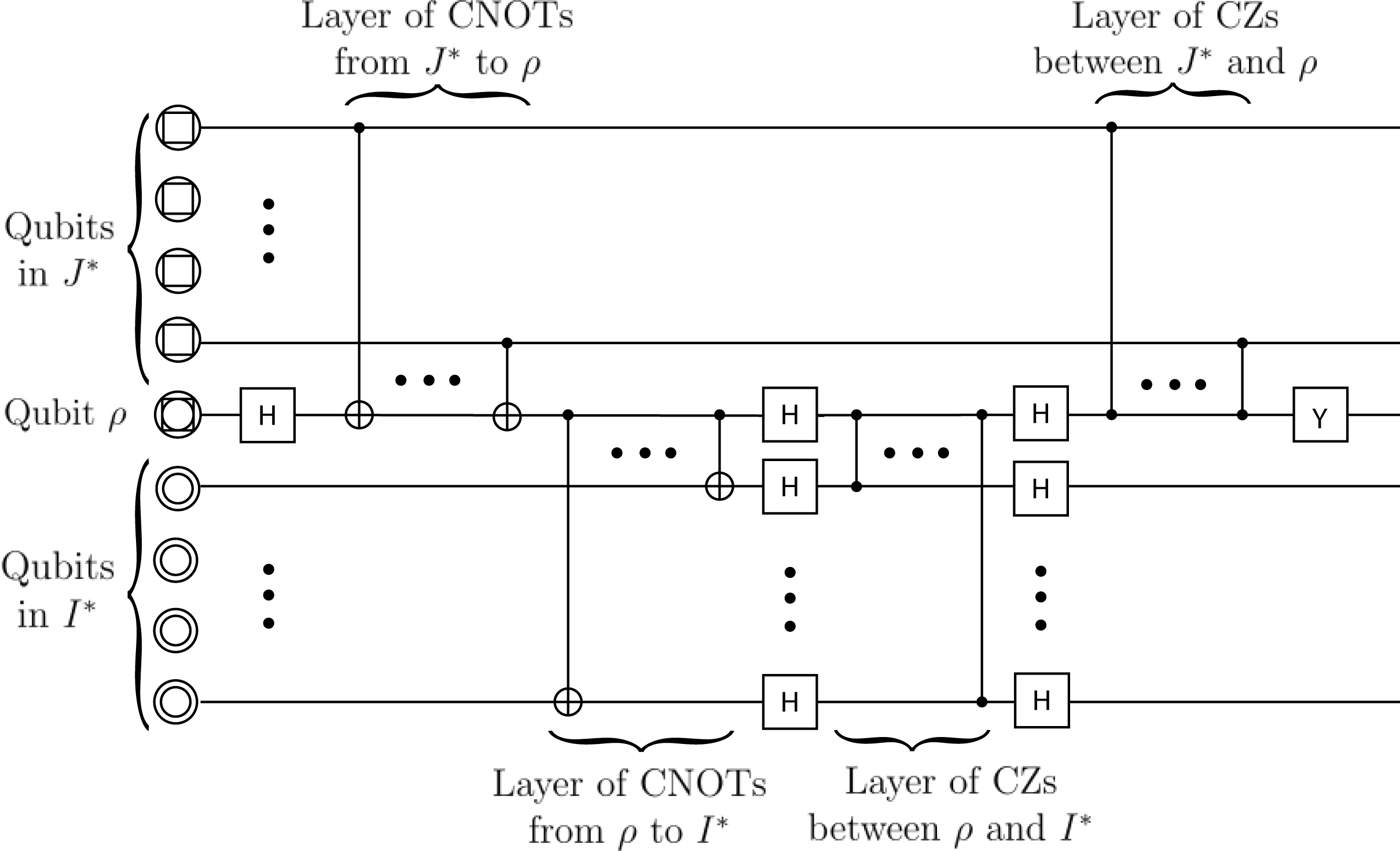}
		\centering
		\caption{A conceptual visualization of the logical Hadamard circuit. Here $J^* = J\setminus\{\rho\}, I^* = I \setminus\{\rho\}$.}
		\label{fig:hadamard_circuit_visual}
	\end{figure}
\end{center}
\vspace*{.1in}
Finally we look at the parameters of the circuit. The circuit acts only on the set of qubits indexed by $I\cup J$ and so $\chi = |a_i|+|b_h|-1$. Furthermore, qubit $\rho$ is operated on the most in the circuit and appears in $2(|I|+|J|)+2$ gates, which is the depth $\delta$ of the circuit.  
Since $\{a_i\}$ is an SLT basis of $\ker(H_a)$ and $\{b_h\}$ is an SLT basis of $\ker(H_b)$,  
by Theorem \ref{thm:logical_paulis_hp}, $|a_i| = |b_h|= \Theta(\sqrt{n})$, and we obtain for both 
parameters $\chi, \delta$ the order of $\sqrt{n}$ estimate.
	\end{proof}

    \subsubsection{Targeted Logical CNOT Gate}\label{subsub:cnot_HP}
In the next two sections we present constructions of two-qubit logical gates. We shall consider two cases
according as the two logical qubits are in the same or in different sectors. 

Let $\overline{q}_{i_1,h_1,A_1}$ and $\overline{q}_{i_2,h_2,A_2}$ be two distinct logical qubits, with $A_1,A_2 \in \{L,R\}$ and let $g=\overline{CNOT}_{(i_1,h_1,A_1)\rightarrow (i_2,h_2,A_2)}$ denote the logical CNOT operation with $(i_1,h_1,A_1)$ acting as the control and $(i_2,h_2,A_2)$ acting as the target. Again assume that $g$ fixes 
the stabilizers pointwise. The mapping $\Gamma_g$ of the logical Paulis is described 
as follows:
	\begin{equation}
		\begin{split}
			\Gamma_g(\overline{X}_{i_1,h_1,A_1}) &= \overline{X}_{i_1,h_1,A_1}\overline{X}_{i_2,h_2,A_2},\\
			\Gamma_g(\overline{Z}_{i_2,h_2,A_2}) &= \overline{Z}_{i_1,h_1,A_1}\overline{Z}_{i_2,h_2,A_2},\\
			\Gamma_g(\overline{X}_{{i},{h},A}) &= \overline{X}_{{i},{h},A}\hspace*{.2in} \text{for all}\hspace*{.1in} ({i},{h},A) \ne (i_1,h_1,A_1),\\
			\Gamma_g(\overline{Z}_{{i},{h},A}) &= \overline{Z}_{{i},{h},A}\hspace*{.2in} \text{for all}\hspace*{.1in} ({i},{h},A) \ne (i_2,h_2,A_2),A \in \{L,R\}
		\end{split}
	\end{equation}
	\begin{proposition}\label{prop:logical_cnot}
		For $\cQ(C_a,C_b)$, the two-qubit targeted logical CNOT with $(i_1,h_1,L)$ as the control and $(i_2,h_2,L)$ as the target has the following symplectic representation:
		$$F^{CNOT}_{(i_1,h_1,L)\rightarrow (i_2,h_2,L)} = \begin{bmatrix}
			I_{\tilde{n}\times \tilde{n}}+\tilde{D}^{\intercal} & 0_{\tilde{n} \times \tilde{m}} & 0_{\tilde{n} \times \tilde{n}} & 0_{\tilde{n} \times \tilde{m}}\\
			0_{\tilde{m} \times \tilde{n}} & I_{\tilde{m} \times \tilde{m}} & 0_{\tilde{m} \times \tilde{n}} & 0_{\tilde{m} \times \tilde{m}}\\
			0_{\tilde{n} \times \tilde{n}} & 0_{\tilde{n} \times \tilde{m}} &I_{\tilde{n}\times \tilde{n}}+\tilde{D} & 0_{\tilde{n} \times \tilde{m}}\\
			0_{\tilde{m} \times \tilde{n}} & 0_{\tilde{m} \times \tilde{m}} & 0_{\tilde{m}\times\tilde{n}} & I_{\tilde{m}\times\tilde{m}}
		\end{bmatrix}$$
		where \begin{align*}
			\tilde{D}^\intercal  =(a_{i_1} \otimes e_{h_1})^\intercal  (e_{i_2} \otimes b_{h_2}).
		\end{align*}
        
		Similarly, the targeted logical CNOT with $(i,h,L)$ as the control and $(j,l.R)$ as the target has the following symplectic representation:
		$$F^{CNOT}_{(i,h,L)\rightarrow (j,l,R)} = \begin{bmatrix}
			I_{\tilde{n}\times \tilde{n}} & \tilde{D}^{\intercal} & 0_{\tilde{n} \times \tilde{n}} & 0_{\tilde{n} \times \tilde{m}}\\
			0_{\tilde{m} \times \tilde{n}} & I_{\tilde{m} \times \tilde{m}} & 0_{\tilde{m} \times \tilde{n}} & 0_{\tilde{m} \times \tilde{m}}\\
			0_{\tilde{n} \times \tilde{n}} & 0_{\tilde{n} \times \tilde{m}} & I_{\tilde{n}\times \tilde{n}} & 0_{\tilde{n} \times \tilde{m}}\\
			0_{\tilde{m} \times \tilde{n}} & 0_{\tilde{m} \times \tilde{m}} & \tilde{D} & I_{\tilde{m}\times\tilde{m}}
		\end{bmatrix}$$
		where \begin{align*}
			\tilde{D}^\intercal  =(a_{i} \otimes e_{h})^\intercal  (\alpha_j\otimes e_l).
		\end{align*}
	\end{proposition}
	\begin{proof}
		By Corollary \ref{cor:tranvections}, the logical CNOT gate with the $(i_1,h_1,L)$-th logical qubit as the control and $(i_2,h_2,L)$-th logical qubit as the target is given by $$T_{[u_{i_2,h_2,L},0]}T_{[u_{i_2,h_2,L},v_{i_1,h_1,L}]}T_{[0 ,v_{i_1,h_1,L}]}$$ where $u_{i_2,h_2,L} = [e_{i_2} \otimes b_{h_2}, 0], v_{i_1,h_1,L} = [a_{i_1} \otimes e_{h_1} , 0]$. Hence, $$F^{CNOT}_{(i_1,h_1,L)\rightarrow (i_2,h_2,L)}  = I + \begin{bmatrix}
			v_{i_1,h_1,L}^\intercal u_{i_2,h_2,L} & 0\\
			0 & u_{i_2,h_2,L}^\intercal v_{i_1,h_1,L}
		\end{bmatrix}.$$ The case where the two logical qubits are in different sectors can be proven similarly.
	\end{proof}
	\begin{theorem}\label{thm:circuit_cnot}
		Let $\cQ(C_a,C_b)$ be an HGP code. 
        
        (a) There exists a Clifford circuit $\sfC$ that implements $\overline{CNOT}_{(i_1,h_1,L)\rightarrow (i_2,h_2,L)}$ for logical qubits $\overline{q}_{i_1,h_1,L}$ and $\overline{q}_{i_2,h_2,L}$ in the same sector. 
        
        (b) There exists a Clifford circuit $\sfC$ that implements $\overline{CNOT}_{(i,h,L)\rightarrow (j,l,R)}$ for logical qubits $\overline{q}_{i,h,L}$ and $\overline{q}_{j,l,R}$ in different sectors.

        In both cases the support $\chi(\sfC)$ and depth $\delta(\sfC)$ are of order $\Theta(\sqrt{n})$
	\end{theorem}
	\begin{proof}
		We shall show that for both cases one can find elementary row operations from Table \ref{table:circuit_algo} which, when applied to the identity matrix, produce the desired symplectic matrices of Proposition \ref{prop:logical_cnot}. 
        
        (a) Consider the case where the two qubits are in different sectors and let $v = (a_i\otimes e_h), u = (\alpha_j \otimes e_l)$. The target matrix is:
		$$F = \begin{bmatrix}
			I_{\tilde{n}\times \tilde{n}} & v^{\intercal}u & 0_{\tilde{n} \times \tilde{n}} & 0_{\tilde{n} \times \tilde{m}}\\
			0_{\tilde{m} \times \tilde{n}} & I_{\tilde{m} \times \tilde{m}} & 0_{\tilde{m} \times \tilde{n}} & 0_{\tilde{m} \times \tilde{m}}\\
			0_{\tilde{n} \times \tilde{n}} & 0_{\tilde{n} \times \tilde{m}} & I_{\tilde{n}\times \tilde{n}} & 0_{\tilde{n} \times \tilde{m}}\\
			0_{\tilde{m} \times \tilde{n}} & 0_{\tilde{m} \times \tilde{m}} & u^{\intercal}v & I_{\tilde{m}\times\tilde{m}}
		\end{bmatrix}$$
	Perform the following row operations on $I_{2n \times 2n}$:
	\begin{enumerate}
		\item Select a row index $x \in \Supp(v)$. For all $y \in \Supp(v)\setminus\{x\}$, add row $x$ to row $y$ and add row $y+n$ to row $x+n$.
		\item For all $\bar{y} \in \Supp(u)$, add row $\bar{y} + \tilde{n}$ to row $x$ and add row $x+n$ to row $\bar{y} + \tilde{n} +n$. 
		\item For all $y \in \Supp(v)\setminus\{x\}$, add row $x$ to row $y$ and add row $y+n$ to row $x+n$.
	\end{enumerate}
		
	(b)	Now consider the case when both logical qubits are in the same sector and let $v = (a_{i_1}\otimes e_{h_1}), u = (e_{i_2}\otimes b_{h_2})$. Due to the fact that $\{a_i\}, \{\alpha_j\}, \{b_h\}, \{\beta_l\}$ are SLT bases, we have that $\Supp(u) \cap \Supp(v) = \emptyset$ whenever $i_1\ne i_2$ or $h_1\ne h_2$. The target matrix is:
		$$F = \begin{bmatrix}
			I_{\tilde{n}\times \tilde{n}}+v^{\intercal}u & 0_{\tilde{n} \times \tilde{m}} & 0_{\tilde{n} \times \tilde{n}} & 0_{\tilde{n} \times \tilde{m}}\\
			0_{\tilde{m} \times \tilde{n}} & I_{\tilde{m} \times \tilde{m}} & 0_{\tilde{m} \times \tilde{n}} & 0_{\tilde{m} \times \tilde{m}}\\
			0_{\tilde{n} \times \tilde{n}} & 0_{\tilde{n} \times \tilde{m}} & I_{\tilde{n}\times \tilde{n}} & 0_{\tilde{n} \times \tilde{m}}\\
			0_{\tilde{m} \times \tilde{n}} & 0_{\tilde{m} \times \tilde{m}} & 0_{\tilde{m} \times \tilde{n}} & I_{\tilde{m}\times\tilde{m}}+u^{\intercal}v
		\end{bmatrix}$$
Perform the following row operations on $I_{2n \times 2n}$:
		\begin{enumerate}
		\item Select a row index $x \in \Supp(v)$. For all $y \in \Supp(v)\setminus\{x\}$, add row $x$ to row $y$ and add row $y+n$ to row $x+n$.
		\item For all $\bar{y} \in \Supp(u)$, add row $\bar{y}$ to row $x$ and add row $x+n$ to row $\bar{y} +n$.
		\item For all $y \in \Supp(v)\setminus\{x\}$, add row $x$ to row $y$ and add row $y+n$ to row $x+n$.
		\end{enumerate}
Note that the only change from the previous case is in the second step. It can be easily verified that in both cases performing  operations (1)-(3) on $I_{2n\times 2n}$ produces the desired matrix $F$. We collect the details of the calculations in Appendix \ref{appndx:cnot}. 
		
		The next task is to translate the row operations to physical gates, which is easily done using Table \ref{table:circuit_algo}. Each of the steps above corresponds to performing a series of physical CNOT gates on specific qubits. Combining the two cases, we give a generic algorithm, Algorithm~\ref{alg:algo_logical_cnot}, for constructing the circuit that performs a CNOT between any pair of logical qubits of an HGP code. Let $I_{{\sf c}}$ be the set of physical qubits that are in the support of the $Z$ logical Pauli of the control  qubit and let $I_{{\sf t}}$ be set of qubits in the support of the $X$ logical Pauli of the target  qubit. Note that due to the SLT property, $I_{{\sf c}}$ and $I_{{\sf t}}$ always have empty overlap irrespective of
        their relative location (the same sector or not).
		\begin{algorithm}
			\caption{Logical CNOT gate circuit design algorithm}
			\label{alg:algo_logical_cnot}
			\hspace*{-5.0in} \textbf{Input:} $I_{{\sf c}}$ and $I_{{\sf t}}$
			\begin{algorithmic}[1]
				\State{Select any qubit $x \in I_{{\sf c}}$}
                \State \textbf{CNOT layer}    
                \Statex \hspace{\algorithmicindent} For {$y \in I_{{\sf c}}\setminus\{x\}$}, apply $CNOT$ from $y$ to $x$
                \State \textbf{CNOT layer}    
                \Statex \hspace{\algorithmicindent} For {$y \in I_{{\sf t}}$}, apply $CNOT$ from $x$ to $y$
               \State \textbf{CNOT layer}    
                \Statex \hspace{\algorithmicindent} For {$y \in I_{{\sf c}}\setminus\{x\}$}, apply $CNOT$ from $y$ to $x$
			\end{algorithmic}
		\end{algorithm}
  
  Let us check that the constructed circuits preserve the stabilizers with their phases.
In this case this is straightforward because HGP codes belong to the CSS family, 
and hence the generators of the stabilizer group are composed of either $X$ operators only or $Z$ operators only. A circuit composed purely of CNOT gates cannot introduce any phase change to Paulis composed purely of either $X$ or $Z$ operators. Hence, the stabilizer generators of the code are preserved with their phases.
  
		Clearly, the circuit acts only on the set of qubits indexed by $I_{{\sf c}}\cup I_{{\sf t}}$ and so $\chi = |a_{i_1}|+|b_{h_2}|$. The circuit depth is at most $\delta = |b_{h_2}|+2|a_{i_1}|-2$.  Since $\{a_i\}$ is an SLT basis of $\ker(H_a)$ and $\{b_h\}$ is an SLT basis of $\ker(H_b)$, by Theorem \ref{thm:logical_paulis_hp}, $|a_{i_1}| = |b_{h_2}|= \Theta(\sqrt{n})$, and we conclude that both $\chi, \delta$ for the circuit are of order $\Theta(\sqrt{n})$. 
\end{proof}

    \subsubsection{Targeted Logical CZ Gate}\label{subsub:cz_HP}
Although the logical Hadamard, Phase, and CNOT gates are enough to generate the logical Clifford group, 
we additionally give a construction of the CZ gate as a possible alternative to CNOT. Again 
we have to consider two cases depending on whether the logical qubits involved are in the same sector 
or not. 

Let $\bar q_{i_1,h_1,A_1}$ and $\bar q_{i_2,h_2,A_2}$ be two distinct logical qubits with $A_1,A_2 \in \{L,R\}$ and let 
$g=\overline{CZ}_{(i_1,h_1,A_1)\leftrightarrow (i_2,h_2,A_2)}$ denote the logical CZ between these 
qubits, with the arrow indicating that it is symmetric with respect to the two 
logical qubits involved. For this operator, the logical mapping of Paulis $\Gamma_g$ is described as follows:
	\begin{equation*}
		\begin{split}
			\Gamma_g(\overline{X}_{i_1,h_1,A_1}) &= \overline{X}_{i_1,h_1,A_1}\overline{Z}_{i_2,h_2,A_2},\\
			\Gamma_g(\overline{X}_{i_2,h_2,A_2}) &= \overline{Z}_{i_1,h_1,A_1}\overline{X}_{i_2,h_2,A_2},\\
			\Gamma_g(\overline{X}_{i,{h},A}) &= \overline{X}_{i,{h},A}\hspace*{.2in} \text{ for all }\hspace*{.1in} (i,{h},A) \ne (i_1,h_1,A_1),(i_2,h_2,A_2),\\
			\Gamma_g(\overline{Z}_{i,{h},A}) &= \overline{Z}_{i,{h},A}\hspace*{.2in} \text{ for all }\hspace*{.1in} (i,{h},A), A \in \{L,R\}.
		\end{split}
	\end{equation*}
    \begin{proposition}\label{prop:logical_cz}
Let $\cQ(C_a,C_b)$ be an HGP code. The two-qubit targeted logical CZ between $(i_1,h_1,L)$ and $(i_2,h_2,L)$ has the following symplectic representation:
		$$F^{CZ}_{(i_1,h_1,L)\leftrightarrow (i_2,h_2,L)}  = \begin{bmatrix}
			I_{\tilde{n} \times \tilde{n}} & 0_{\tilde{n} \times \tilde{m}} & \tilde{B} & 0_{\tilde{n} \times \tilde{m}}\\
			0_{\tilde{m} \times \tilde{n}} & I_{\tilde{m} \times \tilde{m}} & 0_{\tilde{m} \times \tilde{n}} & 0_{\tilde{m} \times \tilde{m}}\\
			0_{\tilde{n} \times \tilde{n}} & 0_{\tilde{n} \times \tilde{m}} &I_{\tilde{n} \times \tilde{n}} & 0_{\tilde{n} \times \tilde{m}}\\
			0_{\tilde{m} \times \tilde{n}} & 0_{\tilde{m} \times \tilde{m}} & 0_{\tilde{m}\times\tilde{n}} & I_{\tilde{m}\times\tilde{m}},
		\end{bmatrix}$$
		where \begin{align*}
			\tilde{B} &= (a_{i_1} \otimes e_{h_1})^\intercal  (a_{i_2} \otimes e_{h_2})+(a_{i_2} \otimes e_{h_2})^\intercal  (a_{i_1} \otimes e_{h_1}).
		\end{align*}
		Similarly, the targeted logical CZ between qubits $(i,h,L)$ and $(j,l.R)$ has the following symplectic representation:
		$$F^{CZ}_{(i,h,L) \leftrightarrow (j,l,R)}  = \begin{bmatrix}
			I_{\tilde{n}\times \tilde{n}} &  0_{\tilde{n} \times \tilde{m}} & 0_{\tilde{n} \times \tilde{n}} & \tilde{B}\\
			0_{\tilde{m} \times \tilde{n}} & I_{\tilde{m} \times \tilde{m}} & \tilde{B}^\intercal  & 0_{\tilde{m} \times \tilde{m}}\\
			0_{\tilde{n} \times \tilde{n}} & 0_{\tilde{n} \times \tilde{m}} & I_{\tilde{n}\times \tilde{n}} & 0_{\tilde{n} \times \tilde{m}}\\
			0_{\tilde{m} \times \tilde{n}} & 0_{\tilde{m} \times \tilde{m}} & 0_{\tilde{m} \times \tilde{n}} & I_{\tilde{m}\times\tilde{m}},
		\end{bmatrix}$$
		where \begin{align*}
			\tilde{B} =(a_{i} \otimes e_{h})^\intercal  (e_j \otimes \beta_l).
		\end{align*}
	\end{proposition}
	\begin{proof}
We prove the claim for the case of different sectors. 
By Corollary \ref{cor:tranvections}, a logical CZ gate between logical qubits $(i,h,L)$ and $(j,l,R)$ is given by 
    $$
    T_{[0,v_{i,h,L}]}T_{[0,v_{i,h,L}+v_{j,l,R}]}T_{[0 ,v_{j,l,R}]},
    $$ 
where $ v_{i,h,L} = [a_i \otimes e_{h} , 0], v_{j,l,R} = [0, e_j\otimes \beta_l],$. Hence, 
    $$
    F^{CZ}_{(i,h,L) \leftrightarrow (j,l,R)}  = I + \begin{bmatrix}
			0 & v_{i,h,L}^\intercal v_{j,l,R}+ v_{j,l,R}^\intercal v_{i,h,L}\\
			0 & 0
		\end{bmatrix}.\qedhere
        $$ 
	\end{proof}
    
	\begin{theorem}\label{thm:circuit_cz} Let $\cQ(C_a,C_b)$ be an HGP code.
		
        (a) There exists a Clifford circuit $\sfC$ that implements $\overline{CZ}_{(i_1,h_1,L)\leftrightarrow (i_2,h_2,L)}$ for logical qubits $\overline{q}_{i_1,h_1,L}$ and $\overline{q}_{i_2,h_2,L}$ in the same sector. 
        
        (b) There exists a Clifford circuit $\sfC$ that implements $\overline{CZ}_{(i,h,L)\leftrightarrow (j,l,R)}$ for logical qubits $\overline{q}_{i,h,L}$ and $\overline{q}_{j,l,R}$ in different sectors.

In both cases the support $\chi(\sfC)$ and depth $\delta(\sfC)$ are of order $\Theta(\sqrt{n})$. 
	\end{theorem}
	\begin{proof}
    We will again show that both for cases (a) and (b) we can find a combination of elementary row operations from Table \ref{table:circuit_algo} which when performed on the identity matrix produce the desired symplectic matrices of Proposition \ref{prop:logical_cz}. 
    
    (a) First let us consider the case where the two qubits are in different sectors and let $v = (a_i\otimes e_h), u = (e_j\otimes \beta_l)$. The target matrix is:
		$$
        F = \begin{bmatrix}
			I_{\tilde{n}\times \tilde{n}} &  0_{\tilde{n} \times \tilde{m}} & 0_{\tilde{n} \times \tilde{n}} & v^{\intercal}u\\
			0_{\tilde{m} \times \tilde{n}} & I_{\tilde{m} \times \tilde{m}} & u^{\intercal}v  & 0_{\tilde{m} \times \tilde{m}}\\
			0_{\tilde{n} \times \tilde{n}} & 0_{\tilde{n} \times \tilde{m}} & I_{\tilde{n}\times \tilde{n}} & 0_{\tilde{n} \times \tilde{m}}\\
			0_{\tilde{m} \times \tilde{n}} & 0_{\tilde{m} \times \tilde{m}} & 0_{\tilde{m} \times \tilde{n}} & I_{\tilde{m}\times\tilde{m}}
		\end{bmatrix}
        $$
		Perform the following row operations on a $2n \times 2n$ identity matrix in order:
		\begin{enumerate}
			\item Select a row index $x \in \Supp(v)$. For all $y \in \Supp(v)\setminus\{x\}$, add row $x$ to row $y$ and add row $y+n$ to row $x+n$.
			\item Select a row index $\bar{x} \in \Supp(u)$. For all $\bar{y} \in \Supp(u)\setminus\{\bar{x}\}$, add row $\bar{x}+\tilde{n}$ to row $\bar{y}+\tilde{n}$ and add row $\bar{y}+\tilde{n}+n$ to row $\bar{x}+n$.
			\item Add row $\bar{x}+\tilde{n}+n$ to row $x$ and add row $x+n$ to row $\bar{x}+\tilde{n}$. 
			\item For all $y \in \Supp(v)\setminus\{x\}$, add row $x$ to row $y$ and add row $y+n$ to row $x+n$.
			\item For all $\bar{y} \in \Supp(u)\setminus\{\bar{x}\}$, add row $\bar{x}+\tilde{n}$ to row $\bar{y}+\tilde{n}$ and add row $\bar{y}+\tilde{n}+n$ to row $\bar{x}+n$.
		\end{enumerate}
		
		(b) Now consider the case when both logical qubits are in the same sector and let $v = (a_{i_1}\otimes e_{h_1}), u = (a_{i_2}\otimes e_{h_2})$ and the target matrix is:
		$$F = \begin{bmatrix}
			I_{\tilde{n}\times \tilde{n}} & 0_{\tilde{n} \times \tilde{m}} & v^{\intercal}u+u^{\intercal}v & 0_{\tilde{n} \times \tilde{m}}\\
			0_{\tilde{m} \times \tilde{n}} & I_{\tilde{m} \times \tilde{m}} & 0_{\tilde{m} \times \tilde{n}} & 0_{\tilde{m} \times \tilde{m}}\\
			0_{\tilde{n} \times \tilde{n}} & 0_{\tilde{n} \times \tilde{m}} & I_{\tilde{n}\times \tilde{n}} & 0_{\tilde{n} \times \tilde{m}}\\
			0_{\tilde{m} \times \tilde{n}} & 0_{\tilde{m} \times \tilde{m}} & 0_{\tilde{m} \times \tilde{n}} & I_{\tilde{m}\times\tilde{m}}
		\end{bmatrix}.$$ 
		In this case, the two vectors $u$ and $v$ may have a non-trivial overlap. However, we note that due to the fact that $\{a_i\},\{b_h\}$ are SLT bases for $\ker(H_a), \ker(H_b)$ respectively, we have $v_{\tilde{n}(i_1-1)+h_1} = 1, u_{\tilde{n}(i_1-1)+h_1}=0$ and $u_{\tilde{n}(i_2-1)+h_2} = 1, v_{\tilde{n}(i_2-1)+h_2}=0$. Note that this property was trivially satisfied by the vectors in the previous case because the logical qubits were in different sectors and so the supports of the two logical $Z$ Pauli operators were disjoint. With this property in mind, consider the following row operations on $I_{2n \times 2n}$ in order:
			\begin{enumerate}
			\item For all $y \in \Supp(v)\setminus\{\tilde{n}(i_1-1)+h_1\}$, add row $\tilde{n}(i_1-1)+h_1$ to row $y$ and add row $y+n$ to row $\tilde{n}(i_1-1)+h_1+n$.
			\item For all $\bar{y} \in \Supp(u)\setminus\{\tilde{n}(i_2-1)+h_2\}$, add row $\tilde{n}(i_2-1)+h_2$ to row $\bar{y}$ and add row $\bar{y}+n$ to row $\tilde{n}(i_2-1)+h_2$.
			\item Add row $\tilde{n}(i_1-1)+h_1+n$ to row $\tilde{n}(i_2-1)+h_2$ and add row $\tilde{n}(i_2-1)+h_2+n$ to row $\tilde{n}(i_1-1)+h_1$. 
			\item For all $y \in \Supp(v)\setminus\{\tilde{n}(i_1-1)+h_1\}$, add row $\tilde{n}(i_1-1)+h_1$ to row $y$ and add row $y+n$ to row $\tilde{n}(i_1-1)+h_1+n$.
			\item For all $\bar{y} \in \Supp(u)\setminus\{\tilde{n}(i_2-1)+h_2\}$, add row $\tilde{n}(i_2-1)+h_2$ to row $\bar{y}$ and add row $\bar{y}+n$ to row $\tilde{n}(i_2-1)+h_2$.
		\end{enumerate}
        The proof of correctness of the two sequences of row operations is not difficult and is given in Appendix \ref{appndx:cz}. We translate the row operations to physical gates using Table \ref{table:circuit_algo}. The row operations in both cases correspond to CNOT and CZ gates. The two cases have been combined by carefully choosing the qubits $x$ and $\bar{x}$ which is always possible as we have argued. Hence, the generic algorithm for constructing the circuit for performing CZ between arbitrary logical qubits for any HP codes is given in Algorithm \ref{alg:algo_logical_cz}. Let $I_1$ and $I_2$ be the sets of physical qubits that are in the support of the $Z$ logical Paulis corresponding to the two logical qubits. 
		\begin{algorithm}
			\caption{Logical CZ gate circuit design algorithm}
			\label{alg:algo_logical_cz}
			\hspace*{-5.0in} \textbf{Input:} $I_1$ and $I_2$
			\begin{algorithmic}[1]
				\State{Select a qubit $x \in I_1\setminus I_2$ and select a qubit $\bar{x} \in I_2\setminus I_1$}
                \State \textbf{CNOT layer}    
                \Statex \hspace{\algorithmicindent} For {$y \in I_1\setminus\{x\}$}, apply $CNOT$ from $y$ to $x$
                \State \textbf{CNOT layer}    
                \Statex \hspace{\algorithmicindent} For {$\bar{y} \in I_2\setminus\{\bar{x}\}$}, apply $CNOT$ from $\bar{y}$ to $\bar{x}$
                \State{$CZ$ between qubits $x$ and $\bar{x}$} 
                \State \textbf{CNOT layer}    
                \Statex \hspace{\algorithmicindent} For {$y \in I_1\setminus\{x\}$}, apply $CNOT$ from $y$ to $x$
                \State \textbf{CNOT layer}    
                \Statex \hspace{\algorithmicindent} For {$\bar{y} \in I_2\setminus\{\bar{x}\}$}, apply $CNOT$ from $\bar{y}$ to $\bar{x}$
			\end{algorithmic}
		\end{algorithm}
  
  We now show that the circuit given in Algorithm \ref{alg:algo_logical_cz} preserves the stabilizers with phases. We shall do this by tracking the changes of any $X$- or $Z$-type stabilizer as it progresses along the circuit. Consider the three stages of the circuit: the first stage composed purely of CNOT gates, the second stage of only a single CZ gate, and the third stage composed again of CNOT gates. Any $Z$-type stabilizer, when acted upon by the successive stages remains a $Z$ type stabilizer with no possible phase change at any point. On the other hand an $X$ type stabilizer can be converted to a mix of $X$ and $Z$ type due to the CZ gate between qubits $x$ and $\bar{x}$ in the second stage. Even if this happens, i.e, the second stage introduces a $Z$ operator on qubit $x$ and/or $\bar{x}$, we know that this change will be reversed by the third stage of CNOT gates by introduction of another $Z$ operator on the corresponding qubit(s) due to the fact that the circuit is guaranteed to fix all the $X$ stabilizers up to their support. These operations do not introduce any negative signs as $XZ=Y, YZ =X, Z^2 =I$. Hence all stabilizers are preserved with their phase.

		The circuit acts only on the set of qubits indexed by $I_1\cup I_2$, hence $\chi = |a_{i}|+|\beta_l|$, while the circuit depth is $\delta = 2(|a_{i}|+|\beta_l|)+1$. The parameter $\tau$ is easily calculated to be $\max\{|a_{i}|,|\beta_l|\}$. Since $\{a_i\}$ is an SLT basis of $\ker(H_a)$ and $\{\beta_l\}$ is an SLT basis of $\ker(H_b^\intercal )$ then by Theorem \ref{thm:logical_paulis_hp},  $|a_i| = |\beta_l|= \Theta(\sqrt{n})$, and we have that $\chi, \delta, \tau$ for the circuit are all of order $\Theta(\sqrt{n})$. Choosing the two logical qubits to be in the same sector only requires minor modification to the algorithm above while the circuit parameters remain of the same order.
	\end{proof}
    
\subsubsection{Examples for toric codes}\label{sec:toric_logical}
Toric codes form a particular case of the HGP family obtained if the constituent classical codes are repetition codes;
see \cite[Sec.3]{Tillich2014} for the details of this specification.
To exemplify the general construction presented above, we explicitly construct circuits for logical Cliffords for the $\cT_3[[18,2,3]]$ toric code. Note that 
   $$
   H_a =H_b = H_a^\intercal  = H_b^\intercal  = H = \begin{bmatrix}
		1 & 1 & 0\\
		1 & 0 & 1\\
		0 & 1 & 1
	\end{bmatrix}; \hspace*{.2in}\ker(H) =\begin{pmatrix}
		1  1  1
	\end{pmatrix},$$
and 
    $\sfP(\begin{bmatrix}
		1 & 1 & 1
	\end{bmatrix}) = \{3\}$. We use the two-grid notation introduced in Section 
	\ref{subsec:HP} to denote the physical qubits by $(i,h,L)$ for the left sector and $(j,l,R)$ for the right sector with $i,h,j,l =1,2,3$, ignoring the third index whenever there is no possibility of confusion. The code encodes two logical qubits, which we call the left and the right logical qubit below.
	
\vspace*{.1in}\par
\begin{figure}[H]
     \begin{subfigure}[b]{0.45\textwidth}
         \centering
    \includegraphics[width=\linewidth]{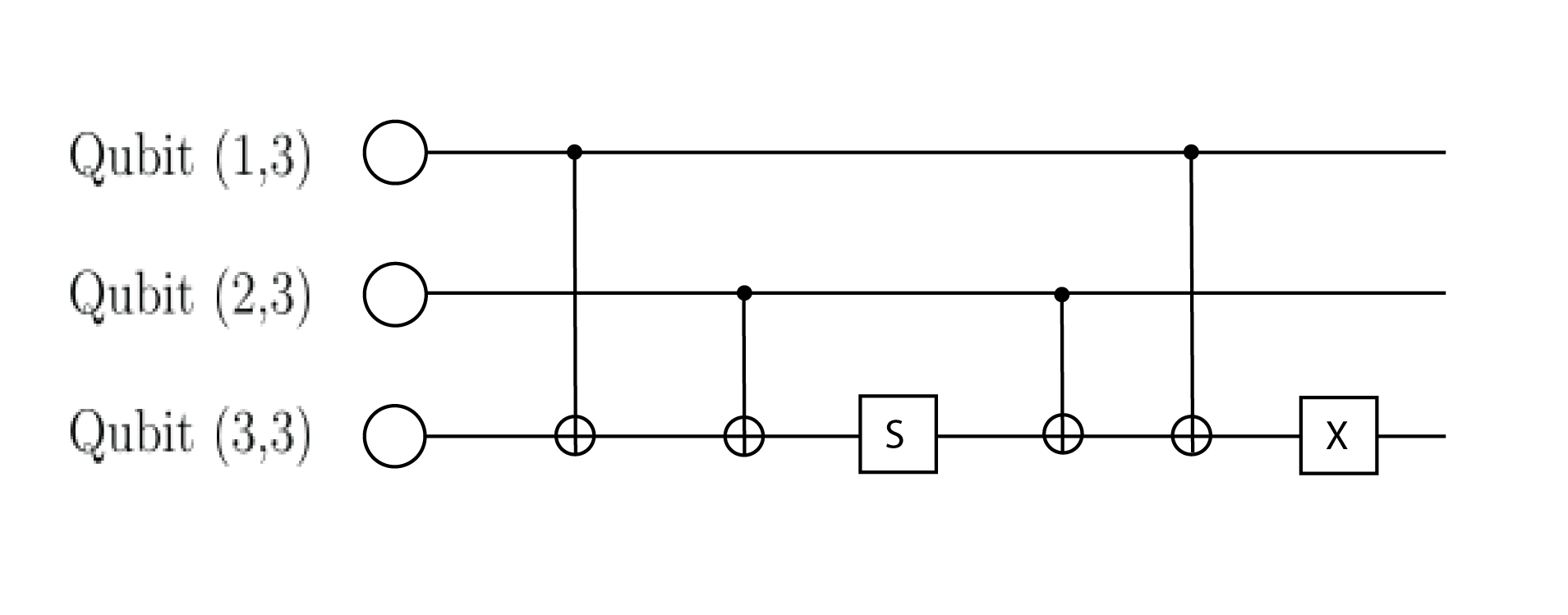}
         \caption{Circuit implementation using Algorithm \ref{alg:algo_logical_phase_new}}
         \label{fig:toric_phase a}
     \end{subfigure}
     \hfill
     \begin{subfigure}[b]{0.45\textwidth}
         \centering
         \includegraphics[width=\linewidth]{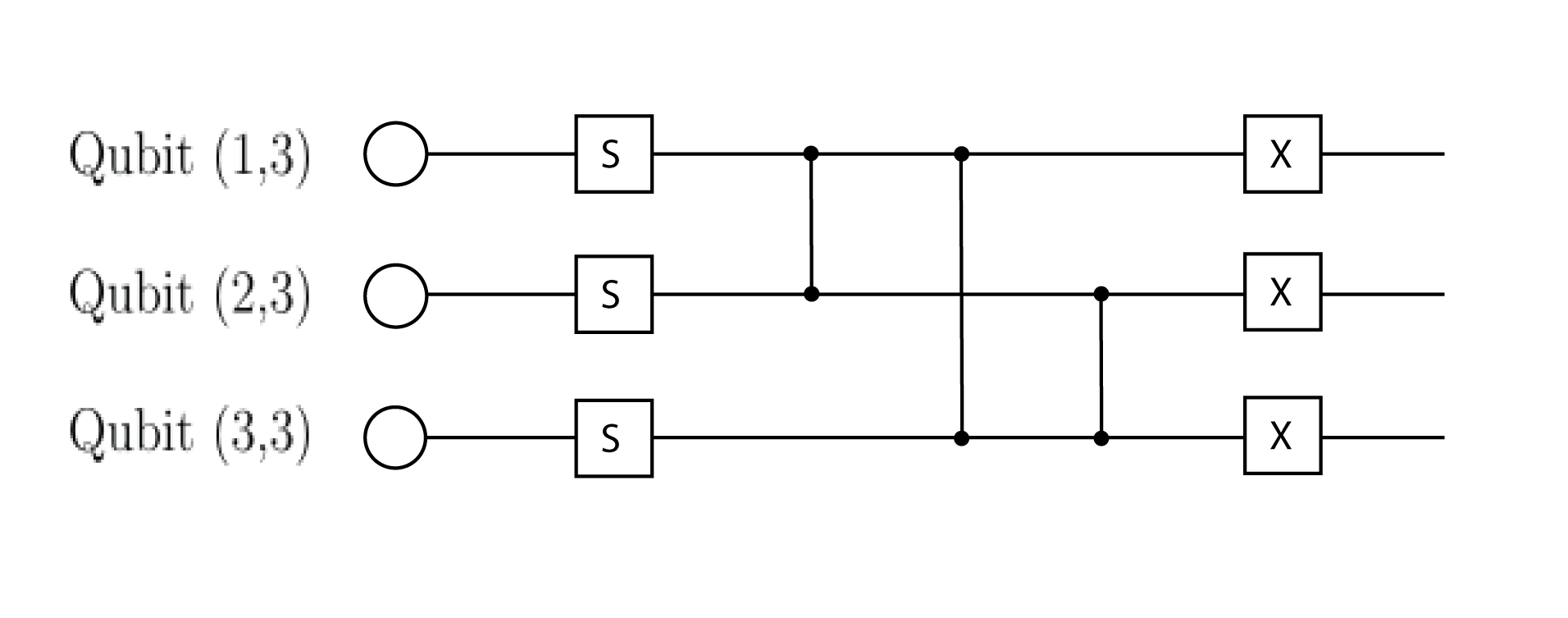}
         \caption{Circuit implementation using Algorithm \ref{alg:algo_logical_phase}}
         \label{fig:toric_phase b}
     \end{subfigure}
     \hfill
            \caption{Logical Phase gate $\overline{S}_L$ for the code $\cT_3$}
        \label{fig:toric_phase}
\end{figure}

\noindent\textbf{Logical Phase gate on left logical qubit:} By Proposition \ref{prop:logical_P},  the symplectic matrix that performs logical Phase on the left logical qubit is given by:
	   $$
       F_{\overline{S}_L} = \begin{bNiceArray}{cc|cc}
		\Block{2-2}<\Large>{I_{18}} && v^\intercal v & 0\\
		&& 0 & 0\\
		\hline
		0 & 0 & \Block{2-2}<\Large>{I_{18}} \\
		0 & 0 	
	\end{bNiceArray}
    $$ 
    with $v = [
		0  0  1  0  0  1  0  0  1].$ 
Using Algorithm \ref{alg:algo_logical_phase_new}, we obtain a circuit that implements the logical Phase $\overline{S}_L$, shown in Fig. \ref{fig:toric_phase a}. An alternative circuit, using Algorithm \ref{alg:algo_logical_phase}, is given in Fig. \ref{fig:toric_phase b}.

	\par\noindent\textbf{Logical Hadamard gate on left logical qubit:} By Proposition \ref{prop:logical_H}, the symplectic matrix that performs 
the logical Hadamard operation on the left logical qubit is given by:
	$$
    F_{\overline{H}_L} = \begin{bNiceArray}{cc|cc}
		I_9 + v^\intercal u & 0 & v^\intercal v & 0\\
		0 & I_9 & 0 & 0\\
		\hline
		u^\intercal u & 0 & I_9 + u^\intercal v & 0\\
		0 & 0 & 0 & I_9
	\end{bNiceArray}$$ 
    with $u =(0 0 0 0 0 0 1 1 1)
	$ and $v =(0 0 1 0 0 1 0 0  1)$. Using Algorithm \ref{alg:algo_logical_hadamard}, we obtain the circuit in Fig. \ref{fig:toric_log_hadamard} that implements the logical Hadamard operation:
	\begin{center}
		\begin{figure}[h]
			\includegraphics[scale=0.6]{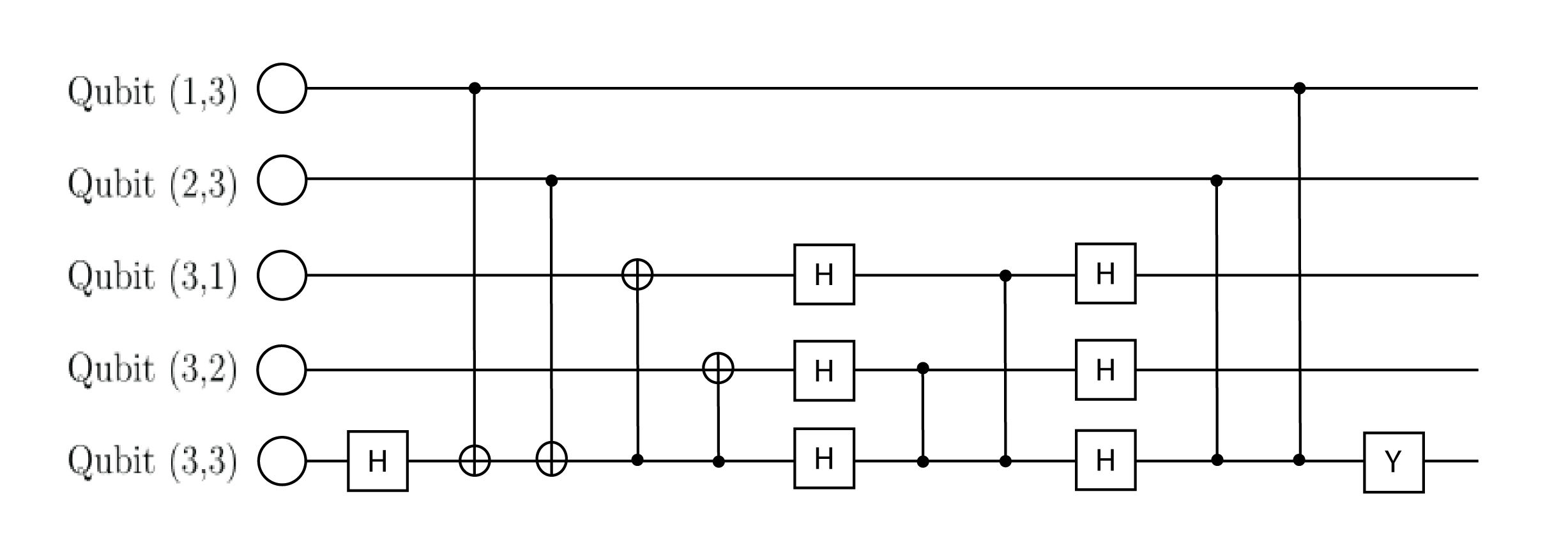}
			\centering
			\caption{Logical Hadamard gate $\overline{H}_L$ for the code $\cT_3$}
			\label{fig:toric_log_hadamard}
		\end{figure}
	\end{center}
    
	\par\noindent\textbf{Logical left-to-right qubit CNOT gate:} 
By Proposition \ref{prop:logical_cnot}, the symplectic matrix that performs logical CNOT with left and right logical qubits as control and target, respectively,
has the form
	$$F_{\overline{H}_L} = \begin{bNiceArray}{cc|cc}
		I_9 & v^\intercal v & 0 & 0\\
		0 & I_9 & 0 & 0\\
		\hline
		0 & 0 & I_9 & 0\\
		0 & 0 & v^\intercal v & I_9
	\end{bNiceArray}$$ 
    with $v = (0 0 1 0 0 1 0 0 1).$
Using Algorithm \ref{alg:algo_logical_cnot}, we 
    obtain the circuit in Fig. \ref{fig:toric_log_cnot} for the logical CNOT:
	\begin{center}
		\begin{figure}[h]
			\includegraphics[scale=0.5]{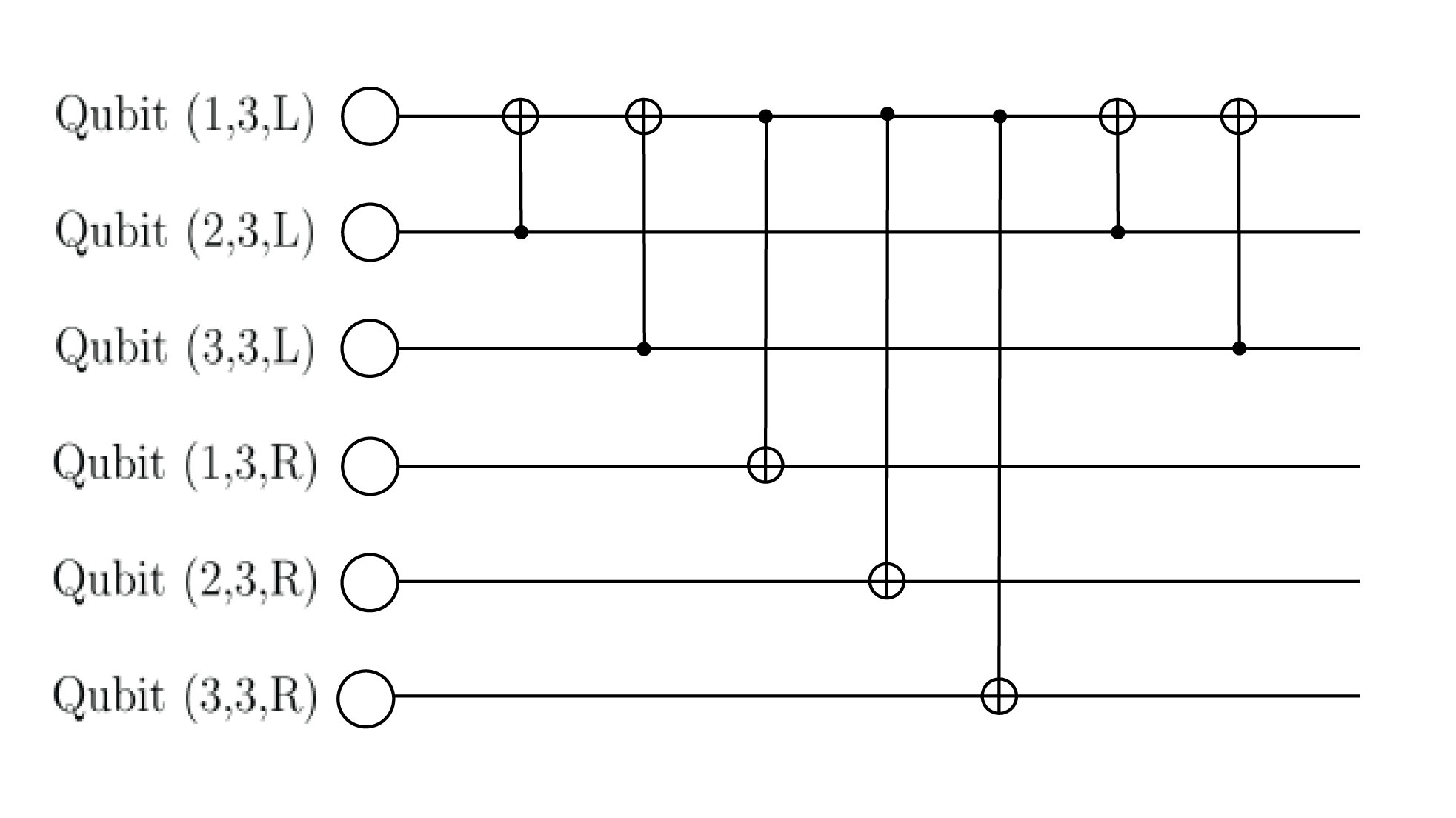}
			\centering
			\caption{Logical CNOT gate $\overline{CNOT}_{L \rightarrow R}$ for the code $\cT_3$}
			\label{fig:toric_log_cnot}
		\end{figure}
	\end{center}
    
	\par
\begin{center}
		\begin{figure}[h]
			\includegraphics[scale=0.5]{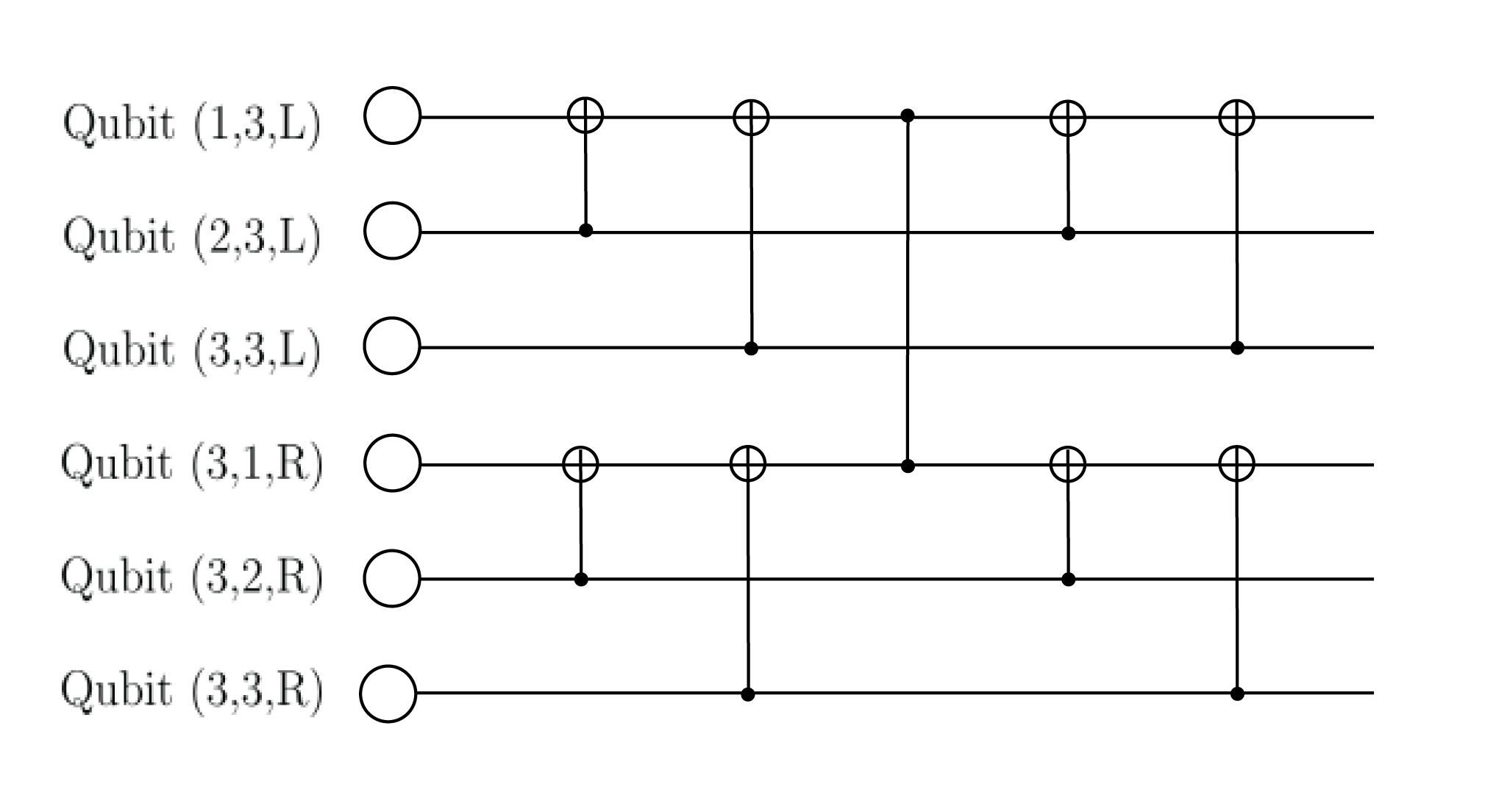}
			\centering
			\caption{Logical CZ gate $\overline{CZ}_{L,R}$ for the code $\cT_3$}
			\label{fig:toric_log_cz}
		\end{figure}
	\end{center}

\noindent\textbf{Logical CZ gate between the left and right logical qubit:} 
    By Proposition \ref{prop:logical_cz}, the symplectic matrix that performs logical CZ between the left and right logical qubits has the form
    Using Proposition \ref{prop:logical_cz}, we have the symplectic matrix performing logical CZ between the left and right logical qubit:
	$$
    F_{\overline{H}_L} = \begin{bNiceArray}{cc|cc}
		I_9 & 0 & 0 & v^\intercal u\\
		0 & I_9 & u^\intercal v & 0\\
		\hline
		0 & 0 & I_9 & 0\\
		0 & 0 & 0 & I_9
	\end{bNiceArray}
    $$ 
    with 
    $u = (0 0 0 0 0 0 1 1 1)$ and $v = (0 0 1 0 0 1 0 0  1)$. Using Algorithm \ref{alg:algo_logical_cz}, we see that the circuit in Fig. \ref{fig:toric_log_cz} implements the logical operation. 

Concluding this section, we also mention surface codes \cite{fowler2012surface}, which are obtained from toric codes by simply discarding the periodicity of the lattice, and represent a practically popular variant of toric codes. By
construction, surface codes encode a single logical qubit.  For this reason, two-qubit target logicals
do not apply, and the single-qubit logical Phase and Hadamard gates can be constructed similar to their construction for toric codes presented above.
 
\section{Concluding remarks}
The circuit design procedures of this paper do not account for the fault tolerance property. To make them more attuned to practical implementations, we can rely on a variety of techniques known in the literature. A general idea of designing fault-tolerant circuits is to resort to the framework of pieceable fault tolerance \cite{Knill1998,Yoder2016} which consists of inserting intermediate stabilizer measurements in between physical layers of the circuit to determine/detect possible errors occurring during physical gate operations. This technique was used for HGP codes in  \cite{Quintavalle2023,malcolm2025computing} to design fault-tolerant gates. Intermediate measurements between the layers of a Clifford circuit were utilized in \cite{delfosse2023spacetime, Martiel2025} to advance the framework of Space-Time codes, which involves tracking the modification of stabilizers 
under the action of the physical gates on the state. 
Note that, since our circuits use only physical Clifford gates, throughout the layers of the circuit, the code remains a stabilizer code, and hence intermediate Pauli measurements are sufficient for detecting errors. However, it appears difficult to keep track of 
the weights of these intermediate stabilizers and logical Pauli operators, which can
cause problems for intermediate error correction. 

Whenever the HGP codes have distance 3 or less, one can replace each of the two-qubit
gates of the circuit with fault-tolerant gadgets using flag qubits \cite{chao2018fault,chen2024}.
These gadgets require a constant overhead, and so the overhead for fault tolerance is a constant factor of the number of two-qubit gates in the circuit. This includes the $[[18,2,3]]$ toric code which we
used to exemplify the general circuit design procedure of this paper.

Thinking of generalizations of our results, recent constructions of qLDPC codes \cite{PK21b,LZ2022,DHLV2022}
reach into the range of parameters superior to HGP codes, and designing logical gates for them
is therefore of significant interest. Since our approach of Sec.~\ref{sec:symplectic_framework} applies to any CSS code, it also applies to these code families. To complete the description of the logical gates, we only need to find an explicit basis for the logical Pauli operators. 
Many of the qLDPC code families are based on different kinds of code products, so if we manage to find a basis of logical Paulis for them and use it to construct symplectic matrices, this can lead to feasible circuit construction procedures. In particular, since homological product codes \cite{bravyi2014homological} are almost
equivalent to HGP codes, for them this path looks within reach.

Finally, our general result, Theorem \ref{thm:symplectic_general}, relies on a simplifying assumption that the circuit acts trivially on all the stabilizers. This assumption was adopted to simplify the presentation, but it is not essential for the implementation of logical Clifford operators. Lifting this assumption in Theorem~\ref{thm:symplectic_general} might lead to lower-depth and/or transversal circuits for HGP codes or other related code families, improving the parameters of our designs.

\section*{Acknowledgment} This research was supported in part by US NSF grant CCF-2330909.
    
	\bibliographystyle{abbrvurl}
	\bibliography{references}
\newpage
   	\appendix
		
	\section{Examples, Proofs, Extensions}\label{sec:appnx}
\subsection{Example for circuit design}\label{sec:exmpl:symplectic_circuit}
\begin{example}\label{exmpl:symplectic_circuit} (Refer to Sec.~\ref{sec: F I F})
  Let $g$ be a Clifford operator on 3 qubits and let $$F_g = \begin{bmatrix}
      1 & 1 & 0 & 0 & 0 & 1\\
      0 & 1 & 0 & 0 & 0 & 0\\
      0 & 0 & 1 & 1 & 0 & 0\\
      0 & 0 & 0 & 1 & 0 & 0\\
      0 & 1 & 0 & 1 & 1 & 0\\
      0 & 0 & 0 & 0 & 0 & 1
  \end{bmatrix}.$$ We can reduce this matrix to the identity by the following row operations:
  \begin{enumerate}
    \item Exchange rows 2 and 5.
      \item Add row 5 to row 2.
      \item Exchange rows 2 and 5. 
      \item Add row 6 to row 1 and add row 4 to row 3.
      \item Add row 2 to row 1 and add row 4 to row 5.
  \end{enumerate}
  that is
  $
  	A_4A_3A_1A_2A_1F_g = I,
  $
  where 
  {\small \begin{align*}
      A_4 = \left[\begin{array}{@{\hspace*{0in}}c*5{@{\hspace*{0.1in}}c}@{\hspace*{0in}}}
      1 & 1 & 0 & 0 & 0 & 0\\
      0 & 1 & 0 & 0 & 0 & 0\\
      0 & 0 & 1 & 0 & 0 & 0\\
      0 & 0 & 0 & 1 & 0 & 0\\
      0 & 0 & 0 & 1 & 1 & 0\\
      0 & 0 & 0 & 0 & 0 & 1
  \end{array}\right]\!, A_3 = \left[\begin{array}{@{\hspace*{0in}}c*5{@{\hspace*{0.1in}}c}@{\hspace*{0in}}}
      1 & 0 & 0 & 0 & 0 & 1\\
      0 & 1 & 0 & 0 & 0 & 0\\
      0 & 0 & 1 & 1 & 0 & 0\\
      0 & 0 & 0 & 1 & 0 & 0\\
      0 & 0 & 0 & 0 & 1 & 0\\
      0 & 0 & 0 & 0 & 0 & 1
  \end{array}\right]\!,A_2=\left[\begin{array}{@{\hspace*{0in}}c*5{@{\hspace*{0.1in}}c}@{\hspace*{0in}}}
      1 & 0 & 0 & 0 & 0 & 0\\
      0 & 1 & 0 & 0 & 1 & 0\\
      0 & 0 & 1 & 0 & 0 & 0\\
      0 & 0 & 0 & 1 & 0 & 0\\
      0 & 0 & 0 & 0 & 1 & 0\\
      0 & 0 & 0 & 0 & 0 & 1
  \end{array}\right]\!,A_1=\left[\begin{array}{@{\hspace*{0in}}c*5{@{\hspace*{0.1in}}c}@{\hspace*{0in}}}
      1 & 0 & 0 & 0 & 0 & 0\\
      0 & 0 & 0 & 0 & 1 & 0\\
      0 & 0 & 1 & 0 & 0 & 0\\
      0 & 0 & 0 & 1 & 0 & 0\\
      0 & 1 & 0 & 0 & 0 & 0\\
      0 & 0 & 0 & 0 & 0 & 1
  \end{array}\right]\!.
  \end{align*}
}
Hence, $F_g = A_1^{-1}A_2^{-1}A_1^{-1}A_3^{-1}A_4^{-1}$, and translating each of the matrices to the corresponding gate, we obtain the circuit for $g$ shown in Figure \ref{fig:symplectic_circuit}.
  \begin{center}
		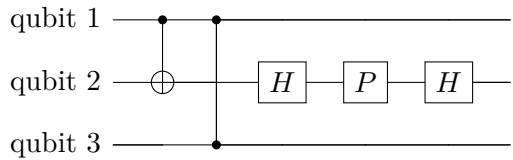
\begin{figure}[H]
			\begin{quantikz}[thin lines] 
            \lstick{qubit 1} & \qw & \qw & \qw & \ctrl{2}  & \ctrl{1} & \qw  \\
            \lstick{qubit 2} & \gate{H} & \gate{S}    &\gate{H} & \qw & \targ{} & \qw   \\
            \lstick{qubit 3 }& \qw        &\qw & \qw &\ctrl{-2} & \qw & \qw
    \end{quantikz}
			\caption{Circuit for Example \ref{exmpl:symplectic_circuit}}
			\label{fig:symplectic_circuit}
		\end{figure}
	\end{center}
\begin{figure}[H]

\end{figure}
    
 \end{example}

    \subsection{Omitted steps in the proof of Theorem \ref{thm:circuit_P}}
    \subsubsection{Correctness proof of row operations in Theorem 4.5.}\label{appndx:phase_rowop}
    Recall that we want to show that for any vector $v \in \ffF_2^{\tilde{n}}$, performing the following sequence of row operations:
    \begin{enumerate}
    	\item Choose $x \in \Supp(v)$. For every $y \in \Supp(v)\setminus\{x\}$, add row $x$ to row $y$ and add row $(y+n)$  to row $(x+n)$.
    	\item Add row $(x+n)$ to row $x$.
    	\item For every $y \in \Supp(v)\setminus\{x\}$, add row $x$ to row $y$ and add row $(y+n)$  to row $(x+n)$.
    \end{enumerate}
    on $I_{2n \times 2n}$ gives rise to the symplectic matrix:
    $$F = \begin{bmatrix}
    	I_{\tilde{n}\times \tilde{n}} & 0_{\tilde{n} \times \tilde{m}} & v^{\intercal}v & 0_{\tilde{n} \times \tilde{m}}\\
    	0_{\tilde{m} \times \tilde{n}} & I_{\tilde{m} \times \tilde{m}} & 0_{\tilde{m} \times \tilde{n}} & 0_{\tilde{m} \times \tilde{m}}\\
    	0_{\tilde{n}\times \tilde{n}} & 0_{\tilde{n} \times \tilde{m}} &I_{\tilde{n}\times \tilde{n}} & 0_{\tilde{n} \times \tilde{m}}\\
    	0_{\tilde{m} \times \tilde{n}} & 0_{\tilde{m} \times \tilde{m}} & 0_{\tilde{m}\times\tilde{n}} & I_{\tilde{m}\times\tilde{m}}
    \end{bmatrix}.$$
    Since $\Supp(v) \subset [\tilde{n}]$, i.e., we are operating only on the qubits of the left sector, for notational simplicity it is convenient to restrict $F$ to these qubits only. The resulting submatrix is:
    $$
    F = \begin{bmatrix}
    	I_{\tilde{n}\times\tilde{n}} & v^{\intercal}v \\
    	 0_{\tilde{n}\times \tilde{n}} & I_{\tilde{n}\times\tilde{n}} 
    \end{bmatrix}.
    $$
    Let $\bar{v} = v+e_x$ and consider the following two $2\tilde n\times 2\tilde n$ matrices:
    \begin{align*}
    F_1 = \begin{bmatrix}
    	I_{\tilde{n}\times \tilde{n}}+\bar{v}^{\intercal}e_x& 0_{\tilde{n}\times \tilde{n}}\\
    	0_{\tilde{n}\times \tilde{n}} & I_{\tilde{n}\times \tilde{n}}+e_x^{\intercal}\bar{v}		
    \end{bmatrix},\hspace*{0.2in}F_2 = \begin{bmatrix}
    	I_{\tilde{n}\times \tilde{n}}& e_x^{\intercal}e_x\\
    	0_{\tilde{n}\times \tilde{n}} & I_{\tilde{n}\times \tilde{n}}			
    \end{bmatrix},
    \end{align*}
    corresponding respectively to the row operations of steps (1) and (2) above. Note that the row operations in step (3) are the same as those of step (1). We have
    \begin{align*}
    	F_1F_2F_1 &= \begin{bmatrix}
    		I_{\tilde{n}\times \tilde{n}}+\bar{v}^{\intercal}e_x& 0_{\tilde{n}\times \tilde{n}}\\
    		0_{\tilde{n}\times \tilde{n}} & I_{\tilde{n}\times \tilde{n}}+e_x^{\intercal}\bar{v}		\end{bmatrix}
            \begin{bmatrix}
    	I_{\tilde{n}\times \tilde{n}}& e_x^{\intercal}e_x\\
    	0_{\tilde{n}\times \tilde{n}} & I_{\tilde{n}\times \tilde{n}}			
    	\end{bmatrix}\begin{bmatrix}
    	I_{\tilde{n}\times \tilde{n}}+\bar{v}^{\intercal}e_x& 0_{\tilde{n}\times \tilde{n}}\\
    	0_{\tilde{n}\times \tilde{n}} & I_{\tilde{n}\times \tilde{n}}+e_x^{\intercal}\bar{v}				\end{bmatrix}\\
    	&= \begin{bmatrix}
    		I_{\tilde{n}\times \tilde{n}}+\bar{v}^{\intercal}e_x& e_x^{\intercal}e_x+\bar{v}^{\intercal}e_x\\
    		0_{\tilde{n}\times \tilde{n}} & I_{\tilde{n}\times \tilde{n}}+e_x^{\intercal}\bar{v}	
    	\end{bmatrix}\begin{bmatrix}
    	I_{\tilde{n}\times \tilde{n}}+\bar{v}^{\intercal}e_x& 0_{\tilde{n}\times \tilde{n}}\\
    	0_{\tilde{n}\times \tilde{n}} & I_{\tilde{n}\times \tilde{n}}+e_x^{\intercal}\bar{v}		
    	\end{bmatrix}\\
    	&= \begin{bmatrix}
    		I_{\tilde{n}\times \tilde{n}}+\bar{v}^{\intercal}e_x& v^{\intercal}e_x\\
    		0_{\tilde{n}\times \tilde{n}} & I_{\tilde{n}\times \tilde{n}}+e_x^{\intercal}\bar{v}	
    	\end{bmatrix}\begin{bmatrix}
    		I_{\tilde{n}\times \tilde{n}}+\bar{v}^{\intercal}e_x& 0_{\tilde{n}\times \tilde{n}}\\
    		0_{\tilde{n}\times \tilde{n}} & I_{\tilde{n}\times \tilde{n}}+e_x^{\intercal}\bar{v}	
    	\end{bmatrix}\\
    	&= \begin{bmatrix}
    		I_{\tilde{n}\times \tilde{n}} & v^{\intercal}v\\
    		0_{\tilde{n}\times \tilde{n}} & I_{\tilde{n}\times \tilde{n}}			
    	\end{bmatrix} = F
    \end{align*} 
    and this completes the proof.
    
    \subsubsection{Correctness proof of Algorithm \ref{alg:algo_logical_phase_new}}\label{appndx:phase_stab}
    For $i=2,3,4,5$, let $U_i$ be the Clifford unitary applied to perform steps $1\le j\le i$ of Algorithm \ref{alg:algo_logical_phase_new}. Recalling the notation of Section \ref{subsec:HP}, we see that $I$ is contained in column $h$ of the left sector. The $X$ stabilizer generator $S_X(h',j)$ is supported on column $h'$ of the left sector and row $j$ of the right sector, while the $Z$ stabilizer generator $S_Z(i',l)$ is supported on row $i'$ of the left sector and column $l$ of the right sector. First consider an $X$ stabilizer $S_X(h',j)$ and let $\sfX$ be its support.
\begin{itemize}\setlength\itemsep{1em}
	\item[Case I:] If $h'\ne h$, then $I \cap \sfX = \emptyset$ and hence $\sfC$ preserves $S_X(h',j)$.
	\item[Case II:] If $h'=h$, then $I$ has even overlap with $\sfX$ because $S_X(h,j)$ commutes with $\overline{Z}_{i,h,L}$. The layer of CNOT gates applies an $X$ to qubit $x$ for every $y \in I\cap\sfX$. We have two subcases:
    \begin{itemize}\setlength\itemsep{1em}
        \item[IIa]: If $x\notin I\cap \sfX$, an even number of $X$s are applied to qubit $x$ and so  
	$$U_2 S_X(h,j) U_2^{\dagger}= S_X(h,j).$$
	Therfore the Phase gate on qubit $x$ does not change the stabilizer, and it is also unaffected by the following layer of CNOTs by the same logic. The $X$ type stabilizer is also unchanged by the $X$ gate on qubit $x$. Hence 
	$$U_5 S_X(h,j) U_5^{\dagger}= S_X(h,j).$$
    \item[IIb:] If, on the other hand, $x\in I\cap \sfX$, an odd number of $X$s are applied to qubit $r$ and so 
	 $$U_2 S_X(h,j) U_2^{\dagger}= \bigotimes_{r\in I \cap\sfX\setminus\{x\}}X_r\bigotimes_{r\in \sfX\setminus I}X_r.$$
	Therefore the Phase gate on qubit $x$ does not change the stabilizer further, and since the following layer of CNOTs again applies an odd number $X$s to qubit $x$, we have
	$$U_4 S_X(h,j) U_4^{\dagger}= S_X(h,j).$$
	Finally the stabilizer is unchanged by the last $X$ operator on qubit $x$. 
    \end{itemize}	
\end{itemize}
Hence any $X$ stabilizer does not gain a phase due to the series of CZ and phase gates. 

Next consider the $Z$ stabilizer $S_Z(i',l)$ and let its support be $\sfZ$. Note that $\sfZ$ can have at most one overlap with $I$ on the qubit $(i',h)$. 
\begin{itemize}\setlength\itemsep{1em}
	\item Case I: If $x \notin I\cap \sfZ$, then the stabilizer $S_Z(i',l)$ is unaffected by the layer of CNOTs, the phase gate on qubit $x$ and the following layer of CNOTs and the $X$ gate on qubit $x$. 
	\item Case II: If $x \in I\cap \sfZ$, then we have
	$$U_2 S_Z(i',l) U_2^{\dagger}= \bigotimes_{r\in I \cup \sfZ}Z_r$$
	and so 
	$$U_3 S_Z(i',l) U_3^{\dagger}= -\bigotimes_{r\in I \cup \sfZ}Z_r$$
	due to the action of the single Phase gate on qubit $x$. Due to the next layer of CNOTs, we have
	$$U_4 S_Z(i',l) U_4^{\dagger}= -\bigotimes_{r\in \sfZ}Z_r$$
	and finally due to the $X$ gate on qubit $x$
	$$U_5 S_Z(i',l) U_5^{\dagger}= S_Z(i',l)$$
\end{itemize}
Hence the circuit in Algorithm \ref{alg:algo_logical_phase_new} preserves all the stabilizers with phases.    

    \subsection{An alternative proof for Theorem \ref{thm:circuit_P}}\label{appndx:alt_proof}
    Let $v = (a_i \otimes e_h)$. We shall construct the matrix 
		\begin{equation}
	F = \begin{bmatrix}
		I_{\tilde{n}\times \tilde{n}} & 0_{\tilde{n} \times \tilde{m}} & v^{\intercal}v & 0_{\tilde{n} \times \tilde{m}}\\
		0_{\tilde{m} \times \tilde{n}} & I_{\tilde{m} \times \tilde{m}} & 0_{\tilde{m} \times \tilde{n}} & 0_{\tilde{m} \times \tilde{m}}\\
		0_{\tilde{n}\times \tilde{n}} & 0_{\tilde{n} \times \tilde{m}} &I_{\tilde{n}\times \tilde{n}} & 0_{\tilde{n} \times \tilde{m}}\\
		0_{\tilde{m} \times \tilde{n}} & 0_{\tilde{m} \times \tilde{m}} & 0_{\tilde{m}\times\tilde{n}} & I_{\tilde{m}\times\tilde{m}}
               \end{bmatrix}
		\end{equation}
by performing the following row operations on the matrix $I_{2n\times 2n}$:
	\begin{enumerate}
		\item For every $x,y \in \Supp(v),x <y$, add row $(x+n)$ to row $y$ and add row $(y+n)$  to row $x$.
        \item For every $x \in \Supp(v)$, add row $(x+n)$ to row $x$.
	\end{enumerate}
	To see that these operations on $I_{2n \times 2n}$ produce the matrix $F$, let $M = v^{\intercal}v$ and note that it can be written
     as 
	    $$
    M = \sum_{r \in \Supp(v)}\sum_{s \in \Supp(v)\setminus\{r\}} e_r^{\intercal}e_s+\sum_{r \in \Supp(v)}e_r^{\intercal}e_r.
        $$ 
	It is easy to see that the row operations in Step (1) above produce the first part of the sum while the row operations in Step (2) produce the second part of the sum.
	
The next step is to translate these row operations to physical gates using the operations in Table \ref{table:circuit_algo}. Let $I$ denote the 
physical qubit indices corresponding to the support of the $\overline Z_{i,h,L}$ logical Pauli. Step (1) of row operations translates to 
applying CZ gates between every pair of qubits in $I$ and step (2) translates to applying a Phase gate to each qubit in $I$. Note that in the 
circuit realization, these steps have to be applied in the opposite order, i.e., the Phase gates of step (2) will precede the CZ gates of step 
(1) in the circuit (cf.~\eqref{eq:symp-decomposition}). Since the circuit realizes a Clifford unitary having $F$ as its symplectic 
representation, the circuit is guaranteed to fix the support of the stabilizer generators and perform the desired transformations on the 
logical Pauli space. However, the circuit can still introduce undesired phase changes to some stabilizers, which may change the codespace. We 
show next that adding a layer of $X$ gates to the qubits in $I$ suffices to preserve the stabilizer space. 

Summarizing, the circuit building procedure can be written as in Algorithm \ref{alg:algo_logical_phase}.

\begin{algorithm}
\caption{Logical Phase gate circuit design algorithm}
\label{alg:algo_logical_phase}
\hspace*{-5.5in}\textbf{Input:} $I$
\begin{algorithmic}[1]
\State \textbf{Phase layer}
\Statex \hspace{\algorithmicindent}  For {$s \in  I$}, apply {Phase on qubit s}
\State \textbf{$CZ$ layer}
\Statex \hspace{\algorithmicindent}  For $r,s \in  I,s < r$, apply $CZ$ between $s$ and $r$
\State \textbf{$X$ layer}
\Statex \hspace{\algorithmicindent}  For $r,s \in  I,s < r$, apply {$X$ on qubit s}
\end{algorithmic}
\end{algorithm}

For $i=1,2,3$, let $U_i$ be the Clifford unitary applied to perform steps $1\le j\le i$ of the Algorithm \ref{alg:algo_logical_phase}. Recalling the notation of Section \ref{subsec:HP}, we see that $I$ is contained in column $h$ of the left sector. The $X$ stabilizer generator $S_X(h',j)$ is supported on column $h'$ of the left sector and row $j$ of the right sector, while the $Z$ stabilizer generator $S_Z(i',l)$ is supported on row $i'$ of the left sector and column $l$ of the right sector. First consider an $X$ stabilizer $S_X(h',j)$ and let $\sfX$ be its support.
\begin{enumerate}
    \item[Case I:] If $h'\ne h$, then $I \cap \sfX = \emptyset$ and hence $\sfC$ preserves $S_X(h',j)$.
    \item[Case II:] If $h'=h$, then $I$ has even overlap with $\sfX$ because $S_X(h,j)$ commutes with $\overline{Z}_{i,h,L}$. The layer of Phase gates in step (1) of \ref{alg:algo_logical_phase} maps every $X$ on qubits in $I \cap \sfX$ to $Y$ and so 
    $$
    U_1 S_X(h,j) U_1^{\dagger}= \bigotimes_{r \in I \cap \sfX}Y_r \bigotimes_{r \in \sfX\setminus I}X_r.
    $$
    Since there is a CZ gate between every pair of $I \cap \sfX$ due to step (2) of Algorithm \ref{alg:algo_logical_phase}, every qubit in $I \cap 
    \sfX$ is operated on by an odd number of CZ gates, which maps $Y_r$ to $X_r$ for $r \in I \cap \sfX$, and so
    $$
    U_2 S_X(h,j) U_2^{\dagger}= S_X(h,j).
    $$
\end{enumerate}
Hence any $X$ stabilizer does not gain a phase due to the series of CZ and phase gates. 

Next consider the $Z$ stabilizer $S_Z(i',l)$ and let its support be $\sfZ$. Note that $\sfZ$ can have at most one overlap with $I$ on the qubit $(i',h)$. Since a single Phase gate changes $Z$ to $-Z$, if $I \cap \sfZ$ is non-empty, i.e., contains the qubit $(i',h)$, then
$$U_1 S_Z(i',l)U_1^{\dagger}= \bigotimes_{r \in I \cap \sfZ}(-Z_r)\bigotimes_{r \in \sfZ\setminus I}Z_r = -S_Z(i',l),$$ and we have
$$U_2S_Z(i',l) U_2^{\dagger}= (-1)^{|I \cap \sfZ|}S_Z(i',l)$$
because CZ gates do not change $Z$ type Paulis. 
This sign is reversed by the layer of $X$ gates on qubits in $I$, implying
$$U_3 S_Z(i',l)U_3^{\dagger}= S_Z(i',l).$$
Hence the circuit in Algorithm \ref{alg:algo_logical_phase} preserves all the stabilizers with phases.    

Finally we check the parameters of the circuit $\sfC$. It only acts on the set of qubits indexed by $I$, whose size equals the Hamming weight of $|a_i|$, hence $\chi = |a_i|$. 
For the depth, note that although there are $\binom{|a_i|}{2} = \frac{|a_i|(|a_i|-1)}{2}$ CZ gates, we can apply $\frac{|a_i|}{2}$ CZ gates (or $\frac{|a_i|-1}{2}$ CZ gates if $|a_i|$ is odd) with non-overlapping supports in parallel in a single layer. Since the Phase gates and $X$ gates add 2 layers, the depth $\delta \le |a_i|+2$. Since $\{a_i\}$ is an SLT basis of $\ker(H_a)$ then by Theorem \ref{thm:logical_paulis_hp}, $|a_i| = \Theta(\sqrt{n})$. Hence we conclude that $\chi = \Theta(\sqrt{n}), \delta = \Theta(\sqrt{n})$.
   \subsection{Omitted steps in the proof of Theorem \ref{thm:circuit_H}} 
	\subsubsection{Correctness proof of row operations in Theorem \ref{thm:circuit_H}}\label{appndx:hadamard_rowop}
	Recall that we want to show that for two vectors $u,v \in \ffF_2^{\tilde{n}}$, $\Supp(u)\cap \Supp(v) = \{\rho\}$, performing the following sequence of row operations:
	\begin{enumerate}
		\item For each $x \in \Supp(v)\setminus \{\rho\}$, add row $x+n$ to row $\rho$ and add row $\rho+n$ to row $x$.
		\item For each $y \in \Supp(u)$, exchange rwo $y$ and row $y+n$.
		\item For each $y \in \Supp(u)\setminus \{\rho\}$, add row $y+n$ to row $\rho$ and add row $\rho+n$ to row $y$.
		\item For each $y \in \Supp(u)$, exchange rwo $y$ with row $y+n$.
		\item For each $y \in \Supp(u)\setminus \{\rho\}$, add row $y$ to row $\rho$ and add row $\rho+n$ to row $y+n$.
		\item For each $x \in \Supp(v)\setminus \{\rho\}$, add row $\rho$ to row $x$ and add row $x+n$ to row $\rho+n$.
		\item Exchange row $\rho$ and row $\rho+n$.
	\end{enumerate}
	 on $I_{2n \times 2n}$ gives rise to the symplectic matrix
	 $$F = \begin{bmatrix}
	 	I_{\tilde{n}\times \tilde{n}}+v^{\intercal}u & 0_{\tilde{n} \times \tilde{m}} & v^{\intercal}v & 0_{\tilde{n} \times \tilde{m}}\\
	 	0_{\tilde{m} \times \tilde{n}} & I_{\tilde{m} \times \tilde{m}} & 0_{\tilde{m} \times \tilde{n}} & 0_{\tilde{m} \times \tilde{m}}\\
	 	u^{\intercal}u & 0_{\tilde{n} \times \tilde{m}} &I_{\tilde{n}\times \tilde{n}}+u^{\intercal}v & 0_{\tilde{n} \times \tilde{m}}\\
	 	0_{\tilde{m} \times \tilde{n}} & 0_{\tilde{m} \times \tilde{m}} & 0_{\tilde{m}\times\tilde{n}} & I_{\tilde{m}\times\tilde{m}}
	 \end{bmatrix}.$$ Since $\Supp(v),\Supp(u) \subset [\tilde{n}]$, i.e., we are operating only on the qubits of the left sector, for notational simplicity it would be convenient to restrict our symplectic matrix to these qubits only. The restricted matrix is:
	$$F = \begin{bmatrix}
		I_{\tilde{n}\times\tilde{n}}+ v^{\intercal}u  & v^{\intercal}v \\
		u^{\intercal}u & I_{\tilde{n}\times\tilde{n}} + u^{\intercal}v 
	\end{bmatrix}.$$
	Let $\bar{u} = u+e_{\rho}, \bar{v} = v+e_{\rho}$ and note that $\bar{u}e_{\rho}^{\intercal} = \bar{v}e_{\rho}^{\intercal}= \bar{u}\bar{v}^{\intercal}=0$. Consider the following matrices in $Sp(2\tilde{n},\ffF_2)$:
	\begin{align*}
    F_1 = \begin{bmatrix}
	I_{\tilde{n}\times \tilde{n}} & e_{\rho}^{\intercal}\bar{v}+\bar{v}^{\intercal}e_{\rho}\\
	0_{\tilde{n}\times \tilde{n}} & I_{\tilde{n}\times \tilde{n}}		
	\end{bmatrix}
	,\hspace*{0.2in}F_2 = \begin{bmatrix}
	I_{\tilde{n}\times \tilde{n}} & 0_{\tilde{n}\times \tilde{n}}\\
	\bar{u}^{\intercal}e_{\rho}+e_{\rho}^{\intercal}\bar{u} & I_{\tilde{n}\times \tilde{n}}				
	\end{bmatrix},\hspace*{0.2in}F_3 = \begin{bmatrix}
		I_{\tilde{n}\times \tilde{n}}+e_{\rho}^{\intercal}\bar{u} & 0_{\tilde{n}\times \tilde{n}}\\
		0_{\tilde{n}\times \tilde{n}} & I_{\tilde{n}\times \tilde{n}}+\bar{u}^{\intercal}e_{\rho}				\end{bmatrix}
	\end{align*}
	\begin{align*}
	F_4 = \begin{bmatrix}
	I_{\tilde{n}\times \tilde{n}}+\bar{v}^{\intercal}e_{\rho} & 0_{\tilde{n}\times \tilde{n}}\\
	0_{\tilde{n}\times \tilde{n}} & I_{\tilde{n}\times \tilde{n}}+e_{\rho}^{\intercal}\bar{v}			
	\end{bmatrix},\hspace*{0.2in}F_5 = \begin{bmatrix}
		I_{\tilde{n}\times \tilde{n}}+e_{\rho}^{\intercal}e_{\rho} & e_{\rho}^{\intercal}e_{\rho}\\
		e_{\rho}^{\intercal}e_{\rho} & I_{\tilde{n}\times \tilde{n}}+e_{\rho}^{\intercal}e_{\rho}
	\end{bmatrix}
	\end{align*}
	corresponding respectively to the row operations of step (1), steps (2-4), step (5), step (6) and step (7). 
	We claim that $F = F_5F_4F_3F_2F_1$. We have:
	\begin{align*}
		F_4F_3 & = \begin{bmatrix}
			I_{\tilde{n}\times \tilde{n}}+\bar{v}^{\intercal}e_{\rho} & 0_{\tilde{n}\times \tilde{n}}\\
			0_{\tilde{n}\times \tilde{n}} & I_{\tilde{n}\times \tilde{n}}+e_{\rho}^{\intercal}\bar{v}			
		\end{bmatrix}\begin{bmatrix}
			I_{\tilde{n}\times \tilde{n}}+e_{\rho}^{\intercal}\bar{u} & 0_{\tilde{n}\times \tilde{n}}\\
			0_{\tilde{n}\times \tilde{n}} & I_{\tilde{n}\times \tilde{n}}+\bar{u}^{\intercal}e_{\rho}			
		\end{bmatrix}\\
		&= \begin{bmatrix}
			I_{\tilde{n}\times \tilde{n}}+\bar{v}^{\intercal}e_{\rho}+ e_{\rho}^{\intercal}\bar{u}+ \bar{v}^{\intercal}e_{\rho}e_{\rho}^{\intercal}\bar{u} & 0_{\tilde{n}\times \tilde{n}}\\
			0_{\tilde{n}\times \tilde{n}} & I_{\tilde{n}\times \tilde{n}}+e_{\rho}^{\intercal}\bar{v}+\bar{u}^{\intercal}e_{\rho}+e_{\rho}^{\intercal}\bar{v}\bar{u}^{\intercal}e_{\rho}
		\end{bmatrix}\\
		&= \begin{bmatrix}
			I_{\tilde{n}\times \tilde{n}}+\bar{v}^{\intercal}u+ e_{\rho}^{\intercal}\bar{u} & 0_{\tilde{n}\times \tilde{n}}\\
			0_{\tilde{n}\times \tilde{n}} & I_{\tilde{n}\times \tilde{n}}+e_{\rho}^{\intercal}\bar{v}+\bar{u}^{\intercal}e_{\rho}
		\end{bmatrix}\\
		&= \begin{bmatrix}
			I_{\tilde{n}\times \tilde{n}}+v^{\intercal}u + e_{\rho}^{\intercal}e_{\rho} & 0_{\tilde{n}\times \tilde{n}}\\
			0_{\tilde{n}\times \tilde{n}} & I_{\tilde{n}\times \tilde{n}}+u^{\intercal}e_{\rho} + e_{\rho}^{\intercal}v
		\end{bmatrix}
	\end{align*}	
	and so
	\begin{align*}
		F_5F_4F_3 & = \begin{bmatrix}
			I_{\tilde{n}\times \tilde{n}}+e_{\rho}^{\intercal}e_{\rho} & e_{\rho}^{\intercal}e_{\rho}\\
			e_{\rho}^{\intercal}e_{\rho} & I_{\tilde{n}\times \tilde{n}}+e_{\rho}^{\intercal}e_{\rho}
		\end{bmatrix}\begin{bmatrix}
			I_{\tilde{n}\times \tilde{n}}+v^{\intercal}u + e_{\rho}^{\intercal}e_{\rho} & 0_{\tilde{n}\times \tilde{n}}\\
			0_{\tilde{n}\times \tilde{n}} & I_{\tilde{n}\times \tilde{n}}+u^{\intercal}e_{\rho} + e_{\rho}^{\intercal}v
		\end{bmatrix}\\
		&=\begin{bmatrix}
			I_{\tilde{n}\times \tilde{n}}+v^{\intercal}u + e_{\rho}^{\intercal}e_{\rho}+e_{\rho}^{\intercal}u & e_{\rho}^{\intercal}v\\
			e_{\rho}^{\intercal}u & I_{\tilde{n}\times \tilde{n}}+u^{\intercal}e_{\rho} 
		\end{bmatrix}\\
		&= \begin{bmatrix}
			I_{\tilde{n}\times \tilde{n}}+v^{\intercal}u + e_{\rho}^{\intercal}\bar{u} & e_{\rho}^{\intercal}v\\
			e_{\rho}^{\intercal}u & I_{\tilde{n}\times \tilde{n}}+u^{\intercal}e_{\rho}
		\end{bmatrix}.
	\end{align*}
	Next
	\begin{align*}
		F_2F_1 &= \begin{bmatrix}
			I_{\tilde{n}\times \tilde{n}} & 0_{\tilde{n}\times \tilde{n}}\\
			\bar{u}^{\intercal}e_{\rho}+e_{\rho}^{\intercal}\bar{u} & I_{\tilde{n}\times \tilde{n}}				
		\end{bmatrix}\begin{bmatrix}
			I_{\tilde{n}\times \tilde{n}} & e_{\rho}^{\intercal}\bar{v}+\bar{v}^{\intercal}e_{\rho}\\
			0_{\tilde{n}\times \tilde{n}} & I_{\tilde{n}\times \tilde{n}}		
		\end{bmatrix}\\
		&= \begin{bmatrix}
			I_{\tilde{n}\times \tilde{n}} & e_{\rho}^{\intercal}\bar{v}+\bar{v}^{\intercal}e_{\rho}\\
			\bar{u}^{\intercal}e_{\rho}+e_{\rho}^{\intercal}\bar{u} & I_{\tilde{n}\times \tilde{n}}+\bar{u}^{\intercal}e_{\rho}e_{\rho}^{\intercal}\bar{v}+e_{\rho}^{\intercal}\bar{u}e_{\rho}^{\intercal}\bar{v}+\bar{u}^{\intercal}e_{\rho}\bar{v}^{\intercal}e_{\rho}+e_{\rho}^{\intercal}\bar{u}\bar{v}^{\intercal}e_{\rho}
		\end{bmatrix}\\
		&= \begin{bmatrix}
			I_{\tilde{n}\times \tilde{n}} & e_{\rho}^{\intercal}\bar{v}+\bar{v}^{\intercal}e_{\rho}\\
			\bar{u}^{\intercal}e_{\rho}+e_{\rho}^{\intercal}\bar{u} & I_{\tilde{n}\times \tilde{n}}+\bar{u}^{\intercal}\bar{v}
		\end{bmatrix}.
	\end{align*}
	Finally (note the order of the factors, corresponding to \eqref{eq:symp-decomposition}),
	\begin{align*}
		F_5F_4F_3F_2F_1 &= \begin{bmatrix}
			I_{\tilde{n}\times \tilde{n}}+v^{\intercal}u + e_{\rho}^{\intercal}\bar{u} & e_{\rho}^{\intercal}v\\
			e_{\rho}^{\intercal}u & I_{\tilde{n}\times \tilde{n}}+u^{\intercal}e_{\rho}
		\end{bmatrix}\begin{bmatrix}
			I_{\tilde{n}\times \tilde{n}} & e_{\rho}^{\intercal}\bar{v}+\bar{v}^{\intercal}e_{\rho}\\
			\bar{u}^{\intercal}e_{\rho}+e_{\rho}^{\intercal}\bar{u} & I_{\tilde{n}\times \tilde{n}}+\bar{u}^{\intercal}\bar{v}
		\end{bmatrix}\\
		&= \begin{bmatrix}
			I_{\tilde{n}\times \tilde{n}}+v^{\intercal}u + e_{\rho}^{\intercal}\bar{u}+ e_{\rho}^{\intercal}\bar{u} & e_{\rho}^{\intercal}\bar{v}+\bar{v}^{\intercal}e_{\rho}+v^{\intercal}\bar{v}+e_{\rho}^{\intercal}v\\
			e_{\rho}^{\intercal}u+ \bar{u}^{\intercal}e_{\rho}+e_{\rho}^{\intercal}\bar{u}+u^{\intercal}\bar{u} &  
			e_{\rho}^{\intercal}\bar{v}+I_{\tilde{n}\times \tilde{n}}+ u^{\intercal}e_{\rho}+\bar{u}^{\intercal}\bar{v}
		\end{bmatrix}\\
		&= \begin{bmatrix}
			I_{\tilde{n}\times\tilde{n}}+ v^{\intercal}u  & v^{\intercal}v \\
			u^{\intercal}u & I_{\tilde{n}\times\tilde{n}} + u^{\intercal}v 
		\end{bmatrix} = F
	\end{align*}
    \subsubsection{Correctness proof of Algorithm \ref{alg:algo_logical_hadamard}}\label{appndx:hadamard_stab}
     As per the notation of Sec. \ref{subsec:HP}, first note that $I$ is a subset of qubits in row $i$ while $J$ is a subset of qubits in column $h$ of the left sector and $I \cap J = \{\rho\}.$ For $i=1,2,\dots,7$, let $U_i$ be the Clifford unitary applied to perform steps $1\le j\le i$ of Algorithm \ref{alg:algo_logical_hadamard}. Recall that an $X$ ($Z$) stabilizer generator is supported on a column (row) of the left sector and a row (column) of the right sector.  Let us consider the following cases for an $X$ stabilizer $S_X(h',j)$ with $\sfX$ denoting its support:
\begin{itemize}
    \item[Case I:] Suppose that $h'\ne h$ and note that $I \cap \sfX$ is either empty or contains the physical qubit $(i,h')$. In the former case, $S_X(h',j)$ is unaffected by the circuit. If, on the other hand, qubit $(i,h')$ is in $I \cap \sfX$, then the $X$ Pauli on that qubit is acted upon by a single CNOT gate (as the target qubit) and a single CZ gate, none of which introduce any 
    global sign changes.
    
    \item[Case II:] Now consider an $X$ stabilizer generator $S_X(h,j)$. Due to the fact that the $X$ type stabilizer commutes with the logical Pauli $Z$ operator, we know that $|J \cap \sfX|$ will be even. Here two possible subcases can happen.
    \begin{itemize}
        \item[II.a:] First suppose that $\rho \notin \sfX$. Due to the even overlap with set $J$, an even number of CNOT gates have their control qubit in $J \cap \sfX$ in Step (3) and an even number of CZ gates have one of their qubits in $J \cap \sfX$ in Step (7) of Algorithm \ref{alg:algo_logical_hadamard}. Hence overall phase does not change.
        \item[II.b:] Now let $\rho \in \sfX.$ We track the modification of the stabilizer via the conjugation relation $U_{i}S_X(h,j)U_{i}^{\dagger}$. First, due to the $H$ gate on qubit $\rho$, we have
        $$U_1S_X(h,j)U_1^{\dagger} = \bigotimes_{r \in \sfX\setminus\{\rho\}}X_r \otimes Z_{\rho}.$$
            After this, the $Z$ Pauli on qubit $\rho$ propagates to all qubits in $J\setminus\{\rho\}$ and the $X$ Paulis on qubits in $J\cap \sfX \setminus\{\rho\}$ propagate to qubit $\rho$. Since $|J\cap \sfX|$ is even, we have an overall phase change
            $$U_2S_X(h,j)U_2^{\dagger} = -\bigotimes_{r \in \sfX\setminus J}X_r\bigotimes_{r \in J\setminus\sfX}Z_r \bigotimes_{r \in \sfX\cap J}Y_r.$$
            In the next stage of CNOT gates, the $Y$ Pauli on qubit $\rho$ propagates an $X$ to every qubit in $I\setminus\{\rho\}$.
            $$U_3S_X(h,j)U_3^{\dagger} = -\bigotimes_{r \in \sfX\cup I\setminus J }X_r\bigotimes_{r \in J\setminus\sfX}Z_r \bigotimes_{r \in \sfX\cap J}Y_r.$$
            Next, due to the Hadamard gates, the $Y$ Pauli on qubit $\rho$ gets a phase change and the $X$ Paulis on qubits in $I\setminus\{\rho\}$ are transformed to $Z$ Paulis. The layer of CZ gates cancels the $Z$ Paulis on qubits in $I\setminus\{\rho\}$. Finally the $Y$ Pauli on qubit $\rho$ acquires another phase change due to the second Hadamard gate. Combining all these changes, we have
            $$
            U_6S_X(h,j)U_6^{\dagger} = -\bigotimes_{r \in \sfX\setminus J }X_r\bigotimes_{r \in J\setminus\sfX}Z_r \bigotimes_{r \in \sfX\cap J}Y_r.
            $$
            After this, the layer of CZ gates introduces a $Z$ Pauli on each qubit in $J\setminus\{\rho\}$ due to the $Y$ operator on qubit $\rho$. This does not introduce any phase change:
            $$U_7S_X(h,j)U_7^{\dagger} = -\bigotimes_{r \in \sfX}X_r = -S_X(h,j).$$
       		Finally the $Y$ gate on qubit $\rho$ maps $-S_X(h,j)$ back to $S_X(h,j)$.
    \end{itemize}
\end{itemize} 
A similar analysis on the $Z$ stabilizer generators reveals that $$U_7S_Z(i',l)U_7^{\dagger} =  -S_Z(i',l)$$ if and only if $i=i'$ and the support of $S_Z(i,l)$ contains qubit $\rho$. The $Y$ gate on qubit $\rho$ maps $-S_Z(i,l)$ back to $S_Z(i,l)$. In conclusion, the circuit $\sfC$, built using Algorithm \ref{alg:algo_logical_hadamard}, preserves
the stabilizers together with their global phases.

    \subsection{Correctness proof of row operations in Theorem \ref{thm:circuit_cnot}}\label{appndx:cnot}
    For $v \in \ffF_2^{\tilde{n}}, u \in \ffF_2^{\tilde{m}}$, we wish to show that performing the following sequence of row operations:
	\begin{enumerate}
		\item Select a row index $x \in \Supp(v)$. For all $y \in \Supp(v)\setminus\{x\}$, add row $x$ to row $y$ and add row $y+n$ to row $x+n$.
		\item For all $\bar{y} \in \Supp(u)$, add row $\bar{y} + \tilde{n}$ to row $x$ and add row $x+n$ to row $\bar{y} + \tilde{n} +n$.
		\item For all $y \in \Supp(v)\setminus\{x\}$, add row $x$ to row $y$ and add row $y+n$ to row $x+n$.
	\end{enumerate}
	 on $I_{2n \times 2n}$ gives rise to the symplectic matrix
	 $$F = \begin{bmatrix}
			I_{\tilde{n}\times \tilde{n}} & v^{\intercal}u & 0_{\tilde{n} \times \tilde{n}} & 0_{\tilde{n} \times \tilde{m}}\\
			0_{\tilde{m} \times \tilde{n}} & I_{\tilde{m} \times \tilde{m}} & 0_{\tilde{m} \times \tilde{n}} & 0_{\tilde{m} \times \tilde{m}}\\
			0_{\tilde{n} \times \tilde{n}} & 0_{\tilde{n} \times \tilde{m}} & I_{\tilde{n}\times \tilde{n}} & 0_{\tilde{n} \times \tilde{m}}\\
			0_{\tilde{m} \times \tilde{n}} & 0_{\tilde{m} \times \tilde{m}} & u^{\intercal}v & I_{\tilde{m}\times\tilde{m}}
		\end{bmatrix}.$$
        Fix an $x$ in $\Supp(v)$ and let $e_x$ be the length $\tilde{n}$ unit vector, let $\bar{v} = v+e_x$. Consider the following symplectic matrices corresponding to the row operations in step (1) and (2) respectively (note that step (3) is the same as step (1))
        \begin{align*}
            F_1 = \begin{bmatrix}
			I_{\tilde{n}\times \tilde{n}}+\bar{v}^{\intercal}e_{x} & 0_{\tilde{n}\times \tilde{{m}}} & 0_{\tilde{n} \times \tilde{n}} & 0_{\tilde{n} \times \tilde{m}}\\
			0_{\tilde{m} \times \tilde{n}} & I_{\tilde{m} \times \tilde{m}} & 0_{\tilde{m} \times \tilde{n}} & 0_{\tilde{m} \times \tilde{m}}\\
			0_{\tilde{n} \times \tilde{n}} & 0_{\tilde{n} \times \tilde{m}} & I_{\tilde{n}\times \tilde{n}}+e_{x}^{\intercal}\bar{v} & 0_{\tilde{n} \times \tilde{m}}\\
			0_{\tilde{m} \times \tilde{n}} & 0_{\tilde{m} \times \tilde{m}} & 0_{\tilde{m}\times \tilde{n}} & I_{\tilde{m}\times\tilde{m}}
		\end{bmatrix}, F_2 = \begin{bmatrix}
			I_{\tilde{n}\times \tilde{n}} & e_x^{\intercal}u & 0_{\tilde{n} \times \tilde{n}} & 0_{\tilde{n} \times \tilde{m}}\\
			0_{\tilde{m} \times \tilde{n}} & I_{\tilde{m} \times \tilde{m}} & 0_{\tilde{m} \times \tilde{n}} & 0_{\tilde{m} \times \tilde{m}}\\
			0_{\tilde{n} \times \tilde{n}} & 0_{\tilde{n} \times \tilde{m}} & I_{\tilde{n}\times \tilde{n}} & 0_{\tilde{n} \times \tilde{m}}\\
			0_{\tilde{m} \times \tilde{n}} & 0_{\tilde{m} \times \tilde{m}} & u^{\intercal}e_x & I_{\tilde{m}\times\tilde{m}}
		\end{bmatrix}.
        \end{align*}
        We have
        \begin{align*}
            F_2F_1 &= \begin{bmatrix}
			I_{\tilde{n}\times \tilde{n}} & e_x^{\intercal}u & 0_{\tilde{n} \times \tilde{n}} & 0_{\tilde{n} \times \tilde{m}}\\
			0_{\tilde{m} \times \tilde{n}} & I_{\tilde{m} \times \tilde{m}} & 0_{\tilde{m} \times \tilde{n}} & 0_{\tilde{m} \times \tilde{m}}\\
			0_{\tilde{n} \times \tilde{n}} & 0_{\tilde{n} \times \tilde{m}} & I_{\tilde{n}\times \tilde{n}} & 0_{\tilde{n} \times \tilde{m}}\\
			0_{\tilde{m} \times \tilde{n}} & 0_{\tilde{m} \times \tilde{m}} & u^{\intercal}e_x & I_{\tilde{m}\times\tilde{m}}
		\end{bmatrix}\begin{bmatrix}
			I_{\tilde{n}\times \tilde{n}}+\bar{v}^{\intercal}e_{x} & 0_{\tilde{n}\times \tilde{{m}}} & 0_{\tilde{n} \times \tilde{n}} & 0_{\tilde{n} \times \tilde{m}}\\
			0_{\tilde{m} \times \tilde{n}} & I_{\tilde{m} \times \tilde{m}} & 0_{\tilde{m} \times \tilde{n}} & 0_{\tilde{m} \times \tilde{m}}\\
			0_{\tilde{n} \times \tilde{n}} & 0_{\tilde{n} \times \tilde{m}} & I_{\tilde{n}\times \tilde{n}}+e_{x}^{\intercal}\bar{v} & 0_{\tilde{n} \times \tilde{m}}\\
			0_{\tilde{m} \times \tilde{n}} & 0_{\tilde{m} \times \tilde{m}} & 0_{\tilde{m}\times \tilde{n}} & I_{\tilde{m}\times\tilde{m}}
		\end{bmatrix}\\
        &= \begin{bmatrix}
			I_{\tilde{n}\times \tilde{n}}+\bar{v}^{\intercal}e_{x} & e_x^{\intercal}u & 0_{\tilde{n} \times \tilde{n}} & 0_{\tilde{n} \times \tilde{m}}\\
			0_{\tilde{m} \times \tilde{n}} & I_{\tilde{m} \times \tilde{m}} & 0_{\tilde{m} \times \tilde{n}} & 0_{\tilde{m} \times \tilde{m}}\\
			0_{\tilde{n} \times \tilde{n}} & 0_{\tilde{n} \times \tilde{m}} & I_{\tilde{n}\times \tilde{n}}+e_{x}^{\intercal}\bar{v} & 0_{\tilde{n} \times \tilde{m}}\\
			0_{\tilde{m} \times \tilde{n}} & 0_{\tilde{m} \times \tilde{m}} & u^{\intercal}v & I_{\tilde{m}\times\tilde{m}}
		\end{bmatrix}
        \end{align*}
         and finally
         \begin{align*}
             F_1F_2F_1 &= \begin{bmatrix}
			I_{\tilde{n}\times \tilde{n}}+\bar{v}^{\intercal}e_{x} & 0_{\tilde{n}\times \tilde{{m}}} & 0_{\tilde{n} \times \tilde{n}} & 0_{\tilde{n} \times \tilde{m}}\\
			0_{\tilde{m} \times \tilde{n}} & I_{\tilde{m} \times \tilde{m}} & 0_{\tilde{m} \times \tilde{n}} & 0_{\tilde{m} \times \tilde{m}}\\
			0_{\tilde{n} \times \tilde{n}} & 0_{\tilde{n} \times \tilde{m}} & I_{\tilde{n}\times \tilde{n}}+e_{x}^{\intercal}\bar{v} & 0_{\tilde{n} \times \tilde{m}}\\
			0_{\tilde{m} \times \tilde{n}} & 0_{\tilde{m} \times \tilde{m}} & 0_{\tilde{m}\times \tilde{n}} & I_{\tilde{m}\times\tilde{m}}
		\end{bmatrix}\\
        &\hspace{2.0in}\times\begin{bmatrix}
			I_{\tilde{n}\times \tilde{n}}+\bar{v}^{\intercal}e_{x} & e_x^{\intercal}u & 0_{\tilde{n} \times \tilde{n}} & 0_{\tilde{n} \times \tilde{m}}\\
			0_{\tilde{m} \times \tilde{n}} & I_{\tilde{m} \times \tilde{m}} & 0_{\tilde{m} \times \tilde{n}} & 0_{\tilde{m} \times \tilde{m}}\\
			0_{\tilde{n} \times \tilde{n}} & 0_{\tilde{n} \times \tilde{m}} & I_{\tilde{n}\times \tilde{n}}+e_{x}^{\intercal}\bar{v} & 0_{\tilde{n} \times \tilde{m}}\\
			0_{\tilde{m} \times \tilde{n}} & 0_{\tilde{m} \times \tilde{m}} & u^{\intercal}v & I_{\tilde{m}\times\tilde{m}}
		\end{bmatrix}\\
        &= \begin{bmatrix}
			I_{\tilde{n}\times \tilde{n}} & v^{\intercal}u & 0_{\tilde{n} \times \tilde{n}} & 0_{\tilde{n} \times \tilde{m}}\\
			0_{\tilde{m} \times \tilde{n}}  & I_{\tilde{m} \times \tilde{m}} & 0_{\tilde{m} \times \tilde{n}} & 0_{\tilde{m} \times \tilde{m}}\\
			0_{\tilde{n} \times \tilde{n}} & 0_{\tilde{n} \times \tilde{m}} & I_{\tilde{n}\times \tilde{n}} & 0_{\tilde{n} \times \tilde{m}}\\
			0_{\tilde{m} \times \tilde{n}} & 0_{\tilde{m} \times \tilde{m}} & u^{\intercal}v & I_{\tilde{m}\times\tilde{m}}
		\end{bmatrix}
         \end{align*}
        and this completes the proof. The proof for the case when both logical qubits are in the same sector is analogous and hence is not repeated.

         \subsection{Correctness proof of row operations in Theorem \ref{thm:circuit_cz}}\label{appndx:cz}
         We prove the correctness for the case when both qubits are in the same sector, as the proof of the case of different sectors is completely analogous. Let $v,u \in \ffF_2^{\tilde{n}}$ be vectors such that there exists $x \in \Supp(v)\setminus\Supp(u)$ and $y \in \Supp(u)\setminus\Supp(v)$. The target matrix is:
		$$F = \begin{bmatrix}
			I_{\tilde{n}\times \tilde{n}} & 0_{\tilde{n} \times \tilde{m}} & v^{\intercal}u+u^{\intercal}v & 0_{\tilde{n} \times \tilde{m}}\\
			0_{\tilde{m} \times \tilde{n}} & I_{\tilde{m} \times \tilde{m}} & 0_{\tilde{m} \times \tilde{n}} & 0_{\tilde{m} \times \tilde{m}}\\
			0_{\tilde{n} \times \tilde{n}} & 0_{\tilde{n} \times \tilde{m}} & I_{\tilde{n}\times \tilde{n}} & 0_{\tilde{n} \times \tilde{m}}\\
			0_{\tilde{m} \times \tilde{n}} & 0_{\tilde{m} \times \tilde{m}} & 0_{\tilde{m} \times \tilde{n}} & I_{\tilde{m}\times\tilde{m}}
		\end{bmatrix}.$$ 
		We wish to prove that the following row operations performed on $I_{2n \otimes 2n}$ in order produce the matrix $F$:
			\begin{enumerate}
			\item For all $y \in \Supp(v)\setminus\{x\}$, add row $x$ to row $y$ and add row $y+n$ to row $x+n$.
			\item For all $\bar{y} \in \Supp(u)\setminus\{y\}$, add row $y$ to row $\bar{y}$ and add row $\bar{y}+n$ to row $y$.
			\item Add row $x+n$ to row $y$ and add row $y+n$ to row $x$. 
			\item For all $y \in \Supp(v)\setminus\{x\}$, add row $x$ to row $y$ and add row $y+n$ to row $x+n$.
			\item For all $\bar{y} \in \Supp(u)\setminus\{y\}$, add row $y$ to row $\bar{y}$ and add row $\bar{y}+n$ to row $y$.
		\end{enumerate}
        Since the operations are completely contained within rows $\{1,\cdots,\tilde{n}\}$ and $\{n+1,\cdots,n+\tilde{n}\}$, for notational simplicity we consider the restricted matrix
        $$F = \begin{bmatrix}
			I_{\tilde{n}\times \tilde{n}} & v^{\intercal}u+u^{\intercal}v \\
			0_{\tilde{n} \times \tilde{n}} & I_{\tilde{n}\times \tilde{n}} 
		\end{bmatrix}.$$
        Let $\bar{v} = v+e_x, \bar{u} = u+e_y$ and note that $e_yv^{\intercal} = e_xu^{\intercal}=0$. Consider the following matrices in $Sp(2\tilde{n},\ffF_2)$ corresponding to the row operations in step (1), (2) and (3) respectively, noting that steps (4) and (5) are the same as steps (1) and (2),
        \begin{align*}
            &F_1 = \begin{bmatrix}
			I_{\tilde{n}\times \tilde{n}}+\bar{v}^{\intercal}e_x &  0_{\tilde{n} \times \tilde{n}}\\
			0_{\tilde{n} \times \tilde{n}} & I_{\tilde{n}\times \tilde{n}} +e_x^{\intercal}\bar{v}
		\end{bmatrix}, \hspace{0.2in} F_2 = \begin{bmatrix}
			I_{\tilde{n}\times \tilde{n}}+\bar{u}^{\intercal}e_y &  0_{\tilde{n} \times \tilde{n}}\\
			0_{\tilde{n} \times \tilde{n}} & I_{\tilde{n}\times \tilde{n}} +e_y^{\intercal}\bar{u}
		\end{bmatrix}, \\
        &\hspace{1.0in}F_3 = \begin{bmatrix}
			I_{\tilde{n}\times \tilde{n}} &  e_x^{\intercal}e_y+e_y^{\intercal}e_x\\
			0_{\tilde{n} \times \tilde{n}} & I_{\tilde{n}\times \tilde{n}} 
		\end{bmatrix}.
        \end{align*}
        We have 
        \begin{align*}
            F_3F_2F_1 &= \begin{bmatrix}
			I_{\tilde{n}\times \tilde{n}} &  e_x^{\intercal}e_y+e_y^{\intercal}e_x\\
			0_{\tilde{n} \times \tilde{n}} & I_{\tilde{n}\times \tilde{n}} 
		\end{bmatrix}\begin{bmatrix}
			I_{\tilde{n}\times \tilde{n}}+\bar{u}^{\intercal}e_y &  0_{\tilde{n} \times \tilde{n}}\\
			0_{\tilde{n} \times \tilde{n}} & I_{\tilde{n}\times \tilde{n}} +e_y^{\intercal}\bar{u}
		\end{bmatrix}\begin{bmatrix}
			I_{\tilde{n}\times \tilde{n}}+\bar{v}^{\intercal}e_x &  0_{\tilde{n} \times \tilde{n}}\\
			0_{\tilde{n} \times \tilde{n}} & I_{\tilde{n}\times \tilde{n}} +e_x^{\intercal}\bar{v}
		\end{bmatrix}\\
        &= \begin{bmatrix}
			I_{\tilde{n}\times \tilde{n}} &  e_x^{\intercal}e_y+e_y^{\intercal}e_x\\
			0_{\tilde{n} \times \tilde{n}} & I_{\tilde{n}\times \tilde{n}} 
		\end{bmatrix}\begin{bmatrix}
			I_{\tilde{n}\times \tilde{n}}+\bar{v}^{\intercal}e_x + \bar{u}^{\intercal}e_y &  0_{\tilde{n} \times \tilde{n}}\\
			0_{\tilde{n} \times \tilde{n}} & I_{\tilde{n}\times \tilde{n}} +e_x^{\intercal}\bar{v}+e_y^{\intercal}\bar{u}
		\end{bmatrix}\\
        &= \begin{bmatrix}
			I_{\tilde{n}\times \tilde{n}}+\bar{v}^{\intercal}e_x + \bar{u}^{\intercal}e_y &  e_x^{\intercal}e_y+e_y^{\intercal}e_x +e_x^{\intercal}\bar{u}+e_y^{\intercal}\bar{v}\\
			0_{\tilde{n} \times \tilde{n}} & I_{\tilde{n}\times \tilde{n}} +e_x^{\intercal}\bar{v}+e_y^{\intercal}\bar{u}
		\end{bmatrix}\\
        &= \begin{bmatrix}
			I_{\tilde{n}\times \tilde{n}}+\bar{v}^{\intercal}e_x + \bar{u}^{\intercal}e_y &  e_x^{\intercal}u+e_y^{\intercal}v\\
			0_{\tilde{n} \times \tilde{n}} & I_{\tilde{n}\times \tilde{n}} +e_x^{\intercal}\bar{v}+e_y^{\intercal}\bar{u}
		\end{bmatrix}.
        \end{align*}
        And finally
        \begin{align*}
            F_1F_2F_3F_2F_1 &= \begin{bmatrix}
			I_{\tilde{n}\times \tilde{n}}+\bar{v}^{\intercal}e_x &  0_{\tilde{n} \times \tilde{n}}\\
			0_{\tilde{n} \times \tilde{n}} & I_{\tilde{n}\times \tilde{n}} +e_x^{\intercal}\bar{v}
		\end{bmatrix}\begin{bmatrix}
			I_{\tilde{n}\times \tilde{n}}+\bar{u}^{\intercal}e_y &  0_{\tilde{n} \times \tilde{n}}\\
			0_{\tilde{n} \times \tilde{n}} & I_{\tilde{n}\times \tilde{n}} +e_y^{\intercal}\bar{u}
		\end{bmatrix}\\
        &\hspace{2.0in}\times\begin{bmatrix}
			I_{\tilde{n}\times \tilde{n}}+\bar{v}^{\intercal}e_x + \bar{u}^{\intercal}e_y &  e_x^{\intercal}u+e_y^{\intercal}v\\
			0_{\tilde{n} \times \tilde{n}} & I_{\tilde{n}\times \tilde{n}} +e_x^{\intercal}\bar{v}+e_y^{\intercal}\bar{u}
		\end{bmatrix}\\
        &= \begin{bmatrix}
			I_{\tilde{n}\times \tilde{n}}+\bar{v}^{\intercal}e_x + \bar{u}^{\intercal}e_y &  0_{\tilde{n} \times \tilde{n}}\\
			0_{\tilde{n} \times \tilde{n}} & I_{\tilde{n}\times \tilde{n}} +e_x^{\intercal}\bar{v}+e_y^{\intercal}\bar{u}
		\end{bmatrix}\\
        &\hspace{2.0in}\times\begin{bmatrix}
			I_{\tilde{n}\times \tilde{n}}+\bar{v}^{\intercal}e_x + \bar{u}^{\intercal}e_y &  e_x^{\intercal}u+e_y^{\intercal}v\\
			0_{\tilde{n} \times \tilde{n}} & I_{\tilde{n}\times \tilde{n}} +e_x^{\intercal}\bar{v}+e_y^{\intercal}\bar{u}
		\end{bmatrix}\\
        &= \begin{bmatrix}
			I_{\tilde{n}\times \tilde{n}} &  e_x^{\intercal}u+e_y^{\intercal}v + \bar{v}^{\intercal}u+\bar{u}^{\intercal}v\\
			0_{\tilde{n} \times \tilde{n}} & I_{\tilde{n}\times \tilde{n}} 
		\end{bmatrix}\\
        &= \begin{bmatrix}
			I_{\tilde{n}\times \tilde{n}} &  v^{\intercal}u+u^{\intercal}v\\
			0_{\tilde{n} \times \tilde{n}} & I_{\tilde{n}\times \tilde{n}} 
		\end{bmatrix}
        \end{align*}
        and this completes the proof.
        
\end{document}